\documentclass[nofootinbib,aps,11pt]{revtex4-1}
\usepackage{graphicx}
\usepackage{hyperref}
\usepackage{cancel}
\usepackage{amssymb}
\usepackage{textcomp}
\usepackage{amsmath}
\usepackage{bm}
\usepackage{times}
\usepackage{epsfig}
\usepackage{color}

\newcommand{\Br}{\textrm{BR}}
\newcommand{\fb}{\textrm{fb}}

\newcommand{\GeV}{{\rm GeV}}

\newcommand{\TeV}{{\rm TeV}}
\newcommand{\eV}{{\rm eV}}

\newcommand{\calO}{{\cal O}}

\begin{document}
\title{\LARGE LHC phenomenology of type II seesaw: nondegenerate case}
\bigskip
\author{Zhi-Long Han~$^{a}$}
\email{hanzhilong@mail.nankai.edu.cn}
\author{Ran Ding~$^{b}$}
\email{dingran@mail.nankai.edu.cn}
\author{Yi Liao~$^{a,c}$}
\email{liaoy@nankai.edu.cn}
\affiliation{
$^a$~School of Physics, Nankai University, Tianjin 300071, China
\\
$^b$ Center for High Energy Physics, Peking University, Beijing 100871, China
\\
$^c$ State Key Laboratory of Theoretical Physics, Institute of Theoretical Physics,
Chinese Academy of Sciences, Beijing 100190, China}
\date{\today}

\begin{abstract}

In this paper, we thoroughly investigate the LHC phenomenology of the type II seesaw mechanism for neutrino masses in the nondegenerate case where the triplet scalars of various charge ($H^{\pm\pm},~H^\pm,~H^0,~A^0$) have different masses. Compared with the degenerate case, the cascade decays of scalars lead to many new, interesting signal channels. In the positive scenario where $M_{H^{\pm\pm}}<M_{H^\pm}<M_{H^0/A^0}$, the four-lepton signal is still the most promising discovery channel for the doubly-charged scalars $H^{\pm\pm}$. The five-lepton signal is crucial to probe the mass spectrum of the scalars, for which, for example, a $5\sigma$ reach at $14~\TeV$ LHC for $M_{H^{\pm}}=430~\GeV$ with $M_{H^{\pm\pm}}=400~\GeV$ requires an integrated luminosity of $76~\fb^{-1}$. And the six-lepton signal can be used to probe the neutral scalars $H^0/A^0$, which are usually hard to detect in the degenerate case. In the negative scenario where $M_{H^{\pm\pm}}>M_{H^\pm}>M_{H^0/A^0}$, the detection of $H^{\pm\pm}$ is more challenging, when the cascade decay $H^{\pm\pm}\to H^{\pm}W^{\pm*}$ is dominant. The most important channel is the associated $H^{\pm}H^0/A^0$ production in the final state $\ell^\pm\cancel{E}_Tb\bar{b}b\bar{b}$, which requires a luminosity of $109~\fb^{-1}$ for a $5\sigma$ discovery, while the final state $\ell^\pm\cancel{E}_Tb\bar{b}\tau^+\tau^-$ is less promising. Moreover, the associated $H^0A^0$ production can give same signals as the standard model Higgs pair production. With a much larger cross section, the $H^0A^0$ production in the final state $b\bar{b}\tau^+\tau^-$ could reach $3\sigma$ significance at 14 TeV LHC with a luminosity of $300~\fb^{-1}$. In summary, with an integrated luminosity $\sim\mathcal{O}(500~\fb^{-1})$, the triplet scalars can be fully reconstructed at 14 TeV LHC in the negative scenario.

\end{abstract}

\maketitle

\section{Introduction}

When the standard model (SM) is considered as a low energy effective field theory, the tiny neutrino masses can be incorporated by higher dimensional operators that are suppressed by new physics above the electroweak scale. Such an operator first appears at dimension five, which is the unique Weinberg operator~\cite{Weinberg:1979sa}, $\calO_5=\big(\overline{F^C_{L}}\epsilon\Phi\big)\big(\Phi^T\epsilon F_{L}\big)$, where $F_{L}$ and $\Phi$ are respectively the left-handed
leptonic and Higgs doublets in SM and $\epsilon$ is an antisymmetric matrix.
If the operator is suppressed for one reason or another, the neutrino masses would be induced by even higher dimensional operators, which are also known to be unique at each dimension,
$\calO_{5+2n}=\calO_5(\Phi^\dagger\Phi)^n$ with $n$ a positive integer \cite{Liao:2010ku}.

It is interesting that there are exactly three possible ways at tree level to realize the Weinberg operator in an underlying theory~\cite{Ma:1998dn}. They correspond to the type I~\cite{type1}, II~\cite{type2}, and III~\cite{Foot:1988aq} seesaw mechanism that introduces respectively fermionic singlets, a scalar triplet of hypercharge $2$, and fermionic triplets of hypercharge zero. Variants of the three seesaws are also suggested based on different theoretical considerations~\cite{Type1-3}. For instance, a combination of type I and III seesaws has a simple realization in the context of grand unified theories~\cite{Type1-3}, while left-right symmetric models~\cite{LR} embrace type I and II naturally. The tiny mass of neutrinos can be more readily induced if the effective operators are radiatively induced \cite{Zee-Babu} or first appear at tree level at a higher dimension \cite{Babu:2001ex}.

To discern the underlying physics that is responsible for neutrino masses, it is vital to produce directly the relevant heavy particles at colliders and study their properties \cite{Han:2006ip,
Han:2007bk,Chiang:2012dk,Perez:2008zc,Perez:2008ha,Garayoa:2007fw,Akeroyd:2007zv,
Franceschini:2008pz,
delAguila:2008cj,Arhrib:2009mz,Han:2012vk,Dev:2013oxa,FileviezPerez:2010ch,Ding:2014nga}. In this paper, we will study the LHC phenomenology of the type II seesaw mechanism. We recall in this circumstance that the previous literature mainly concentrates on the simplified version of it, assuming that the triplet scalars are degenerate, and on the search of the doubly charged scalars $H^{\pm\pm}$~\cite{Gunion:1996pq,Chun:2003ej,Han:2007bk,Perez:2008ha,delAguila:2008cj,Akeroyd:2010ip}. Assuming degeneracy, $H^{\pm\pm}$ have only two decay modes: the lepton number violating (LNV) like-sign dilepton decays $H^{\pm\pm}\to\ell^\pm\ell^\pm$ and the like-sign diboson decays $H^{\pm\pm}\to W^\pm W^\pm$. Given the range of neutrino masses, their relative importance is controlled by the vacuum expectation value (vev) of the triplet, $v_{\Delta}$. The dilepton decays dominate when $v_{\Delta}<10^{-4}~\GeV$, resulting in the characteristic four-lepton signal from pair production $pp\to H^{++}H^{--}\to\ell^+\ell^+\ell^-\ell^-$~\cite{Han:2007bk,delAguila:2008cj,Akeroyd:2010ip} and also the promising three-lepton signal from associated production $pp\to H^{\pm\pm}H^{\mp}\to\ell^\pm\ell^\pm\ell^\mp\nu$ ~\cite{Perez:2008ha,delAguila:2008cj,Akeroyd:2010ip}. For $v_{\Delta}>10^{-4}~\GeV$ instead, the diboson decays are dominant, and the important signal channel is $pp\to H^{++}H^{--}\to W^+W^+W^-W^-\to\ell^\pm\ell^\pm\cancel{E}_T4j$ ~\cite{Han:2007bk} as well as associated production $pp\to H^{\pm\pm}H^{\mp}\to \ell^\pm\ell^\pm\cancel{E}_T4j$~\cite{Perez:2008ha}.

The (dominantly) triplet scalars are generally nondegenerate, and to good accuracy their mass splittings are determined by a coupling constant $\lambda_5$ in the potential, see Eq.~(\ref{Vpotential}). We discriminate two scenarios:
\begin{eqnarray}
\label{masssplitting}
\text{positive scenario }(\lambda_5>0): M_{H^{\pm\pm}}<M_{H^\pm}<M_{H^0/A^0},
\notag\\
\text{negative scenario }(\lambda_5<0): M_{H^{\pm\pm}}>M_{H^\pm}>M_{H^0/A^0}.
\end{eqnarray}
The collider signatures for the nondegenerate case have received relatively less attention. For the positive scenario, Ref.~\cite{Akeroyd:2011zza} computed the enhancement of $H^{++}H^{--}$ pair production due to cascade decays of heavier scalars $H^\pm$ and mentioned further contributions from cascade chain decays of $H^0/A^0$ production. The cascade chain decays can produce a like-sign four-lepton signal whose production rate was estimated in Refs.~\cite{Chun:2012zu,Chun:2013vma} in the narrow width approximation. The authors in Ref.~\cite{Akeroyd:2012nd} investigated the five- and six-lepton signals from cascade decays of $H^{\pm\pm}H^{\mp}$, $H^{\pm}H^0/A^0$, and $H^0A^0$. For the negative scenario, Ref.~\cite{Akeroyd:2005gt} noticed the importance of the cascade decay $H^{\pm\pm}\to H^{\pm}W^{\pm*}$ and compared its branching ratio with dilepton decays $H^{\pm\pm}\to \ell^{\pm}\ell^{\pm}$. The work~\cite{Aoki:2011pz} focused on the signal channels $pp\to H^{+}H^0\to\ell^+\cancel{E}_Tb\bar{b}b\bar{b}$ for the $H^\pm$ search and $pp\to H^{++}H^-\to\ell^+\ell^+\cancel{E}_Tjjb\bar{b}b\bar{b}$ for the $H^{\pm\pm}$ search; see \cite{Yagyu:2012qp} for more details. Nevertheless, to the best of our knowledge, a systematic and comprehensive study of LHC signatures is still lacking. The purpose of this paper is to fill the gap.

The rest of the paper is organized as follows. In Sec.~\ref{Constraints}, we set up our notations in the type-II seesaw model, and review the current experimental constraints on the model parameters. The decay properties of the triplet scalars in the nondegenerate case are investigated in Sec.~\ref{Decays}. This is then followed by the core Sec.~\ref{signal-positive} and Sec.~\ref{signal-negative}, where we study systematically the LHC signatures for the positive and negative scenarios in the nondegenerate case. We estimate the SM backgrounds and develop the strategies to separate the signals from backgrounds in each signal channel. Finally, in Sec.~\ref{Conclusions}, we present our conclusions with critical discussions.

\section{Constraints on parameters in type II seesaw}
\label{Constraints}

We will review in this section the current experimental constraints on the type II seesaw model, so that our later phenomenological analysis at LHC can be more realistic. To set the stage, we first give a concise introduction to the model. The seesaw operates with the help of a scalar triplet $\Delta$ of hypercharge $2$ in addition to the SM scalar doublet $\Phi$ of hypercharge $1$:
\begin{align}
\Phi=\left(
\begin{array}{c}
\phi^+\\
\phi^0
\end{array}\right),\quad \Delta =\left(
\begin{array}{cc}
\delta^+/\sqrt{2} & \delta^{++}\\
\delta^0 & -\delta^+/\sqrt{2}
\end{array}\right),
\end{align}
where the superscripts denote the electric charge. The most general potential is given by
\begin{eqnarray}
\label{Vpotential}
V(\Phi,\Delta)&=&
m^2\Phi^\dagger\Phi+M^2\text{Tr}(\Delta^\dagger\Delta)+\lambda_1(\Phi^\dagger\Phi)^2
+\lambda_2\left(\text{Tr}(\Delta^\dagger\Delta)\right)^2
+\lambda_3\text{Tr}(\Delta^\dagger\Delta)^2\notag\\
&&+\lambda_4(\Phi^\dagger\Phi)\text{Tr}(\Delta^\dagger\Delta)
+\lambda_5\Phi^\dagger\Delta\Delta^\dagger\Phi+\left(\mu \Phi^T i\tau^2\Delta^\dagger \Phi+\text{h.c.}\right),
\end{eqnarray}
where spontaneous symmetry breaking is triggered by assuming $m^2 < 0$ that results in a vev ($v$) for the scalar doublet, while the parameter $M^2$ is assumed to be positive to set the mass scale for the heavy scalars. The $\mu$ term is important, and deserves a few comments. It induces a vev ($v_\Delta$) for the triplet out of that for the doublet. This in turn causes mixing between the doublet and triplet scalars of equal charge. Since the $\mu$ parameter can be taken real without loss of generality, CP is preserved by the potential and mixing. Together with the Yukawa couplings between the leptons and the scalar triplet, it violates the lepton number by two units. The $\mu$ parameter and thus $v_\Delta$ are considered to be naturally small in this sense. The triplet vev causes the deviation of the $\rho$ parameter from unity at tree level, i.e., $\rho\approx 1-2v^2_\Delta/v^2$ for $v_\Delta\ll v$. The precise experimental measurement on $\rho$ then translates to a limit on $v_\Delta$, which we take safely to be $v_\Delta<1~\GeV$ in our numerical analysis.

Separating out the vev's,
\begin{align}
\phi^0 = \frac{1}{\sqrt{2}}(v + \phi + i \chi),~
\delta^0 = \frac{1}{\sqrt{2}}(v_{\Delta}+\delta+i\xi),
\end{align}
the scalars mix as follows (see for instance, \cite{Arhrib:2011uy,Aoki:2012jj})
\begin{align}
&\left(
\begin{array}{c}
\phi^{\pm}\\
\delta^{\pm}
\end{array}\right)=R(\theta_+)\left(
\begin{array}{c}
G^{\pm}\\
H^{\pm}
\end{array}\right),~
\left(
\begin{array}{c}
\chi\\
\xi
\end{array}\right)=R(\alpha)\left(
\begin{array}{c}
G^{0}\\
A^{0}
\end{array}\right),~
\left(
\begin{array}{c}
\phi\\
\delta
\end{array}\right)=R(\theta_0)\left(
\begin{array}{c}
h\\
H^0
\end{array}\right).
\end{align}
Here $R(\omega)$ is the standard rotation matrix in the plane, and the mixing angles are given by
\begin{align}
\tan \theta_+ = \frac{\sqrt{2} v_{\Delta}}{v},~
\tan \alpha = \frac{2 v_{\Delta}}{v},~
\tan 2\theta_0 =\frac{v_{\Delta}}{v} \frac{2v^2(\lambda_4+\lambda_5)-4M_{\Delta}^2}
{2v^2\lambda_1-M_{\Delta}^2-v_\Delta^2(\lambda_2+\lambda_3)}.
\end{align}
The physical scalars thus include the doubly-charged  $H^{\pm\pm}(=\delta^{\pm\pm})$, singly charged $H^\pm$, and the CP-even (-odd) neutral $h,~H^0$ ($A^0$), while $G^{\pm,0}$ are the would-be Goldstone bosons. An auxiliary parameter is introduced for convenience
\begin{align}
M_\Delta^2=\frac{v^2\mu}{\sqrt{2}v_\Delta},
\end{align}
so that the heavy scalars have the masses approximately
\begin{align}
M^2_{H^{\pm\pm}}\approx M^2_{\Delta}-\frac{1}{2}\lambda_5v^2,~
M^2_{H^{\pm}}\approx M^2_{\Delta}-\frac{1}{4}\lambda_5v^2,~
M^2_{H^0}\approx M^2_{A^0} \approx M^2_{\Delta},
\end{align}
while the light scalar has the mass $M^2_h \approx 2\lambda_1 v^2$. The heavy scalars are equidistant in masses squared to good approximation:
\begin{equation}
M^2_{H^{\pm\pm}}-M^2_{H^{\pm}}\approx M^2_{H^{\pm}}-M^2_{\psi^0}\approx -\frac{1}{4}\lambda_5v^2,
\label{eq_massrelation}
\end{equation}
with $\psi^0=H^0,~A^0$, and therefore we distinguish between the two scenarios of masses as shown in Eq. (\ref{masssplitting}) according to the sign of $\lambda_5$ while the degenerate spectrum corresponds to the special case of $\lambda_5=0$. There are various theoretical considerations such as perturbativity, vacuum stability, and unitarity that are employed to constrain the parameters in the potential. A detailed study of the potential $V(\Phi,\Delta)$ shows both scenarios are allowed~\cite{Arhrib:2011uy}. The parameters chosen for our collider simulation will be located in the allowed region.

Since the triplet scalars contribute through radiative corrections to the precisely measured electroweak quantities, their parameters are constrained by the electroweak precision data. A careful analysis of one-loop radiative corrections has been made in Ref. \cite{Aoki:2012jj}, and the result turns out to depend significantly on the renormalization scheme employed. In the so-called scheme I, one employs the effective mixing angle $\theta_\textrm{eff}^e$ defined by the $Zee$ vertex as the fourth input parameter in the gauge sector. This is the scheme also used in Refs. \cite{Kanemura:2012rs,Yagyu:2012qp}. It was found that the measured value of $m_W$ requires a large mass splitting $\Delta M$ in the positive scenario for $v_\Delta\gtrsim 1~\GeV$, $M_{H^{++}}\sim 150-300~\GeV$ and specific values of other parameters, while there is severe tension in the negative scenario for the same ranges of parameters. Here $\Delta M$ is defined as the (positive) mass splitting between $H^\pm$ and the lighter of $H^{\pm\pm}$ and $\psi^0$. In the scheme II, one adopts instead the mixing angle $\alpha$ between $A^0$ and $G^0$ as the fourth input parameter. It was found \cite{Aoki:2012jj} that for $v_\Delta\lesssim 1~\GeV$ the positive and negative scenarios can be accommodated with $\Delta M\lesssim 50,~30~\GeV$ respectively. This result is consistent with the bound from the electroweak $S$, $T$, $U$ parameters~\cite{Chun:2012jw}, which requires $\Delta M \lesssim 40~\GeV$, independently of the doubly charged scalar mass. As we are interested here in the case of small $v_\Delta$, we will assume this latter set of constraints on $\Delta M$ in our physics analysis.

The Yukawa coupling between the scalar triplet and lepton doublets is responsible for neutrino masses:
\begin{eqnarray}
\mathcal{L}_\textrm{Yuk}&=&
-Y_{ij}\overline{F^C_{Li}}\left(i\tau^2\right)\Delta F_{Lj} +\textrm{h.c.},
\end{eqnarray}
where the superscript $C$ denotes charge conjugation and $i,~j$ are generation indices. The matrix $Y$ is generally complex and symmetric, and gives the Majorana neutrino mass matrix
\begin{equation}
\label{Mnu}
M_{\nu} = \sqrt{2} Y v_{\Delta}.
\end{equation}
In the basis where the mass matrix of charged leptons is diagonal, the neutrino mass matrix is digonalized by the unitary PMNS matrix,
$M_{\nu}=V^*_\textrm{PMNS}m_{\nu}V^{\dag}_\textrm{PMNS}$, with $m_\nu$ being diagonal, real, and semi-definite positive. The matrix $Y$ is thus governed by the neutrino spectrum, mixing pattern, and the triplet vev $v_\Delta$, and the relationship is constrained through Yukawa contributions to the leptonic decays of the charged scalars and the lepton flavor violating (LFV) transitions of the charged leptons.

The mixing matrix $V_\textrm{PMNS}$ is generically parameterized in terms of three mixing angles $\theta_{ij}$, a Dirac CP phase $\delta$, and two Majorana CP phases. Assuming vanishing Majorana phases, we use the following best fit values~\cite{Tortola:2012te} of parameters for the normal (inverted in parentheses, if different) hierarchy of masses:
\begin{align}
&\Delta m_{21}^2 = 7.62 \times 10^{-5}~\eV^2, ~|\Delta m_{31}^2|= 2.55~(2.43) \times 10^{-3}~\eV^2;
\nonumber\\
&\sin^2\theta_{12}= 0.320, ~\sin^2\theta_{23}= 0.613~(0.600),~\sin^2\theta_{13}= 0.0246~(0.0250);
\nonumber\\
&\delta=0.80\pi~(-0.03\pi).
\label{nupara}
\end{align}
The absolute neutrino masses remain unknown. Cosmological considerations set an upper bound on the sum of masses~\cite{Lesgourgues:2012uu,Ade:2013zuv}, $\sum_i m_i < 0.23 ~\eV$.
The neutrinoless double-$\beta$ decay is sensitive to the effective Majorana mass~\cite{Bilenky:2012qi}, $\langle m \rangle_{ee}=|\sum_i(V_\textrm{PMNS}^2)_{ei} m_i|$, with the most stringent bound coming from the EXO Colla.~\cite{Auger:2012ar}, $\langle m \rangle_{ee}<0.14-0.38~\eV$. The direct neutrino mass search is based on kinematics and sensitive to the average electron neutrino mass, $m_{\nu_e}^2=\sum_i |(V_\textrm{MNS})_{ei}|^2m_i^2$.
The current record is kept by the Troitsk Colla., $m_{\nu_e}<2.0~\eV$ \cite{Aseev:2011dq}, and the upcoming KATRIN experiment has the potential to reach a level of $0.2~\eV$~\cite{Drexlin:2013lha}.
Considering all of these constraints, we assume the lightest neutrino to be massless in either normal (NH) or inverted hierarchy (IH).

\begin{figure}[!htbp]
\begin{center}
\includegraphics[width=0.45\linewidth]{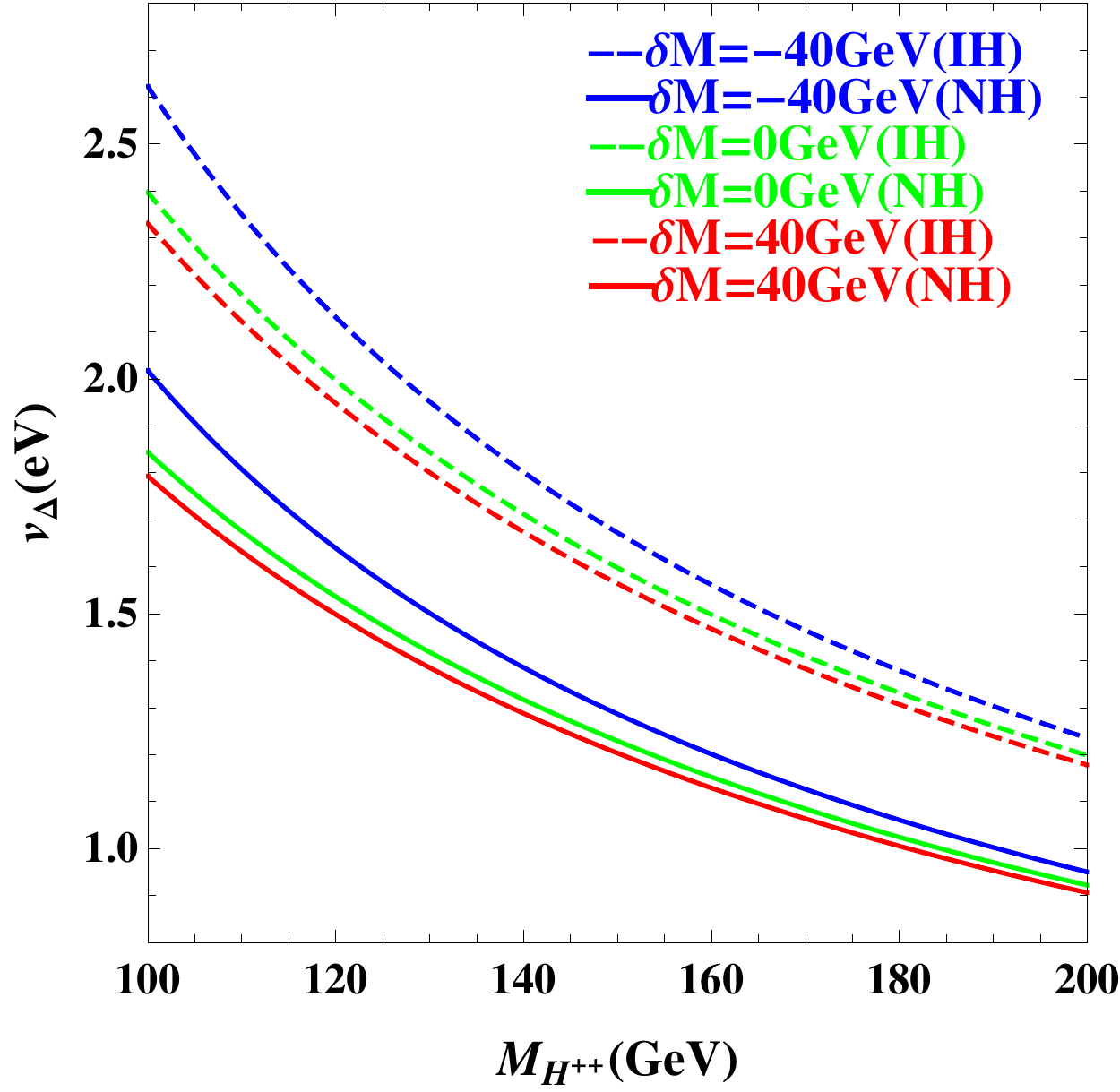}
\includegraphics[width=0.46\linewidth]{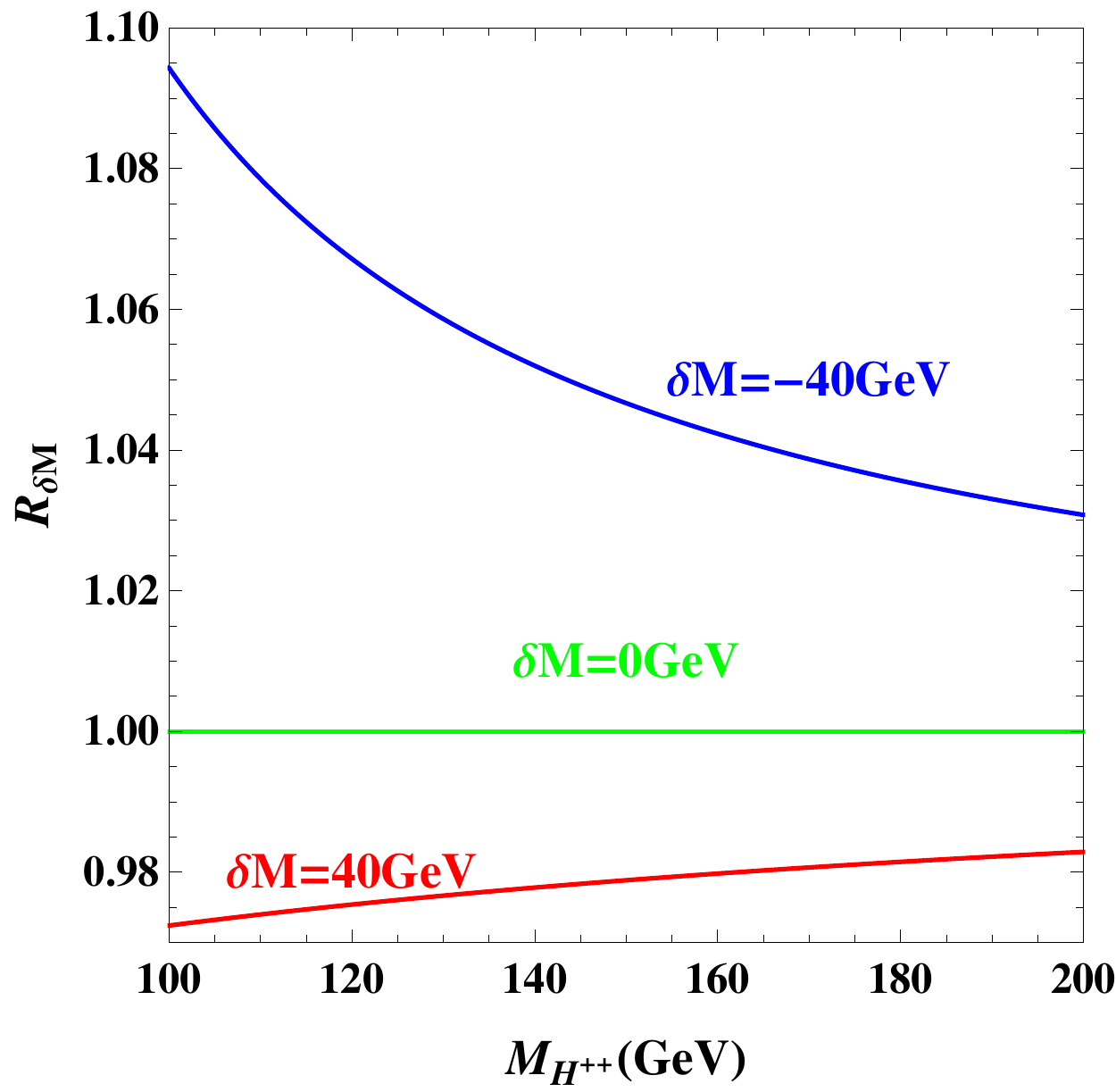}
\end{center}
\caption{Lower bound on $v_{\Delta}$ for nondegenerate case (left panel) and ratio $R_{\delta M}$ (right) as a function of $M_{H^{\pm\pm}}$.
\label{LFV}}
\end{figure}

The Yukawa couplings can mediate low energy LFV processes. The purely leptonic decays $\ell_i\to\bar{\ell}_j\ell_k\ell_l$ proceed at tree level by the exchange of the doubly-charged scalars $H^{\pm\pm}$, while the radiative transitions $\ell_i\to\ell_j\gamma$ gain contributions from virtual $H^{\pm\pm}$ and $H^{\pm}$ loops; see, for instance, Ref.~\cite{Abada:2007ux} for a comprehensive analysis. The parameter space of $M_{\nu}$ to be probed with such processes in ongoing and planned experiments has been discussed in Ref.~\cite{Akeroyd:2009nu}. Using the experimental limits on LFV processes and the muon anomalous magnetic dipole moment and assuming degenerate heavy scalars for simplicity, Ref.~\cite{Fukuyama:2009xk} gives the combined bound, $v_{\Delta}M_{H^{\pm\pm}} \gtrsim 150 ~\eV~\GeV$. Since both $H^{\pm\pm}$ and $H^{\pm}$ contribute to, e.g., $\mu\to e\gamma$,
\begin{eqnarray}
\label{LFV-bound}
\textrm{BR}(\mu\to e\gamma) &=& \frac{\alpha_\textrm{QED}\left|(M_{\nu}^2)_{e\mu}\right|^2}{12\pi G_F^2 v_{\Delta}^4}
\left[\frac{1}{M^2_{H^{\pm\pm}}}+\frac{1}{8M^2_{H^{\pm}}}\right]^2,
\end{eqnarray}
the bound will be different in the nondegenerate case. Using the current most stringent bound by the MEG Colla., $\textrm{BR}(\mu\to e\gamma)<5.7\times10^{-13}$ (90\% C.L.) \cite{Adam:2013mnn}, we draw on the left panel of Fig.~\ref{LFV} the lower bound on $v_\Delta$ as a function of $M_{H^{\pm\pm}}$ for a given mass splitting of the charged scalars $\delta M=M_{H^\pm}-M_{H^{\pm\pm}}$ for both NH and IH. ($\delta M=\Delta M$ in the positive scenario and $|\delta M|\approx\Delta M$ in the negative scenario for a relatively small mass splitting.)
One sees that the bound for IH is generally more stringent than for NH, and that the positive (negative) scenario is more loosely (stringently) constrained than the degenerate case:
\begin{eqnarray}
\delta M=40~\GeV ~ : ~ v_{\Delta}M_{H^{\pm\pm}} &\gtrsim& 180(\mbox{NH}),235(\mbox{IH}) ~\eV~\GeV ;
\nonumber\\
\delta M=0~\GeV ~ : ~ v_{\Delta}M_{H^{\pm\pm}} &\gtrsim& 185(\mbox{NH}),240(\mbox{IH}) ~\eV~\GeV ;
\nonumber\\
\delta M=-40~\GeV ~ : ~ v_{\Delta}M_{H^{\pm\pm}} &\gtrsim& 200(\mbox{NH}),260(\mbox{IH}) ~\eV~\GeV.
\end{eqnarray}
On the right panel of Fig.~\ref{LFV} we show the relative change by the ratio $R_{\delta M} =v_{\Delta}(\delta M)/v_{\Delta}(0)$. The deviation between the positive scenario and degenerate case is small and usually cannot excess 3\%, while it can reach about 10\% for the negative scenario. For $H^{\pm\pm}$ heavier than $200~\GeV$, all deviations can be safely neglected for both scenarios. Finally, we should mention that the above limits also depend on the lightest neutrino mass $m_1$ (NH) or $m_3$ (IH) and Majorana phases that we have assumed to be zero.

We finally turn to the constraints at colliders. The decay $h\to\gamma\gamma$ played a central role in the discovery of $h$. While the tree-level decays of $h$ keep essentially intact in the type II seesaw due to small mixing governed by $v_\Delta\ll v$, the loop-induced decays $h\to\gamma\gamma,\,Z\gamma$ can receive sizable contributions from the new charged scalars $H^{\pm\pm},~H^{\pm}$ \cite{Aoki:2012jj,Chun:2012jw,Kanemura:2012rs,Akeroyd:2012ms,Arbabifar:2012bd,
Dev:2013ff,Chabab:2014ara}. The partial decay widths for these processes
can be expressed in terms of the following ratios:
\begin{eqnarray}
R_{\gamma\gamma}&=&\frac{\sigma_{\rm II}(pp\to h \to \gamma\gamma)}{\sigma_{\rm SM}(pp\to h \to \gamma\gamma)}
= \frac{\sigma_{\rm II}(pp\to h)}{\sigma_{\rm SM}(pp\to h)}\frac{{\rm BR}_{\rm II}(h\to \gamma\gamma)}{{\rm BR}_{\rm SM}(h\to \gamma\gamma)}\, ,
\label{rggdef}
\end{eqnarray}
and similarly for $R_{Z\gamma}$. The current signal strength in the channel $pp\to h\to\gamma\gamma$ is $1.17{\pm0.27}$ at ATLAS \cite{Aad:2014eha} and $1.14^{+0.26}_{-0.23}$ at CMS \cite{Khachatryan:2014ira}. Both ATLAS and CMS are consistent with SM at 1$\sigma$ level, but still have a relatively large uncertainty. Detailed analysis shows that the signal rate $R_{\gamma\gamma}$ is totally determined by the sign of $\lambda_4$ and $\lambda_4+\lambda_5/2$.
$R_{\gamma\gamma}$ is enhanced (suppressed) for negative (positive) $\lambda_4$ with a small variation caused by $\lambda_5$ \cite{Akeroyd:2012ms,Chabab:2014ara} as $\lambda_5$ is tightly constrained by the electroweak precision data. Ref.~\cite{Dev:2013ff} shows that for a given enhancement in the decay rate, an upper bound can be set on the type II seesaw scale; for instance, for an enhancement of $10\%$, the upper limit on the seesaw scale is about $450~\GeV$, which is completely within the reach of the $14~\TeV$ LHC. With this conclusion, dedicated searches at the LHC for the decays $H^{\pm\pm}\to W^{\pm}W^{\pm}$ and $H^{\pm\pm}\to H^{\pm}W^{\pm*}$ with $M_{H^{\pm\pm}}<450~\GeV$ would be strongly motivated.

The current collider limits on the type II seesaw are mainly drawn from the doubly-charged scalar search through its like-sign dilepton signature, $H^{\pm\pm}\to\ell^{\pm}\ell^{\pm}$. At LHC $7~\TeV$, $L=4.7~\fb^{-1}$ and assuming $\rm{BR}(H^{\pm\pm}\to\ell^{\pm}\ell^{\pm})\sim 100\%$, ATLAS \cite{ATLAS:2012hi} excluded $M_{H^{\pm\pm}}$ below $409,~398,~375~\GeV$ at $95\%$ C.L. in the $e^{\pm}e^{\pm}$, $\mu^{\pm}\mu^{\pm}$, $e^{\pm}\mu^{\pm}$ channel respectively. With the same assumption, CMS \cite{Chatrchyan:2012ya} set a lower bound ranging from $204~\GeV$ to $459~\GeV$. Recently, ATLAS has updated its limits to LHC $8~\TeV$, $L=20.3~\fb^{-1}$, and pushed the most stringent lower limit up to $550~\GeV$~\cite{ATLAS:2014kca}. We should emphasize that these bounds do not apply to the nondegenerate case especially in the negative scenario, where two other decay modes of $H^{\pm\pm}$, the like-sign diboson decay $H^{\pm\pm}\to W^{\pm}W^{\pm}$ and the cascade decay $H^{\pm\pm}\to H^{\pm}W^{\pm*}$, can dominate over the like-sign dilepton decays in a large portion of parameter space. To our knowledge, there are no direct experimental limits for these two channels so far. For the like-sign diboson decay, Refs.~\cite{Kanemura:2013vxa,Kanemura:2014goa,Kanemura:2014ipa} concluded that the lower bound on $M_{H^{\pm\pm}}$ can be derived from the like-sign dilepton ATLAS limit, which is $60~\GeV$ for LHC $7~\TeV$, $L=4.7~\fb^{-1}$, and extends to $84~\GeV$ for LHC $8~\TeV$, $L=20.3~\fb^{-1}$. We will show later that the constraints from the cascade decay may be comparable with the one obtained from the like-sign diboson signature. As was also shown in Fig. 5 of \cite{ATLAS:2014kca}, the mass limit on $H^{\pm\pm}$ depends significantly on the branching ratios assumed for the like-sign dilepton decays, and degrades quickly with a decreasing branching ratio. For instance, the benchmark point chosen for our simulation in the positive scenario $M_{H^{\pm\pm}}=400~\GeV$ is located in the safe region, as long as $\Br(H^{\pm\pm} \to \ell^\pm \ell^\pm)\lesssim 0.5$. From Table~\ref{tab:H++BR} shown in Sec.~\ref{positive-doubly}, we see that a potential danger may occur in the channel $H^{\pm\pm} \to e^\pm e^\pm$, which is almost at the edge of the exclusion limit. In the negative scenario, we can choose a much lighter $H^{\pm\pm}$ as our benchmark point, see Eq. (\ref{massspectrum}), because its cascade decays can dominate overwhelmingly over the dilepton modes in a large portion of parameter space; for instance, it can be lighter than $150~\GeV$ if $\Br(H^{\pm\pm} \to \ell^\pm \ell^\pm)<0.02$. In this case, the two triplet-dominating neutral scalars $H^0,~A^0$ can be nearly degenerate with the SM-like Higgs $h$, and could result in interesting phenomena. To put it in short, taking into account the rich decay modes in the nondegenerate case, our benchmark points are compatible with the most recent ATLAS constraints on type II seesaw.

The previous direct searches at LEP also put constraints on new scalars. For the neutral ones, LEP set a lower bound $m_{\psi^0} > 80-90~\GeV$ for general models with an extended scalar sector \cite{Schael:2006cr}. For the charged scalars, searches were performed for pair production; for instance, in minimal supersymmetric standard model (MSSM) or in the more general type-II two-Higgs-doublet model (2HDM), the combined LEP data yielded $m_{H^{\pm}}> 80~\GeV$ or $72.5~\GeV$ \cite{Abbiendi:2013hk}. Although these lower bounds on masses are generally model dependent, they are respected in our later numerical analysis.

\section{Decay properties of the scalars with nondegenerate masses}
\label{Decays}

The decay properties of the new scalars have been studied in the previous literature \cite{Perez:2008ha,Melfo:2011nx,Aoki:2011pz,Yagyu:2012qp}; see, in particular, Refs \cite{Aoki:2011pz,Yagyu:2012qp} for the relevant formulae of the decay widths. Since these properties will be important in devising signal channels at colliders, we summarize them in this section and present a few figures that would help the reader understand our later analysis more readily. We are aware that some similar figures have been plotted in the above literature albeit for different parameters than ours.

As we mentioned earlier, the cascade decay $H^{+}\to H^{++}W^{-*}$ or $H^{++}\to H^{+}W^{+*}$ is possible in the nondegenerate case, which can change significantly the decay patterns even for a small mass splitting $\Delta M$. These decays are classified into three categories: leptonic, gauge boson, and cascade decays, and their branching ratios are respectively most sensitive to the Yukawa coupling $Y$, triplet vev $v_{\Delta}$, and mass splitting $\Delta M$. From the decay phase diagram presented in Ref. \cite{Melfo:2011nx}, one knows that leptonic decays dominate for both $\Delta M$ and $v_{\Delta}$ being small, gauge boson decays dominate in the region of small $\Delta M$ and large $v_{\Delta}$, and cascade decays are dominant for a large $\Delta M$ and a moderate $v_{\Delta}$. In particular, for $v_{\Delta}$ around $10^{-4}~\GeV$, cascade decays dominate in a large portion of the $\Delta M$ parameter space even if $\Delta M$ is as low as $2~\GeV$.

\subsection{Positive Scenario}

\begin{figure}[!htbp]
\begin{center}
\includegraphics[width=0.45\linewidth]{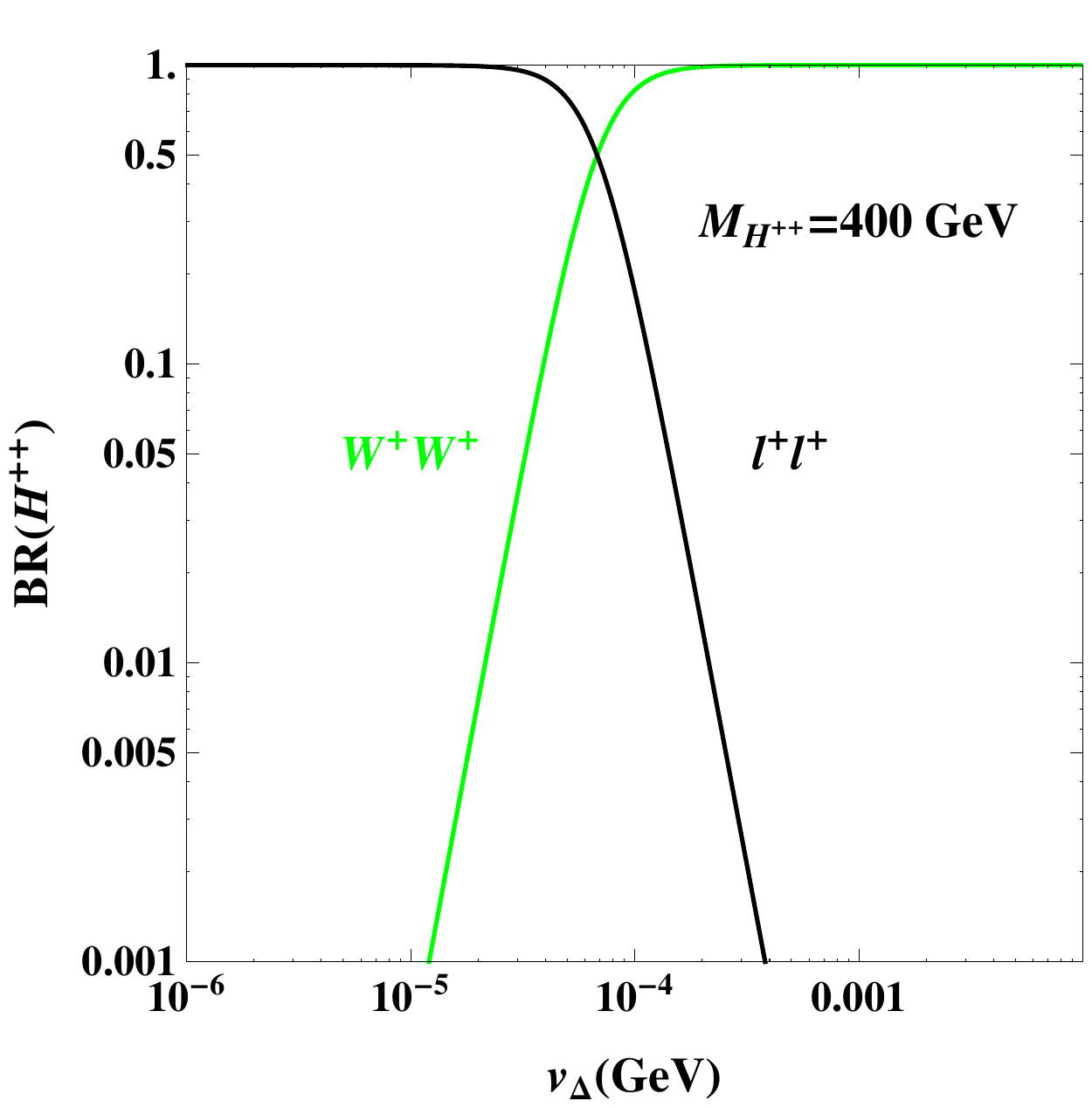}
\includegraphics[width=0.46\linewidth]{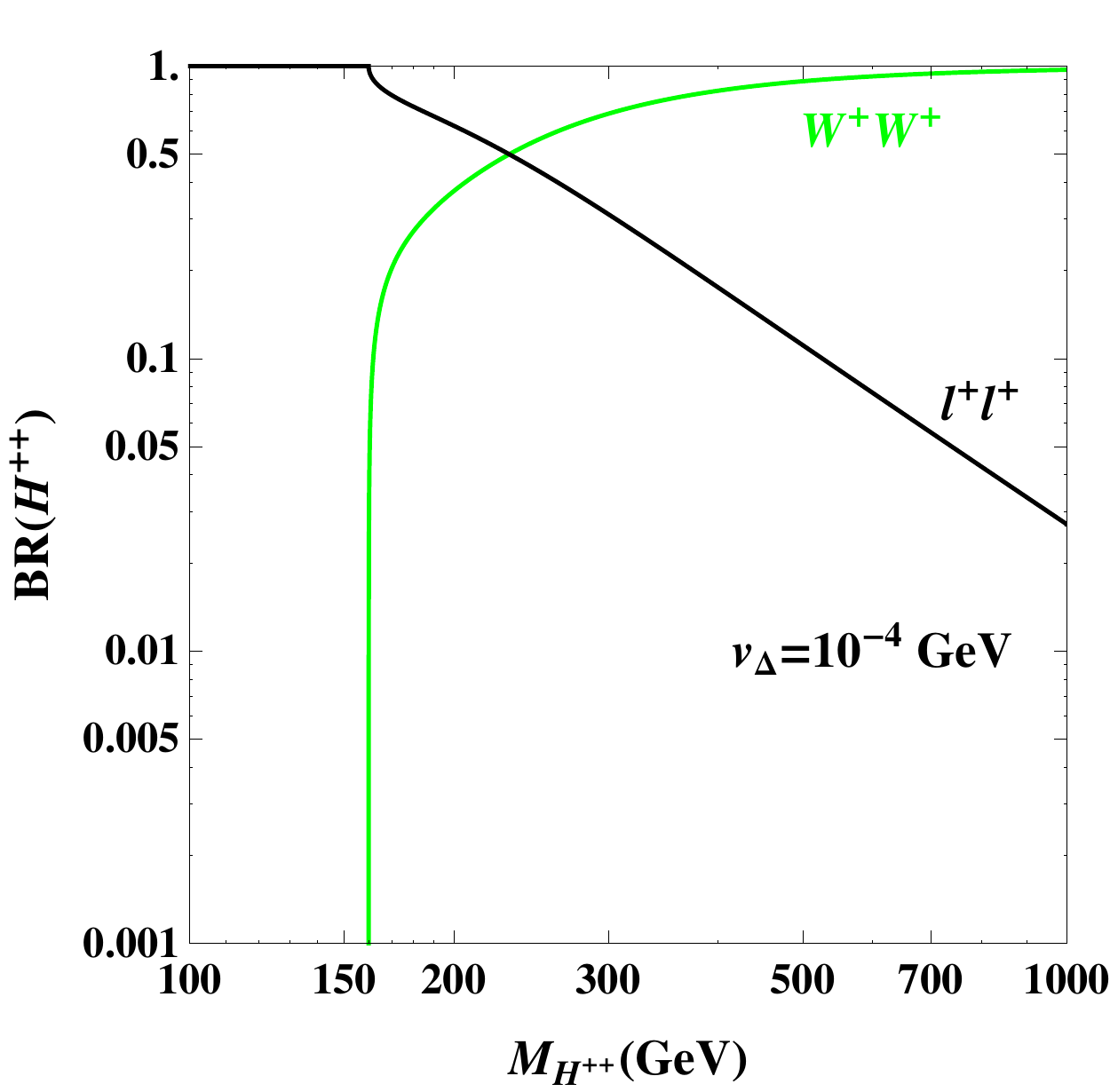}
\end{center}
\caption{Decay branching ratios of $H^{++}$ as a function of $v_{\Delta}$ (left panel) and $M_{H^{++}}$ (right) in the positive scenario.
\label{fig:H++decay}}
\end{figure}

In the positive scenario, $M_{H^{\pm\pm}} < M_{H^{\pm}} < M_{H^0,A^0}$, the doubly charged scalar $H^{++}$ is the lightest and thus shares decay modes similar to the degenerate case, i.e., the like-sign dilepton decay $H^{++}\to \ell^+_i \ell^+_j$ $(\ell= e, \mu, \tau)$ and diboson decay $H^{++} \to W^+W^+$. Their decay amplitudes are proportional to the Yukawa coupling $Y$ and the triplet vev $v_{\Delta}$, respectively. Figure \ref{fig:H++decay} shows their branching ratios as a function of $v_{\Delta}$ at $M_{H^{++}} =400~\GeV$ on the left panel, and as a function of $M_{H^{++}}$ at $v_{\Delta}=10^{-4}~\GeV$ on the right. One can see that the two decays are comparable around $v_{\Delta} \approx  10^{-4}~\GeV$ and $M_{H^{++}}=400~\GeV$. For a given $M_{H^{++}}$, the dilepton decay dominates at $v_{\Delta} < 10^{-5} \GeV$, while the diboson decay takes over at $v_{\Delta} > 10^{-4}~\GeV$; conversely, given $v_{\Delta}$, the dilepton (diboson) decay dominates for a lighter (heavier) $H^{++}$.
A plot similar to the left panel of Fig. \ref{fig:H++decay} can be found in Refs. \cite{Aoki:2011pz,Yagyu:2012qp} at a lower mass.

\begin{figure}[!htbp]
\begin{center}
\includegraphics[width=0.4\linewidth]{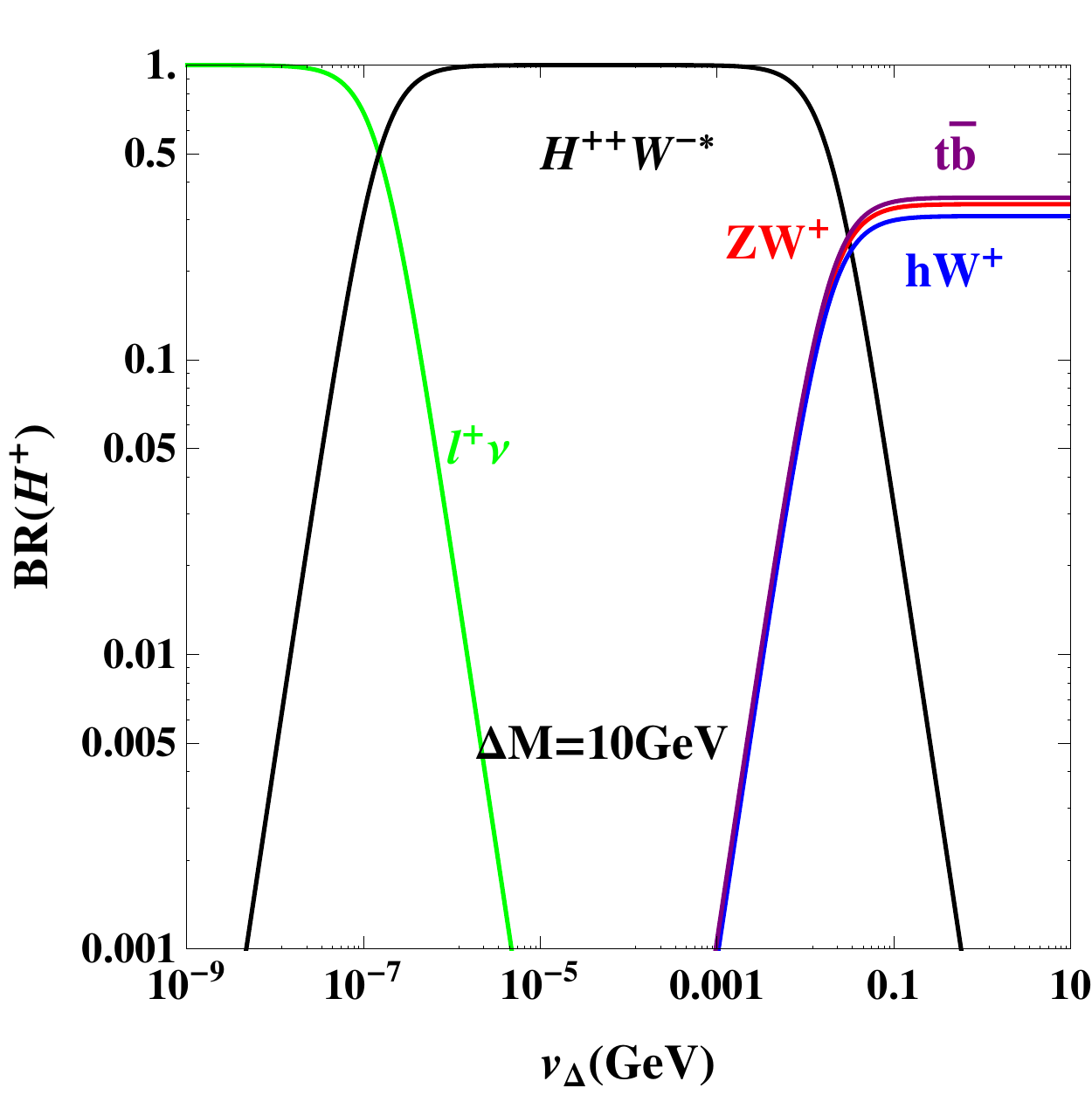}
\includegraphics[width=0.4\linewidth]{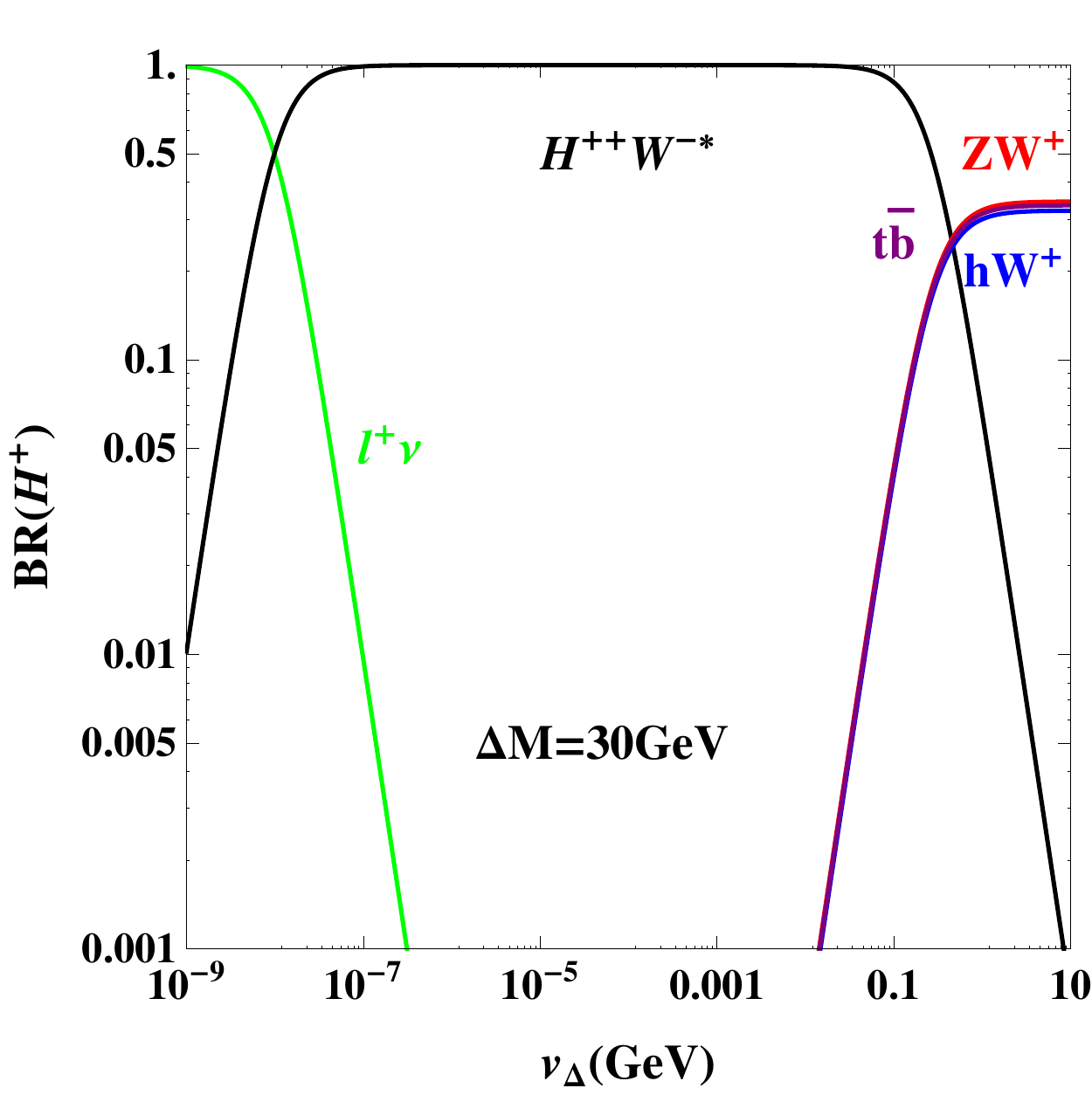}
\includegraphics[width=0.4\linewidth]{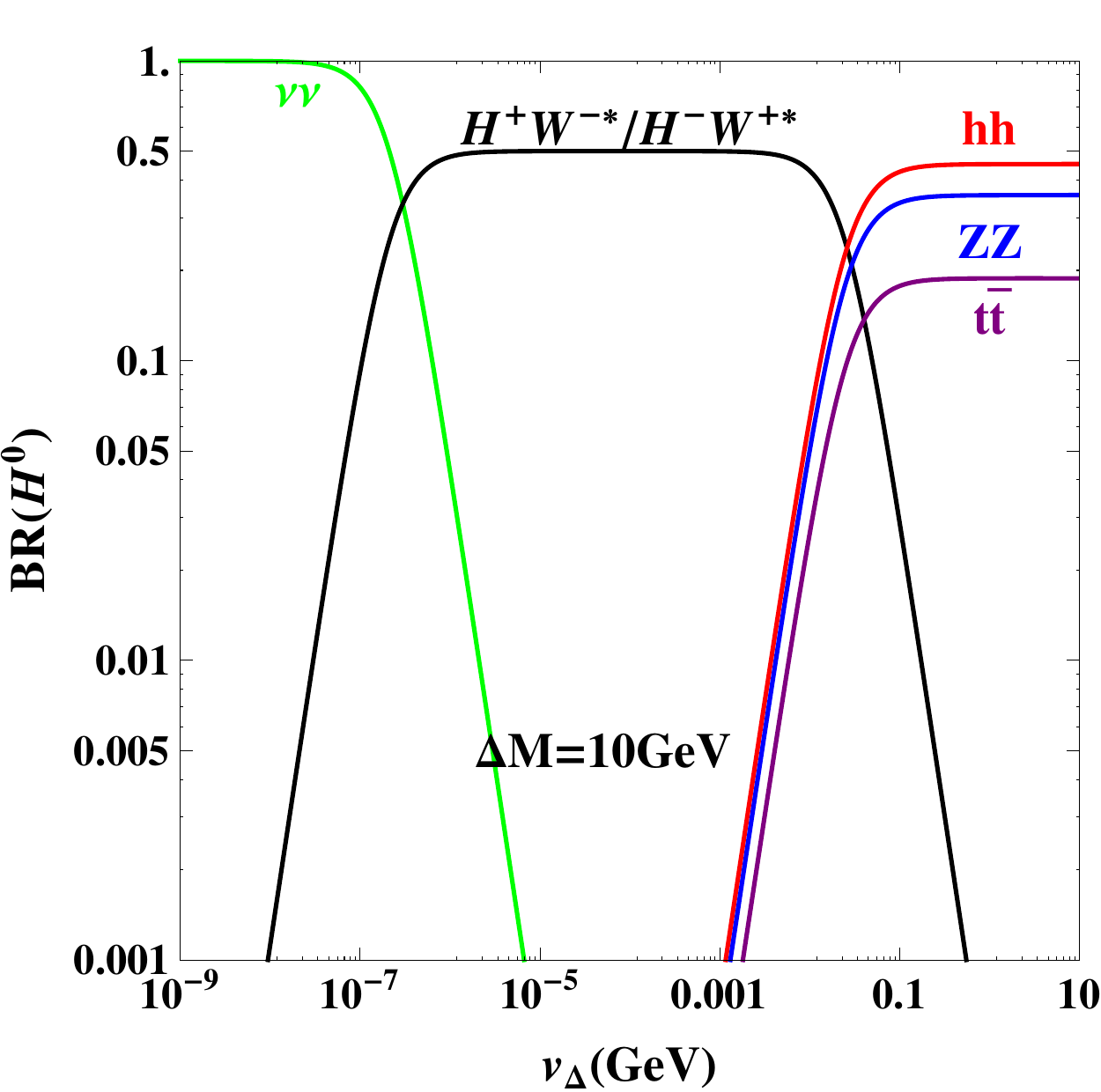}
\includegraphics[width=0.4\linewidth]{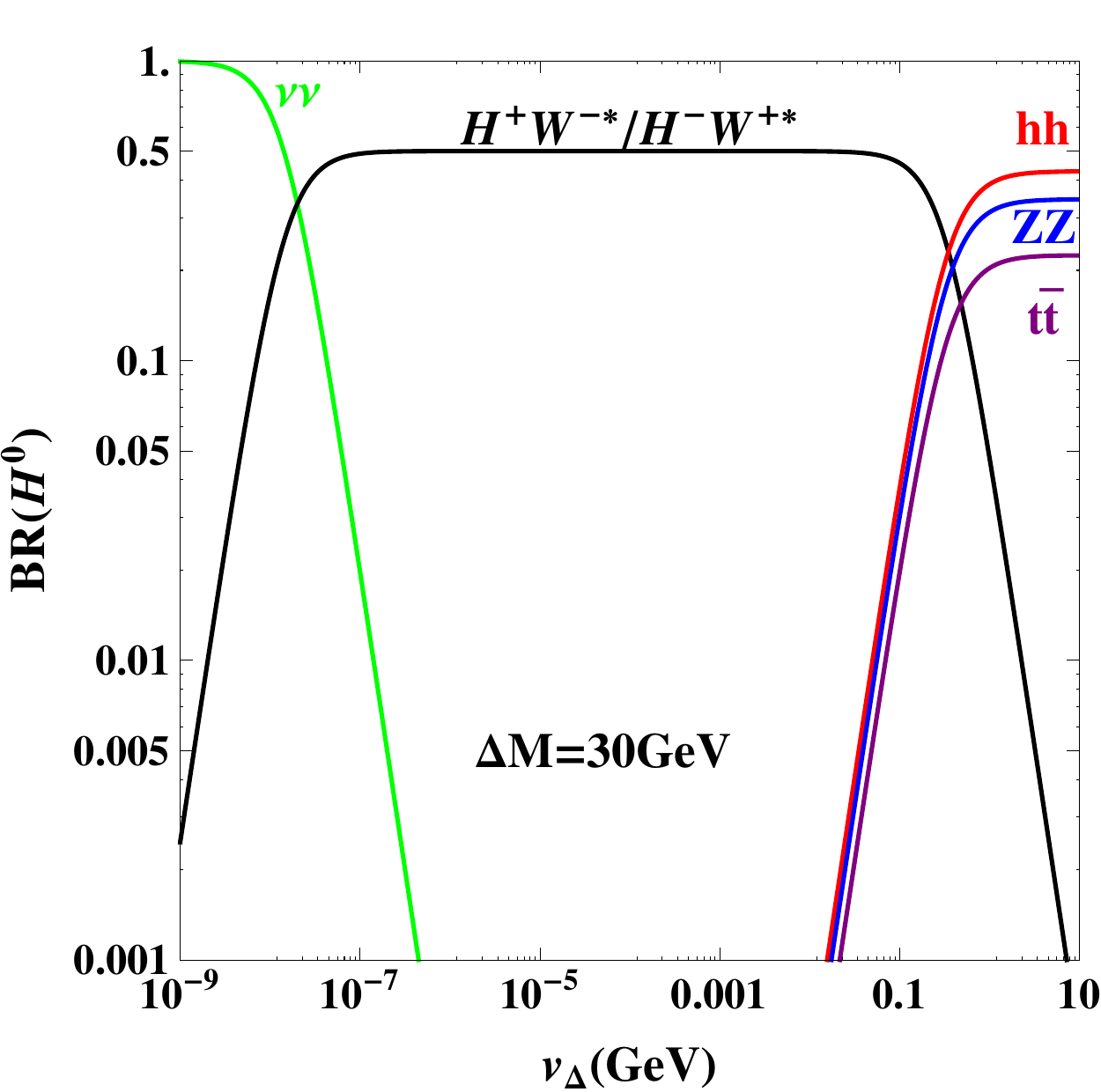}
\includegraphics[width=0.4\linewidth]{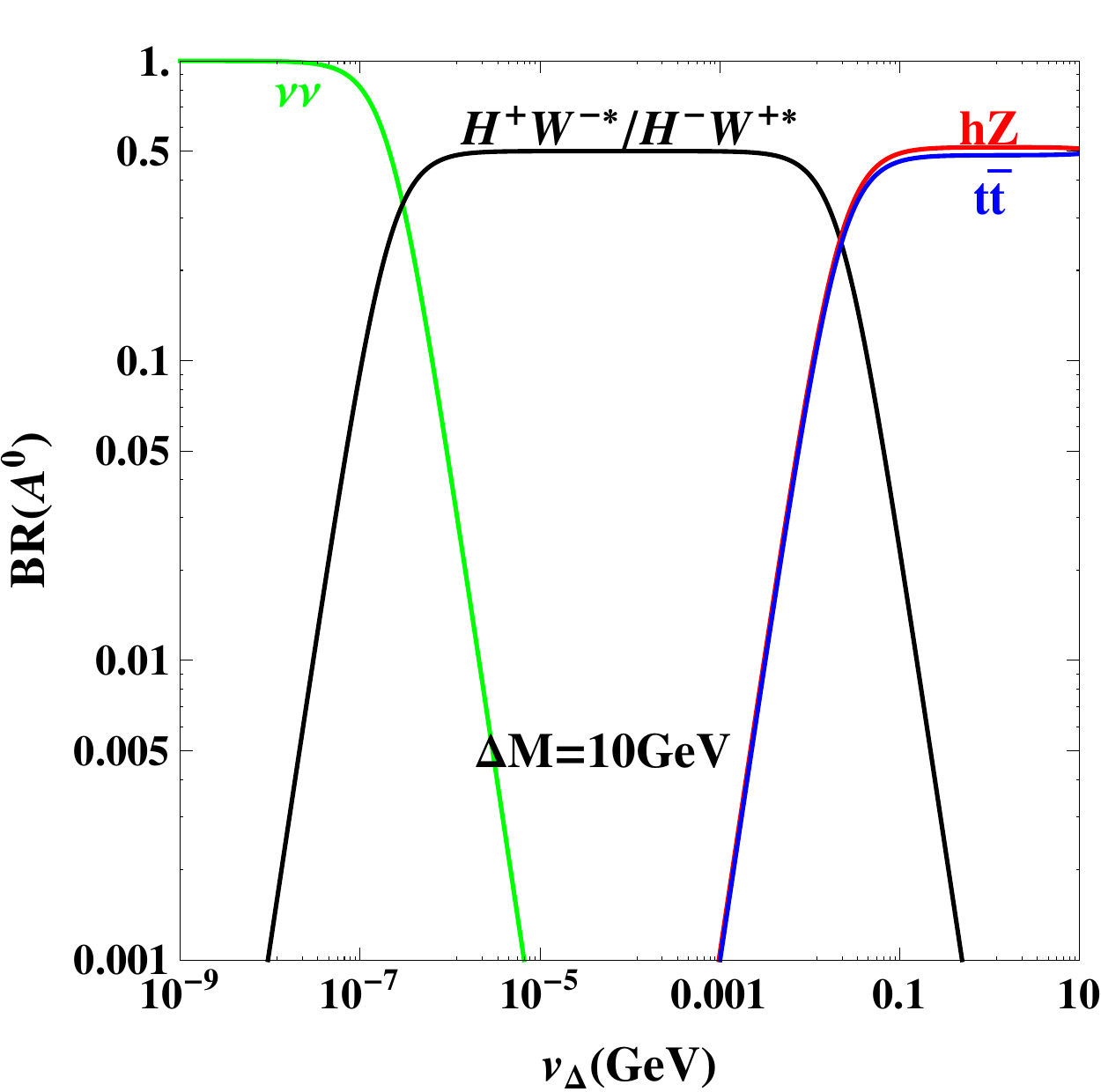}
\includegraphics[width=0.4\linewidth]{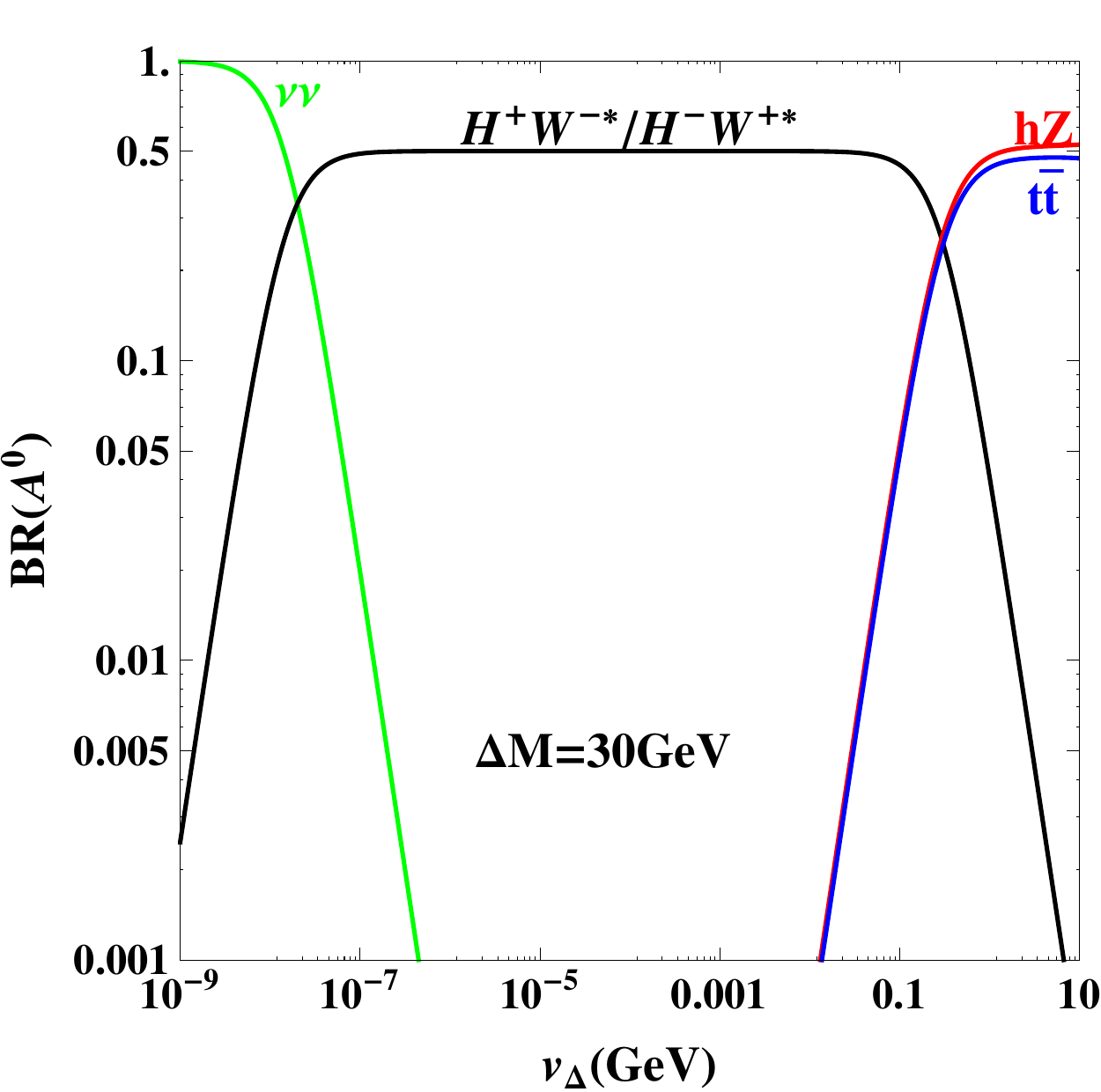}
\end{center}
\caption{Decay branching ratios of $H^{+}$ (upper panel), $H^0$ (middle), and $A^0$ (lower) as a function of $v_{\Delta}$ in the positive scenario, with $M_{H^{++}}=400~\GeV$ and $\Delta M=10~\GeV$ (left panel) and $\Delta M=30~\GeV$ (right).
\label{fig:H+decay}}
\end{figure}

The heaviest scalars $\psi^0=H^0,~A^0$ follow the cascade decay chain $\psi^0\to H^{\pm} W^{\mp*}$ with $H^{\pm}\to H^{\pm\pm} W^{\mp*}$. The branching ratios are plotted in Fig. \ref{fig:H+decay} as a function of $v_{\Delta}$ at $\Delta M=10,~30~\GeV$ and $M_{H^{++}}=400~\GeV$. We notice some interesting features. First, for a given $\Delta M$, the cascade decays of $H^+$, $H^0$, and $A^0$ share a similar region and shape in the plots. Second, for $\Delta M=10~\GeV$, the cascade decays dominate in the region $10^{-7}~\GeV$ $\leq v_{\Delta} \leq 10^{-2}~\GeV$. The leptonic decays dominate for $v_{\Delta}\leq 10^{-8}~\GeV$, which is about three orders of magnitude lower than in the degenerate case. On the other hand, gauge boson decays dominate when $v_{\Delta} \geq 10^{-1}~\GeV$, which is three orders of magnitude higher than in the degenerate case.
For $\Delta M=30~\GeV$, the domination region of cascade decays is further extended to $10^{-8}~\GeV~\leq v_{\Delta}~\leq 10^{-1}~\GeV$, making the other two modes less important in a larger parameter region.

\subsection{Negative Scenario}

We choose the following benchmark masses to illustrate our results in the negative scenario:
\begin{align}
\Delta M=10~\GeV: M_{H^{++}}=149~\GeV,\quad M_{H^{+}}=140~\GeV,\quad M_{H^0,A^0}=130~\GeV;
\notag\\
\Delta M=30~\GeV: M_{H^{++}}=185~\GeV,\quad M_{H^{+}}=160~\GeV,\quad M_{H^0,A^0}=130~\GeV,
\label{eq:spectrum}\end{align}
where $M_{H^{++}}$ is worked out from $M_{\psi^0}$ and $\Delta M$ by Eq. (\ref{eq_massrelation}).
They will also be used in our later collider simulations in Sec~\ref{signal-negative}.

\begin{figure}[!htbp]
\begin{center}
\includegraphics[width=0.4\linewidth]{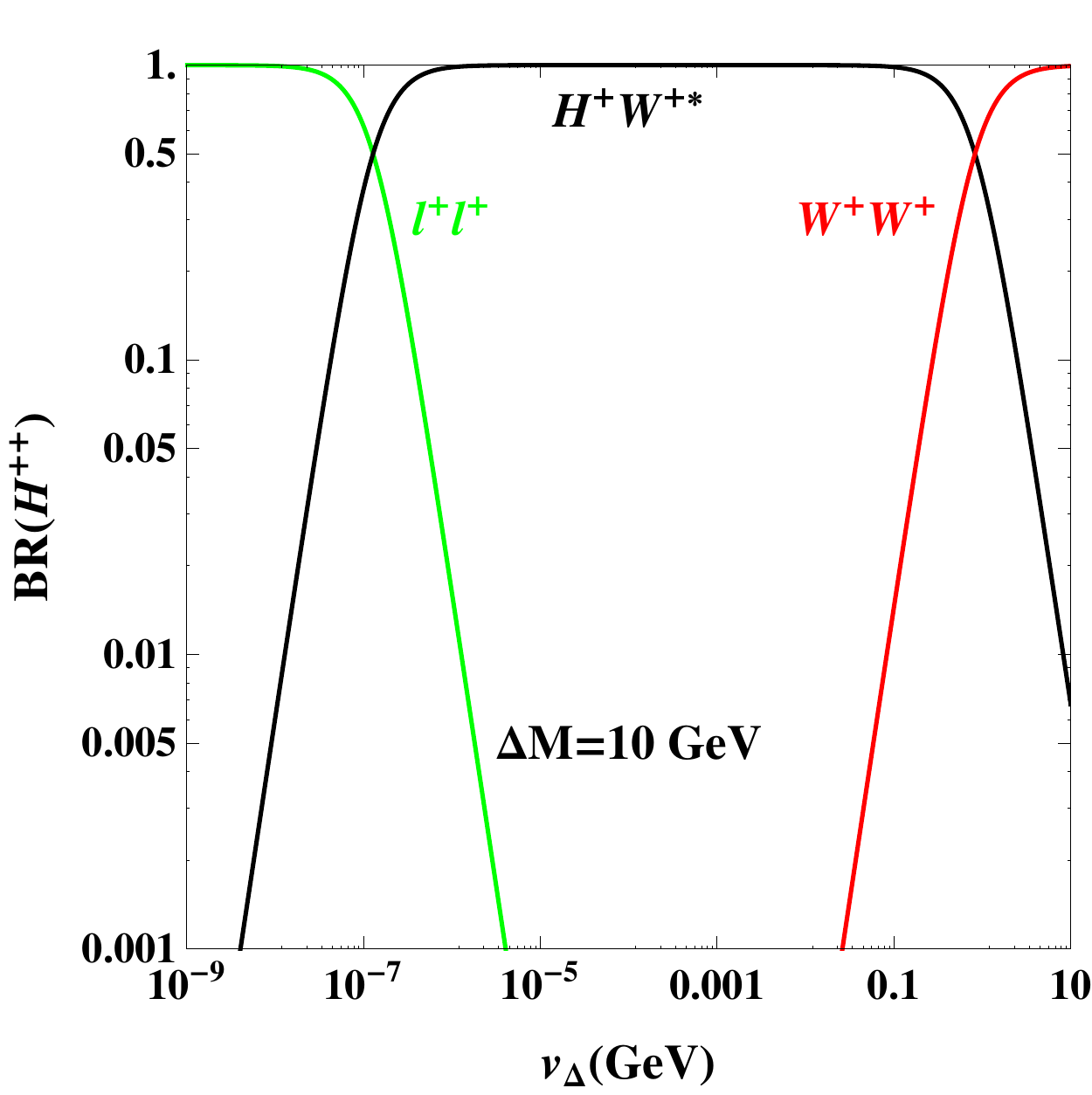}
\includegraphics[width=0.4\linewidth]{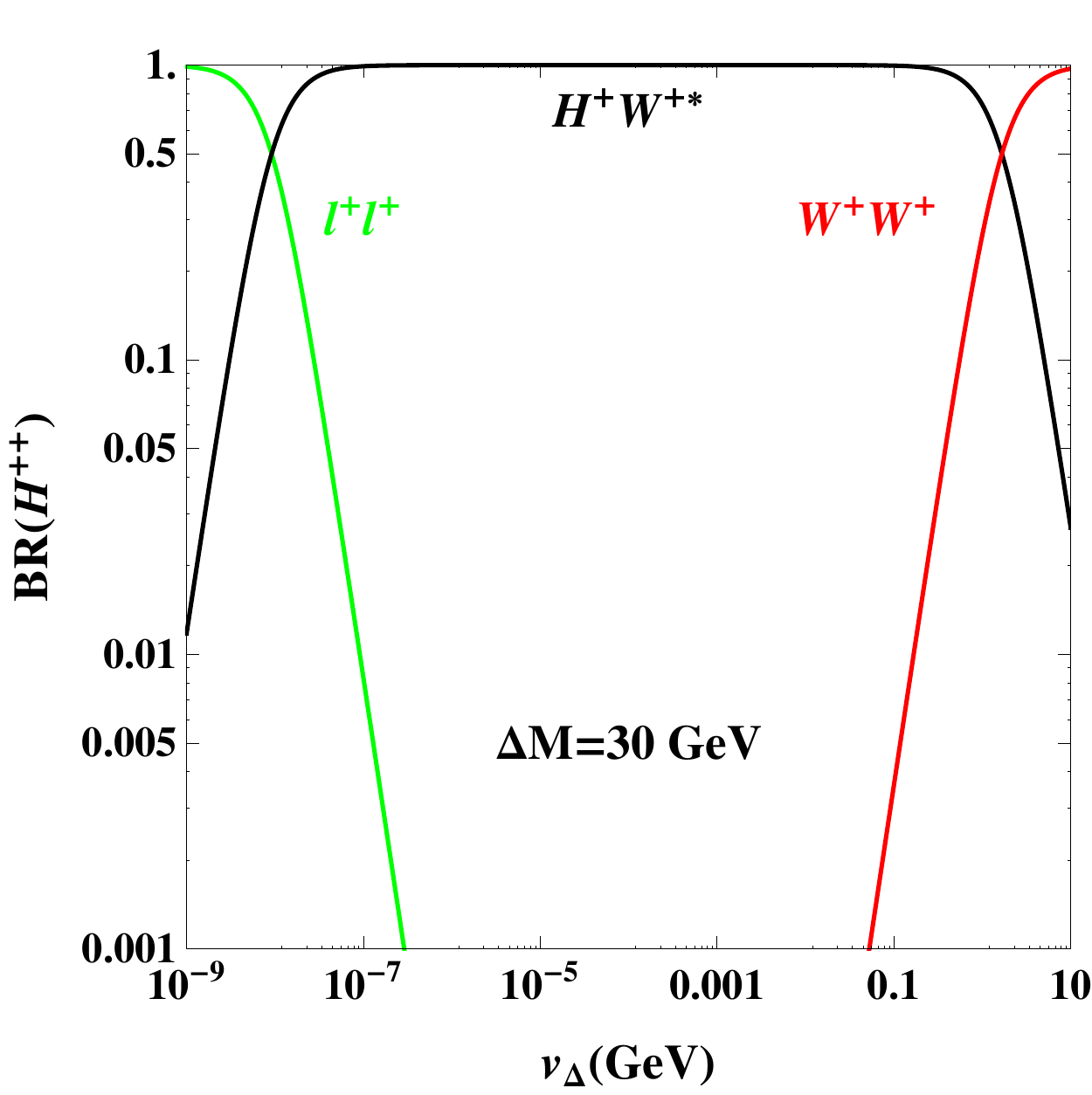}
\includegraphics[width=0.4\linewidth]{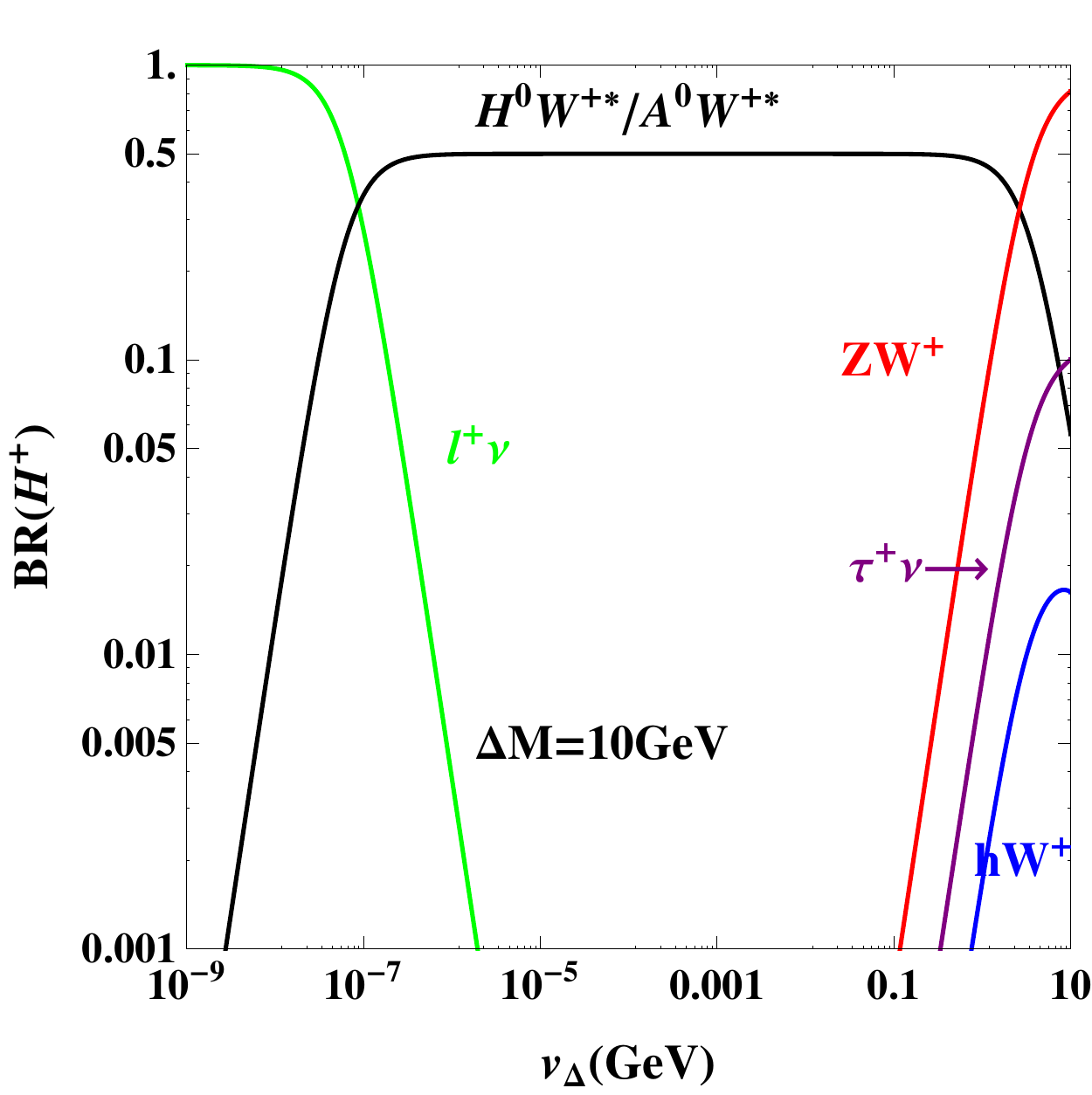}
\includegraphics[width=0.4\linewidth]{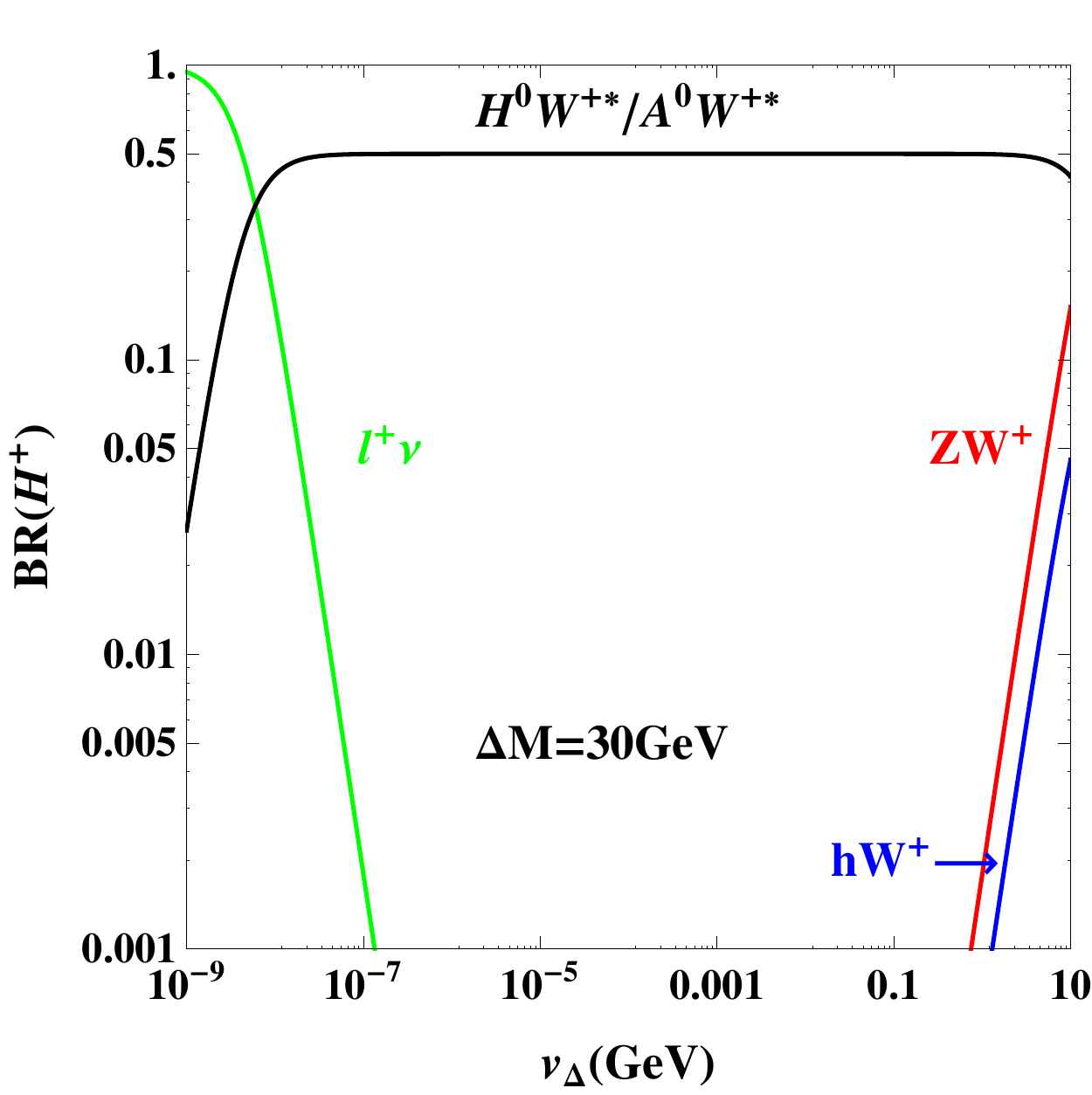}
\includegraphics[width=0.4\linewidth]{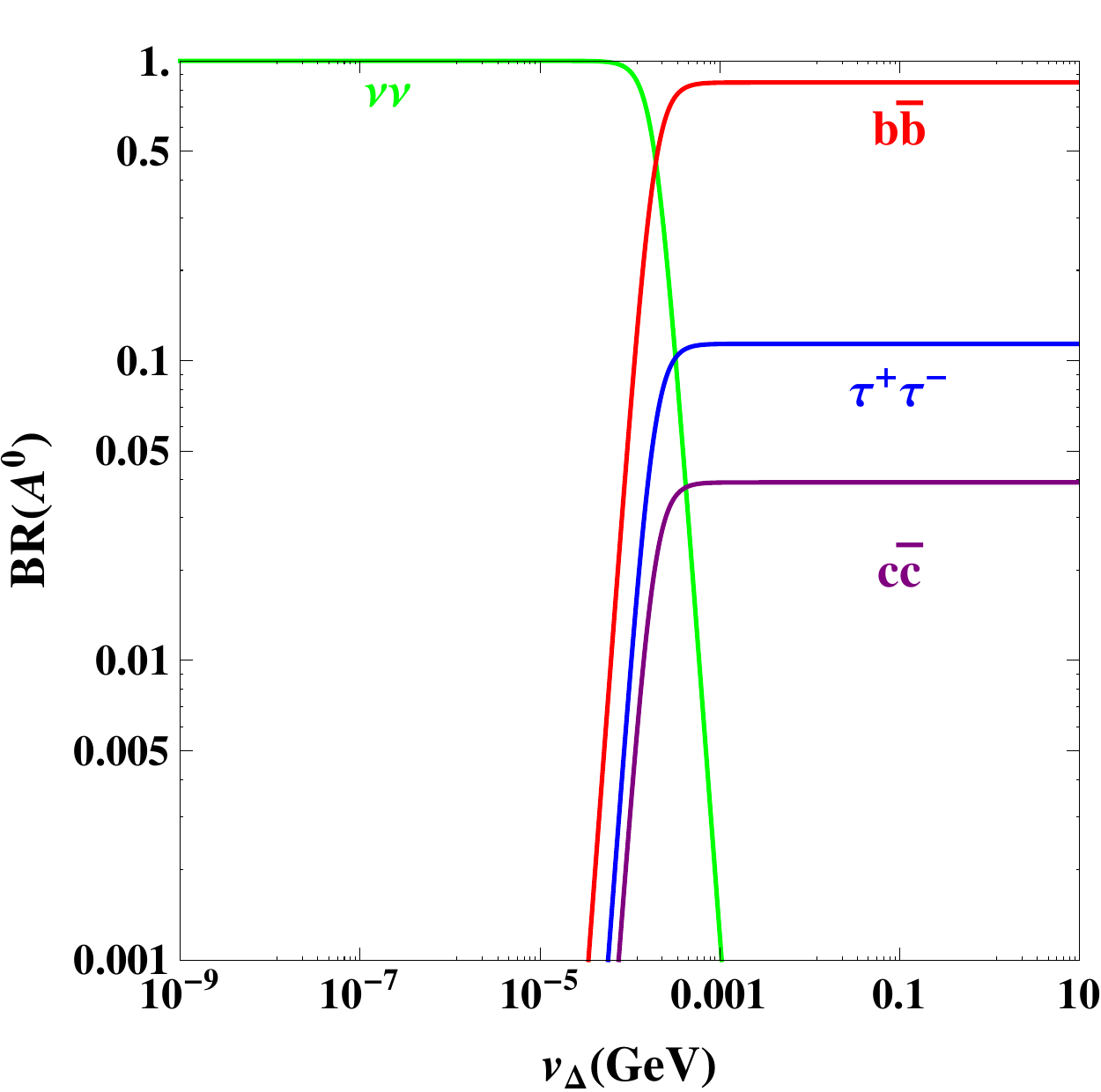}
\includegraphics[width=0.4\linewidth]{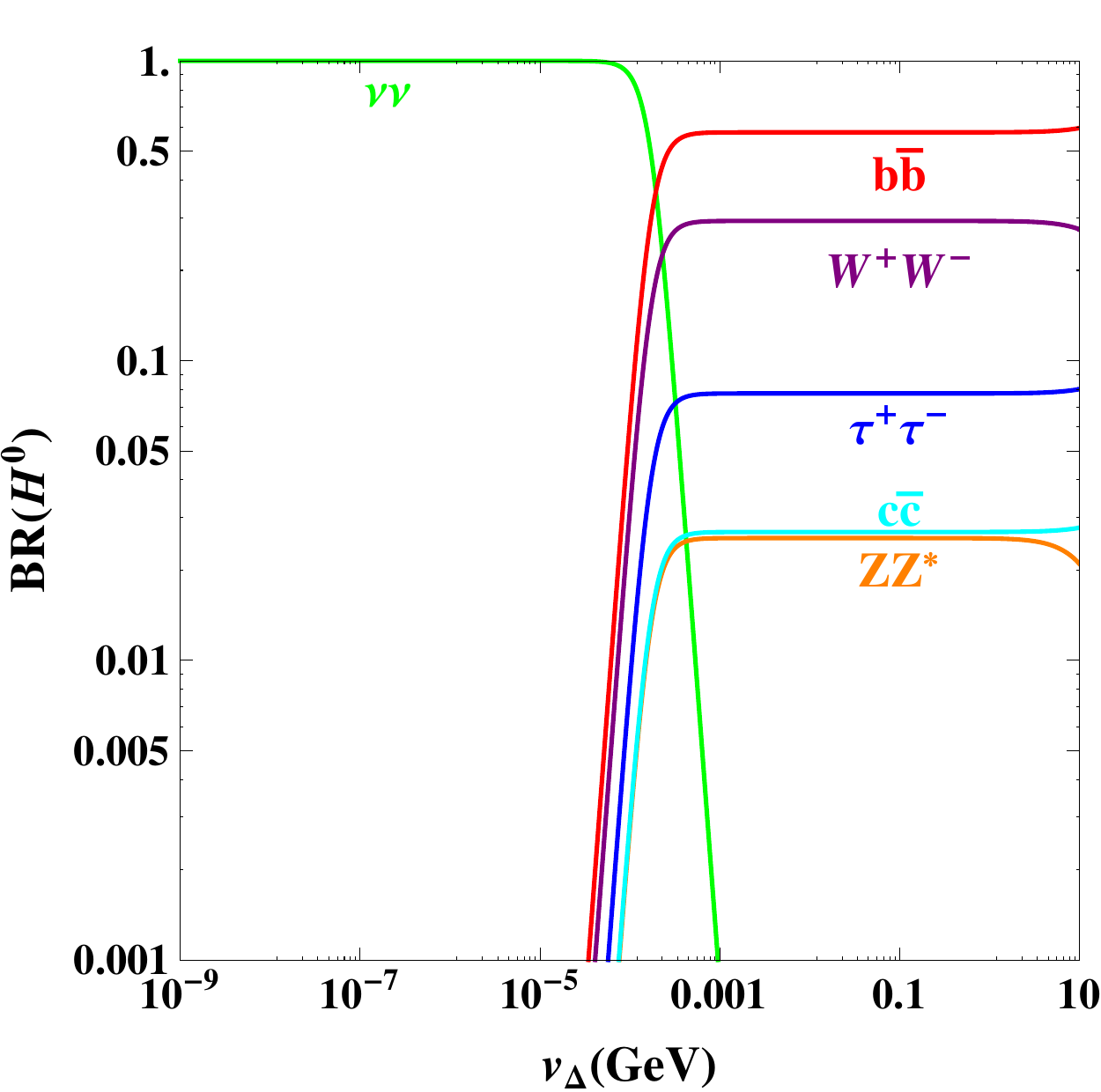}
\end{center}
\caption{Decay branching ratios of $H^{++}$ (upper panel), $H^{+}$ (middle), $H^0$ (lower right) and $A^0$ (lower left) as a function of $v_\Delta$ in the negative scenario, with $M_{\psi^0}=130~\GeV$ and $\Delta M=10~\GeV$ (left panel) and $\Delta M=30~\GeV$ (right).
$H^0,~A^0$ being the lightest, their branching ratios are independent of $\Delta M$.
\label{fig:Phidecay}}
\end{figure}

The decay branching ratios of $H^{++},~H^{+},~H^0,~A^0$ are shown in Fig.~\ref{fig:Phidecay} as a function of $v_{\Delta}$. The same plots were presented previously in Refs. \cite{Aoki:2011pz,Yagyu:2012qp} for either smaller or larger masses.
The decay properties of $H^{++}$ have changed significantly from the degenerate case. The cascade decay $H^{++}\to H^+ W^{+*}$ is dominant in a large region of parameters. For instance, $H^{++}\to H^+ W^{+*}$ dominates in the range $10^{-6}~\GeV\leq v_{\Delta}~\leq1~\GeV$ ($10^{-7}~\GeV~\leq v_{\Delta}~\leq1~\GeV$) for $\Delta M=10~\GeV$ ($\Delta M=30~\GeV$), while the like-sign dilepton and diboson decays are heavily suppressed compared to the degenerate case (see Fig.~\ref{fig:H++decay}). Actually, the like-sign dilepton decays can be safely ignored when $v_{\Delta}~>10^{-5}~\GeV$ ($10^{-7}~\GeV$) for $\Delta M=10~\GeV$ ($30~\GeV$), thus in the majority of parameter space that we are interested in, the current LHC bound on $H^{++}$ can be easily avoided. The dominant regions for $H^{+}\to H^0W^{+*},~A^0W^{+*}$ are similar to those of $H^{++}$.

For the neutral scalars, the relevant decay modes are $H^0 \to b\bar{b},~\tau^+\tau^-,~c\bar{c},~W^+W^-,~ZZ,~hh$
and $A^0 \to b\bar{b},~\tau^+\tau^-,~c\bar{c},~Zh$. (The branching ratios for $H^0 \to hh$ and $A^0 \to Zh$ are too small to show properly in Fig.~\ref{fig:Phidecay}.) The invisible decay $H^0/A^0 \to \nu \bar{\nu}$ dominates in the small $v_{\Delta}$ region, while the hadronic decay $H^0/A^0 \to b\bar{b}$ dominates for a larger $v_{\Delta}$. When $v_{\Delta}>10^{-4}~\GeV$, $\Br(H^0/A^0 \to \tau^+\tau^-)$ can reach about $10\%$, which is a useful channel for probing neutral Higgs scalars as we will discuss in Sec.~\ref{negative-neutral}.

\section{LHC signatures in the Positive Scenario}
\label{signal-positive}

The main goal of this section and the next is to investigate the LHC signatures of type-II seesaw with a nondegenerate spectrum. Since there are many possible final states resulting from cascade decays of scalars and the decay patterns are also different for the positive and negative scenarios, we treat them separately. Before going to the details, we briefly summarize our simulation procedures. We implement the model in the {\em Mathematica} package {\tt FeynRules}~\cite{feynrules}, whose output {\tt UFO}~\cite{Degrande:2011ua} model file is taken by {\tt MadGraph5}~\cite{MG5} to generate parton level events for relevant physical processes. Those events then pass through {\tt Pythia6} \cite{Sjostrand:2006za} to include the initial- and final-state radiation, fragmentation, and hadronization. {\tt Delphes3}~\cite{Delphes} is then used for detector simulation and {\tt MadAnalysis5}~\cite{Conte:2012fm} for analysis. In our simulation, we choose to work with the {\tt CTEQ6L1} parton distribution function (PDF)~\cite{Nadolsky:2008zw}. The tree-level total production cross sections of scalars at 14 TeV LHC (LHC14) have been plotted previously as a function of scalar masses for the $H^{++}H^{--},~H^{\pm\pm}H^{\mp}$ channels in Ref. \cite{Perez:2008ha} and for the $H^{\pm\pm}H^\mp,~H^\pm H^0,~H^\pm A^0,~H^0A^0$ channels in Refs. \cite{Aoki:2011pz,Yagyu:2012qp}. The interested reader should refer to those papers for details.~\footnote{In this work, we consider only the tree-level contributions. The QCD correction to $H^{++}H^{--}$ pair production was computed in~\cite{Muhlleitner:2003me}, with a $K$-factor of about $1.25$, while the contribution from real photon annihilation tends to increase the production by $10\%$, resulting in an overall $K$-factor of $1.35$~\cite{Perez:2008ha}. The $H^{++}H^{-}$ associated production in principle gives a similar $K$-factor $\simeq 1.25$.}

Some of the signal channels to be studied here were considered in previous papers \cite{Aoki:2011pz,Akeroyd:2011zza,Akeroyd:2012nd}, but most of those analyses were based on theoretical estimation or parton level simulation and thus cannot be directly compared with realistic experimental data. In contrast, we simulate the signal channels for both positive and negative scenarios at the detector level and design specific cut criteria for each channel, and this will be more powerful in signal prediction. In the following subsections we present our analysis for each signal channel in the positive scenario, and devote the next section for the negative scenario. To facilitate lepton and charge identification, the leptons here refer only to electrons and muons.

\subsection{Signals for doubly-charged scalars}
\label{positive-doubly}

In the positive scenario, the doubly-charged scalars $H^{\pm\pm}$ are the lightest and decay directly into SM particles. Therefore, their signatures are essentially analogous to the degenerate case, the most promising signal being still the four-lepton channel:
\begin{equation}
\label{h++h--4l}
pp\to H^{++}H^{--}\to \ell^+\ell^+ +  \ell^-\ell^-.
\end{equation}
The branching ratios of the dilepton decays depend on the lepton flavor and neutrino mass hierarchy. They can be easily worked out \cite{Perez:2008ha}, and the numbers are shown in Table~\ref{tab:H++BR}. We observe the following relations,
\begin{eqnarray}
\label{leptonflavor}
&&\Br(H^{\pm\pm}\to e^\pm e^\pm) > \Br(H^{\pm\pm}\to \mu^\pm \mu^\pm) > \Br(H^{\pm\pm}\to \tau^\pm \tau^\pm)\quad \rm{for\;IH},
\nonumber\\
&&\Br(H^{\pm\pm}\to \tau^\pm \tau^\pm) > \Br(H^{\pm\pm}\to \mu^\pm \mu^\pm) \gg \Br(H^{\pm\pm}\to e^\pm e^\pm)\quad \rm{for\;NH},
\end{eqnarray}
which arise as a consequence of the mass hierarchy and mixing pattern~\cite{Garayoa:2007fw,Kadastik:2007yd,Akeroyd:2007zv}. The main irreducible background for this signal channel is
\begin{equation}
ZZ \to \ell^+\ell^- + \ell^+\ell^-,
\end{equation}
and reducible backgrounds are mainly from $t\bar{t},Zb\bar{b},Zt\bar{t}$. We simulate the channel at $14~\TeV$ LHC with an integrated luminosity of 300 fb$^{-1}$ (LHC14@300) and assume $v_{\Delta}=10^{-5}~\GeV,~\Delta M=30~\GeV$. As we mentioned in Sec.~\ref{Constraints}, the latest ATLAS bound on $M_{H^{\pm\pm}}$ can be relaxed to $\sim 400$ GeV for a nondegenerate spectrum, we thus choose the four values of $M_{H^{\pm\pm}}=400,~500,~600,~700~\GeV$ for both NH and IH.
To perform more realistic simulation, we use the same selection criteria as in the ATLAS paper~\cite{ATLAS:2012hi}.

\begin{table} [!htbp]
\begin{center}
\begin{tabular}{|c|c|c|c|c|c|c|}
\hline
 & $ee$ & $e\mu$ & $e\tau$ & $\mu\mu$ & $\mu\tau$ & $\tau\tau$
\\
\hline
~IH~ &~48.7~&0.498&~0.602~&~14.6~&~24.5~&~11.1~
\\
\hline
~NH~ &~0.793~&~4.23~&~0.177~&~27.6~&~30.7~&~36.5~
\\
\hline
\end{tabular}
\end{center}
\caption{Dileptonic branching ratios (in percentage) of $H^{\pm\pm}$ to different flavors.}
\label{tab:H++BR}
\end{table}

\begin{figure}[!htbp]
\begin{center}
\includegraphics[width=0.45\linewidth]{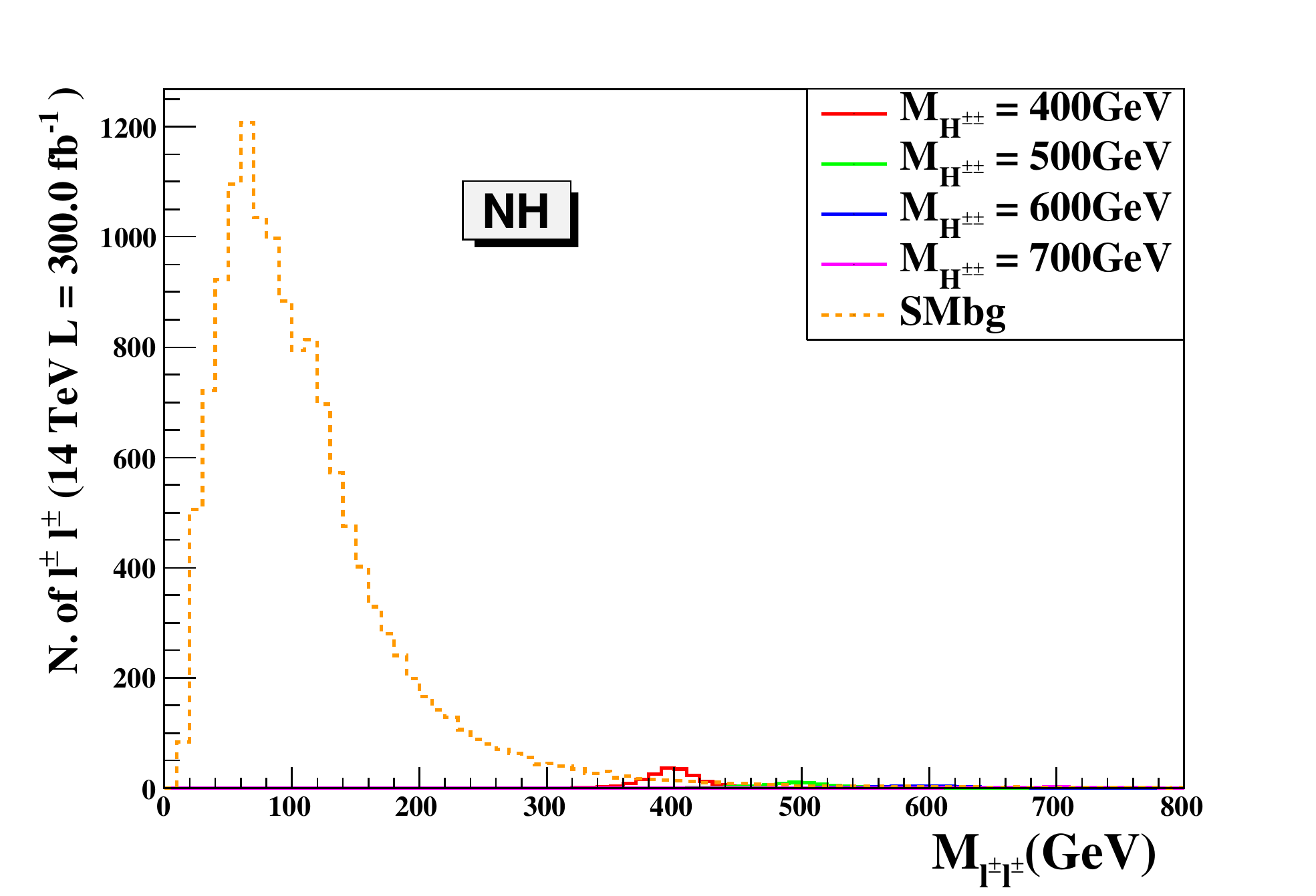}
\includegraphics[width=0.45\linewidth]{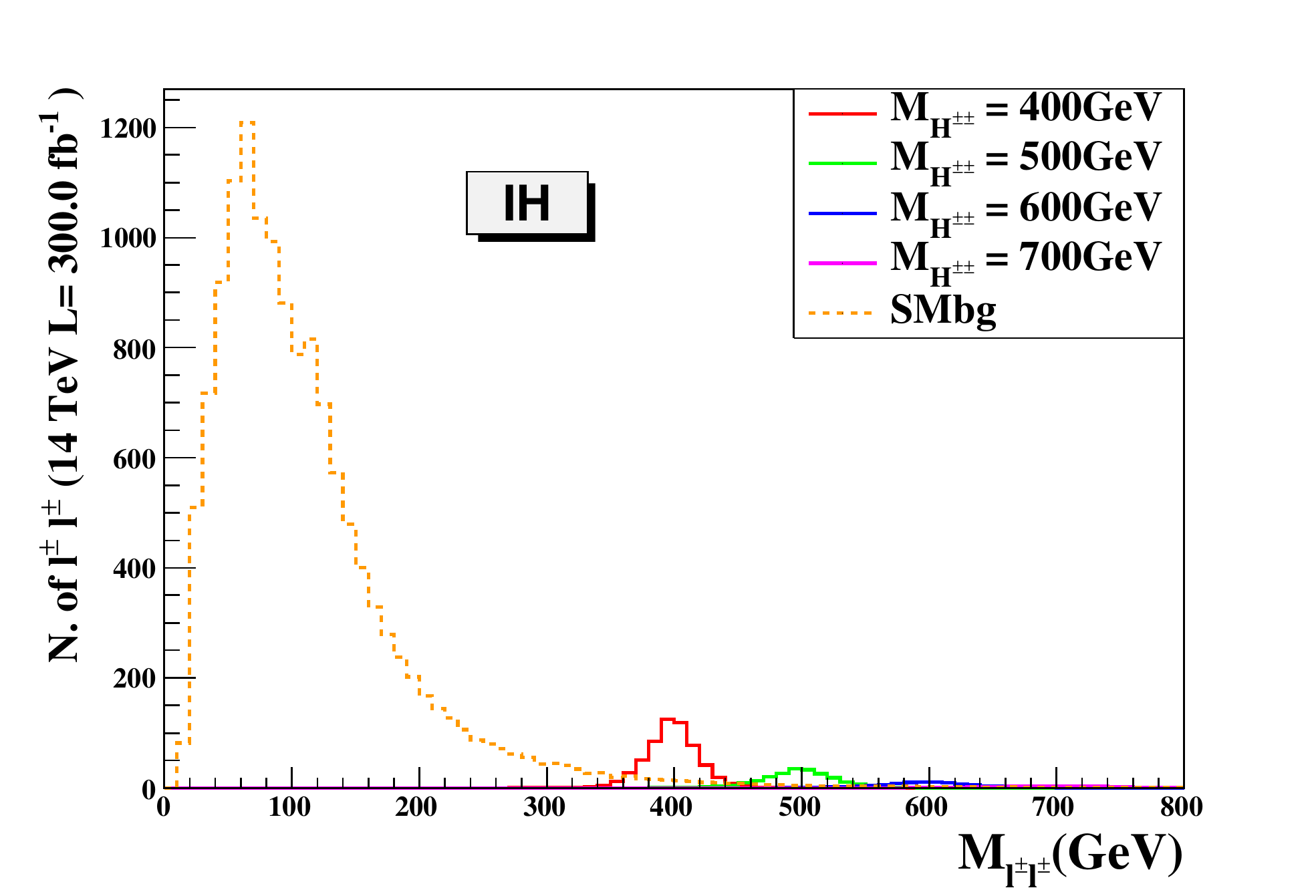}
\end{center}
\caption{Reconstruction of $H^{\pm\pm}$ via dilepton invariant mass $M_{\ell^{\pm}\ell^{\pm}}$ at LHC14@300 and for $M_{H^{\pm\pm}}=400,~500,~600,~700~\GeV$.
\label{fig:h++2ll}}
\end{figure}

\begin{figure}[!htbp]
\begin{center}
\includegraphics[width=0.5\linewidth]{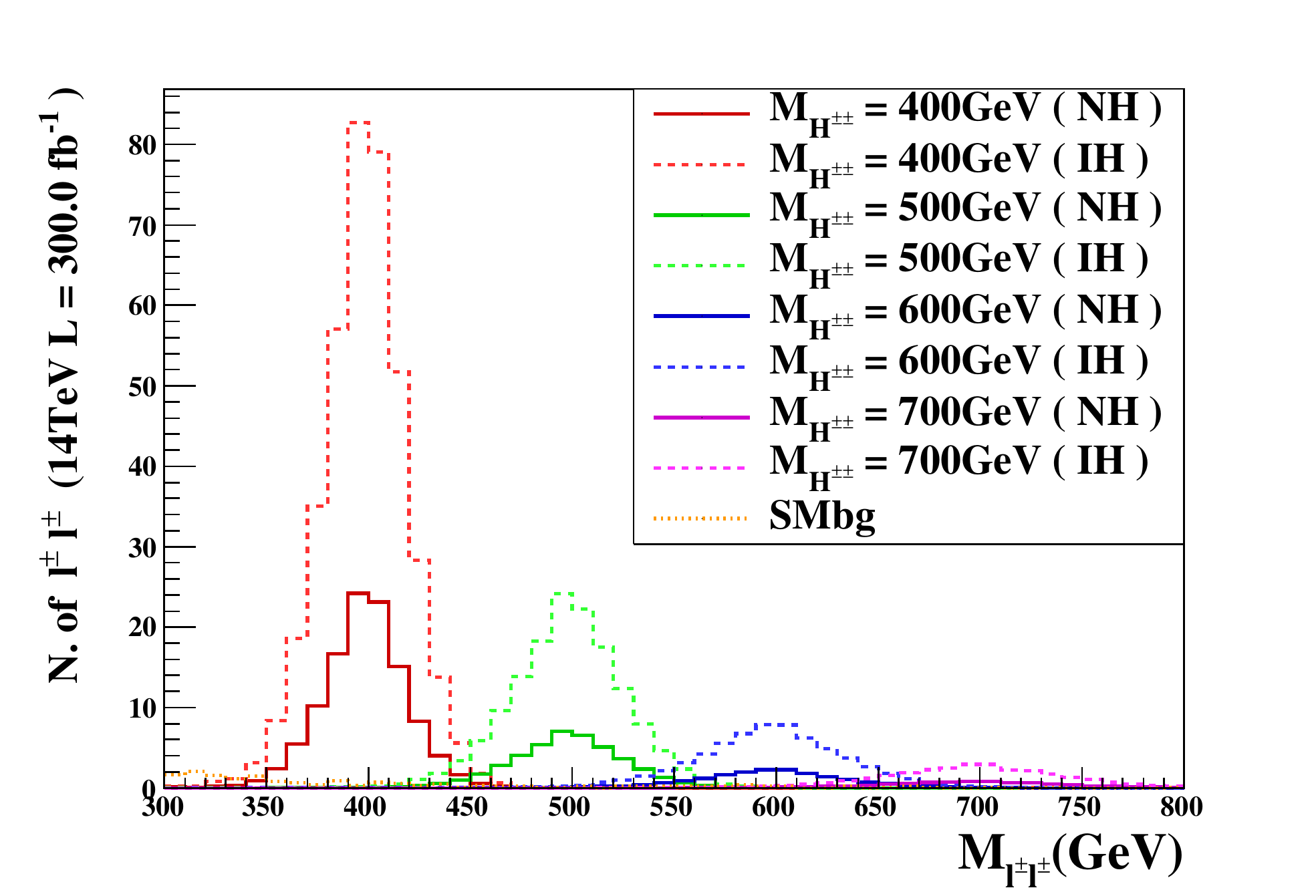}
\end{center}
\caption{Same as Fig.~\ref{fig:h++2ll} but upon imposing the cut (\ref{cutMll}).
\label{fig:4lMll}}
\end{figure}

The distributions in the like-sign dilepton invariant mass $M_{\ell^{\pm}\ell^{\pm}}$ for the signal and background are plotted in Fig.~\ref{fig:h++2ll} for both NH (left panel) and IH (right) cases. We see that the distributions of signal and background are well separated: compared with the signal, the background is mainly located in the small $M_{\ell^\pm\ell^\pm}$ region. Moreover, the IH case has larger signal events, because $H^{\pm\pm}$ have larger branching ratios to the electrons and muons (see Eq.~(\ref{leptonflavor})) which are counted as leptons in simulation. To further purify the signal, we impose the cut
\begin{equation}
M_{\ell^\pm\ell^\pm}>300~\GeV.
\label{cutMll}
\end{equation}
In Fig. \ref{fig:4lMll}, we show the distributions of $M_{\ell^\pm\ell^\pm}$ in the range $300-800~\GeV$ after imposing the cut. As an illustration of details, Table~\ref{tab:4l} shows the evolution of survival numbers of events, statistical significance $S/\sqrt{S+B}$, and signal to background ratio $S/B$ upon imposing the cuts step by step for $M_{H^{\pm\pm}}=400~\GeV$. Here the first three cuts are just the ATLAS ones~\cite{ATLAS:2012hi}, while the final one is Eq. (\ref{cutMll}). One can see that the backgrounds now become negligible. As a consequence, at LHC14@300, one has the potential to probe the doubly-charged scalars with a mass up to $\sim 600~\GeV$ for NH and $\sim 700~\GeV$ for IH.

\begin{table} [htbp]
\begin{center}
\begin{tabular}{|c|c|c|c|c|c|c|c|c|}
\hline
cuts & \multicolumn{2}{|c|}{signal $2\ell^{\pm}2\ell^{\mp}$} & \multicolumn{2}{|c|}{$ZZ$} &
\multicolumn{2}{|c|}{$S/\sqrt{S+B}$}&\multicolumn{2}{|c|}{$S/B$}
\\
\hline
& ~~~IH~~~ & NH & \multicolumn{2}{|c|}{} &~~IH~~ & NH & IH & NH
\\
\hline
no cuts & 468 & 136 & \multicolumn{2}{|c|}{29276 } & 2.71  & 0.798 & 0.016 & 0.00467
\\
$p_T^{\mu}>20\GeV,p_T^{e}>25\GeV,\Delta R_{\ell\ell}>0.4$ & 446 & 131  & \multicolumn{2}{|c|}{ 15921 } &3.49   & 1.03 & 0.0280 & 0.00820
\\
$M_{\ell^\pm\ell^\pm}>15\GeV$ & 446  &  131 & \multicolumn{2}{|c|}{15919} & 3.49  & 1.03 & 0.0280 & 0.00820
\\
$70<M_{e^\pm e^\pm}<110\GeV\;(Z \rm{veto})$ & 446 & 131 & \multicolumn{2}{|c|}{10200} & 4.33  & 1.28 & 0.0438 & 0.0128
\\
\hline
$M_{\ell^\pm \ell^\pm}>300\GeV$ & 194 & 56.8 & \multicolumn{2}{|c|}{3.19} & 13.8  & 7.33  & 60.8 & 17.8
\\
\hline
\end{tabular}
\end{center}
\caption{Survival numbers of four-lepton signal from $H^{++}H^{--}$ with $M_{H^{\pm\pm}}=400~\GeV$ and its main background $ZZ$, statistical significance $S/\sqrt{S+B}$, and signal to background ratio $S/B$ at LHC14@300.
\label{tab:4l}}
\end{table}

\subsection{Signals for singly-charged scalars}
\label{positive-singly}

To illustrate our analysis for the production and detection of singly-charged scalars, we choose $M_{H^{\pm\pm}}=400~\GeV$ and $\Delta M=30~\GeV$ so that $M_{H^0,A^0}=458~\GeV$. In the positive scenario, $H^\pm$ cascade decays into $H^{\pm\pm}$ via the radiation of an off-shell $W^\mp$. Here we consider the five-lepton signal coming from $H^{\pm\pm}H^{\mp}$ associated production \footnote{The five-lepton signal can also originate from $H^{\pm}H^0$, $H^{\pm}A^0$ and $H^0A^0$ associated production \cite{Akeroyd:2011zza}, which can potentially enhance the signal strength. We do not consider these additional channels for simplicity.}:
\begin{equation}\label{h++h-5l}
pp \to H^{\pm\pm}H^{\mp} \to H^{\pm\pm}+H^{\mp\mp} W^{\pm*}
\to \ell^{\pm}\ell^{\pm} + \ell^{\mp}\ell^{\mp} + \ell^{\pm}\cancel{E}_T,
\end{equation}
where the off-shell $W^{\pm*}$ decays leptonically. The main SM background is
\begin{equation}
ZZW^\pm\to \ell^+\ell^- + \ell^+\ell^- + \ell^{\pm}\cancel{E}_T.
\end{equation}
The simulation is performed at LHC14@300. We apply the same cuts as for the four-lepton signal. The numbers of events before and after cut selections for both signal and backgrounds are listed in Table~\ref{tab:5l}. Since BR($Z\to\ell^+\ell^-$) is only about $6\%$, the background $ZZW^{\pm}$ is already much smaller than the signal at the pre-selection level. After imposing the cuts, we have $27.2$ signal events for NH case, with a statistical significance of $4.73$. So for the NH case, a $430~\GeV$ singly-charged scalar $H^{\pm}$ can be actually discovered if we consider additional contributions from $H^{\pm}H^0$, $H^{\pm}A^0$, and $H^0A^0$ production. The IH case is even better. According to our simulation, $104$ signal events pass the selection cuts, with a statistical significance of $9.92$. Thus a $5\sigma$ significance discovery requires $L=76~\fb^{-1}$ at LHC14.

\begin{table} [!htbp]
\begin{center}
\begin{tabular}{|c|c|c|c|c|}
\hline
 & Pre-selection & Post-selection & $S/\sqrt{S+B}$  &$S/B$
\\
\hline
~NH~ &~40.5~&~27.2~&~4.73~&~4.63~
\\
\hline
~IH~ &~154~&~104~&~9.92~&~17.7~
\\
\hline
~$ZZW^{\pm}$~ &~9.00 ~&~5.87~ & --- & ---
\\
\hline
\end{tabular}
\end{center}
\caption{Numbers of events and statistical significance $S/\sqrt{S+B}$, $S/B$ before
and after imposing cuts for $M_{H^{\pm\pm}}=400~\GeV,~M_{H^{\pm}}=430~\GeV$ at LHC14@300.}
\label{tab:5l}
\end{table}

\begin{figure}[!htbp]
\begin{center}
\includegraphics[width=0.5\linewidth]{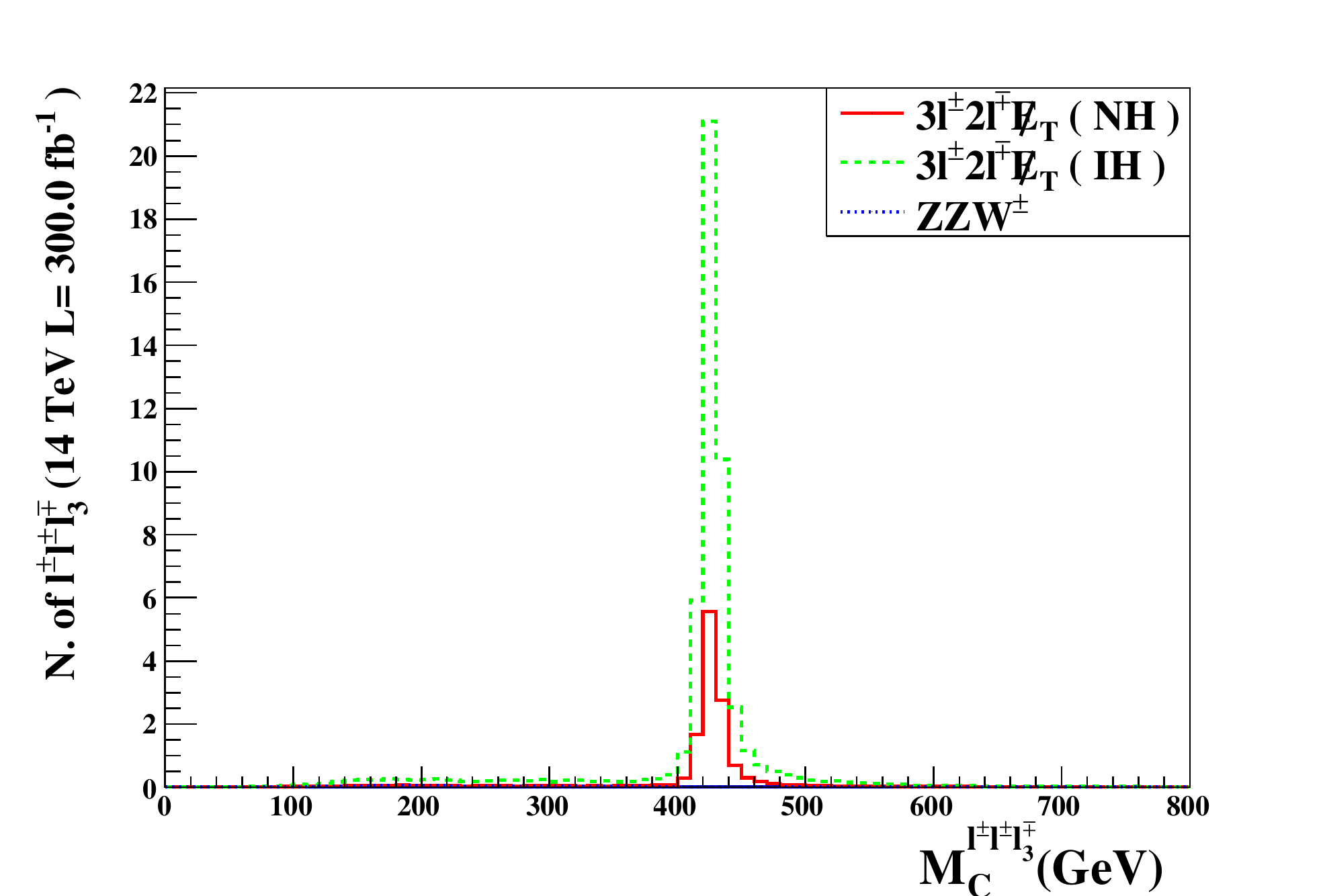}
\end{center}
\caption{Reconstruction of $H^\pm$ via $M_C^{\ell^{\pm}\ell^{\pm}\ell^{\mp}_3}$ for $M_{H^{\pm}}=430~\GeV$ at LHC14@300}.
\label{fig:5lMClll}
\end{figure}

The final states originating from $H^{\pm}$ decays include missing particles, and therefore we cannot fully reconstruct its invariant mass. Since the leptons coming directly from $H^{\pm\pm}$ decays are generically much more energetic than those from off-shell $W$ decays, we can employ this feature to design a kinematical variable. Consider the cluster formed by a like-sign dilepton ($\ell^\pm\ell^\pm$) and the least energetic lepton of opposite charge ($\ell_3^\mp$) and define the cluster transverse mass,
\begin{equation}
M_C^{\ell^{\pm}\ell^{\pm}\ell^{\mp}_3}=
\sqrt{\Big(\sqrt{p_{T,\ell^{\pm}\ell^{\pm}\ell^{\mp}_3}^2
+M^2_{\ell^{\pm}\ell^{\pm}\ell^{\mp}_3}}+\cancel{E}_T \Big)^2
-\Big(\vec{p}_{T,\ell^{\pm}\ell^{\pm}\ell^{\mp}_3}+\vec{\cancel{E}}_{T}\Big)^2},
\end{equation}
where $\vec{p}_{T,\ell^{\pm}\ell^{\pm}\ell^{\mp}_3}$, $M_{\ell^{\pm}\ell^{\pm}\ell^{\mp}_3}$ are respectively the transverse momentum and invariant mass of the $\ell^{\pm}\ell^{\pm}\ell^{\mp}_3$ cluster. [Similar notations will be used below.] The peak structure of this variable then indicates the mass of $H^{\pm}$, as can be clearly seen in its distribution in Fig.~\ref{fig:5lMClll} for both IH and NH cases.

\subsection{Signals for neutral scalars}
\label{positive-neutral}

Now we consider the heaviest particles in the positive scenario, the neutral scalars $H^0,~A^0$. The previous literature has discussed possible like-sign four-lepton signals from associated production involving the neutral scalars ($H^{\pm}H^0,~H^{\pm}A^0$ and $H^0A^0$) which cascade decay into like-sign doubly-charged scalar pair $H^{\pm\pm}H^{\pm\pm}$ plus far off-shell $W^*$, with $H^{\pm\pm}$s being detected in their dilepton channel~\cite{Akeroyd:2011zza,Chun:2012zu,Chun:2013vma}. However, to gain sizable signal events, the mass splitting has to be limited in $\Delta M\sim 1-4~\GeV$~\cite{Chun:2013vma}. Such a splitting may be too small to be directly detectable at a hadron collider, and this signal channel may only be considered as indirect evidence for neutral scalars. In this work, we consider a more promising channel, i.e., the LNV signal:
\begin{equation}
p p \to H^{\pm}\psi^0 \to H^{\pm\pm}W^{\mp*}+W^{\pm*}W^{\pm*}H^{\mp\mp}
                   \to\ell^\pm\ell^\pm jj+\ell^{*\pm}\ell^{*\pm}\cancel{E}_T\ell^\mp\ell^\mp,
\label{eq:sixlepton1}
\end{equation}
where $\psi^0=H^0,~A^0$, and $\ell^{*\pm}$ denotes leptons from off-shell $W$ decays, and the two jets in the final state are too soft to be detected.

We emphasize that the above signal channel has distinct features compared with previous studies. For instance, Refs.~\cite{Akeroyd:2012nd, Chun:2013vma} focused on another LNV channel:
\begin{equation}
p p \to H^{\pm}\psi^0 \to H^{\pm\pm}W^{\mp*}+W^{\mp*}W^{\mp*}H^{\pm\pm}
                   \to\ell^\pm\ell^\pm jj+\ell^{*\mp}\ell^{*\mp}\cancel{E}_T\ell^\pm\ell^\pm.
\label{eq:sixlepton2}
\end{equation}
They also noted that due to significant interference between $H_0$ and $A_0$ for $M_{H^0} \simeq M_{A^0}$, the usual zero width approximation does not apply. The two channels can be labeled by the final leptons from the $H^{\pm\pm}$ decay plus off-shell $W^{*}$s, with the cross sections being
\begin{eqnarray}
\sigma(2\ell^{\pm}2\ell^{\mp}+2W^{*\pm}W^{*\mp})&=&
\sigma(pp \to H^{\pm}\psi^0,\psi^0 \to H^{\mp}W^{\pm *})\Br_{\rm cas.}^2,
\nonumber\\
\sigma(4\ell^{\pm}+3W^{*\mp})&=&
\sigma(pp \to H^{\pm}\psi^0,\psi^0 \to H^{\pm}W^{\mp *})\Br_{\rm cas.}^2,
\end{eqnarray}
where $\Br_{\rm cas.}=\Br(H^{\pm}\to H^{\pm\pm}W^{*\mp})\Br(H^{\pm\pm} \to \ell^{\pm}_i\ell^{\pm}_j)$. One sees that the only difference between the two channels is the cross section $\sigma(pp \to H^{\pm}\psi^0,\psi^0 \to H^{\mp}W^{\pm *})$ versus $\sigma(pp \to H^{\pm}\psi^0,\psi^0 \to H^{\pm}W^{\mp *})$. Following the generalized narrow width approximation introduced in Refs.~\cite{Fuchs:2014ola}, we have calculated the above cross sections including the interference effect, with the result:
\begin{eqnarray}
\sigma(2\ell^{\pm}2\ell^{\mp}+2W^{*\pm}W^{*\mp})&=&\sigma(pp \to H^{\pm}\psi^0)\left[\Br_{H^0}+\Br_{A^0} +2\left(\frac{\Br_{H^0}^2+\Br_{A^0}^2}{\Br_{H^0}+\Br_{A^0}}\right)\right]\Br_{\rm cas.}^2,
\nonumber\\
\sigma(4\ell^{\pm}+3W^{*\mp})&=&\sigma(pp \to H^{\pm}\psi^0)\left[\Br_{H^0}+\Br_{A^0} -2\left(\frac{\Br_{H^0}^2+\Br_{A^0}^2}{\Br_{H^0}+\Br_{A^0}}\right)\right]\Br_{\rm cas.}^2,
\label{eq:final}
\end{eqnarray}
where $\Br_{H^0/A^0}=\Br(H^0/A^0\to H^{\pm} W^{*\mp})$. For rigorously degenerate masses and widths, $M_{H^0}=M_{A^0}$ and $\Br_{H^0}=\Br_{A^0}$, Eq. (\ref{eq:final}) reduces to
\begin{eqnarray}
\sigma(2\ell^{\pm}2\ell^{\mp}+2W^{*\pm}W^{*\mp})&=&2\sigma(pp \to H^{\pm}\psi^0)\cdot2\Br_{H^0/A^0}\cdot\Br_{\rm cas.}^2,
\nonumber\\
\sigma(4\ell^{\pm}+3W^{*\mp})&=&0.
\label{eq:final2}
\end{eqnarray}
Therefore, upon taking into account the interference effect between the intermediate $H_0$ and $A_0$ states, our signal in Eq.~(\ref{eq:sixlepton1}) is enhanced by approximately a factor of two while the signal in Eq.~(\ref{eq:sixlepton2}) tends to disappear. We have confirmed this effect by {\tt Madgraph} simulation, and the vanishing result for the process (\ref{eq:sixlepton2}) is also consistent with the statements in Refs.~\cite{Akeroyd:2012nd, Chun:2013vma}.

\begin{figure}[!htbp]
\begin{center}
\includegraphics[width=0.5\linewidth]{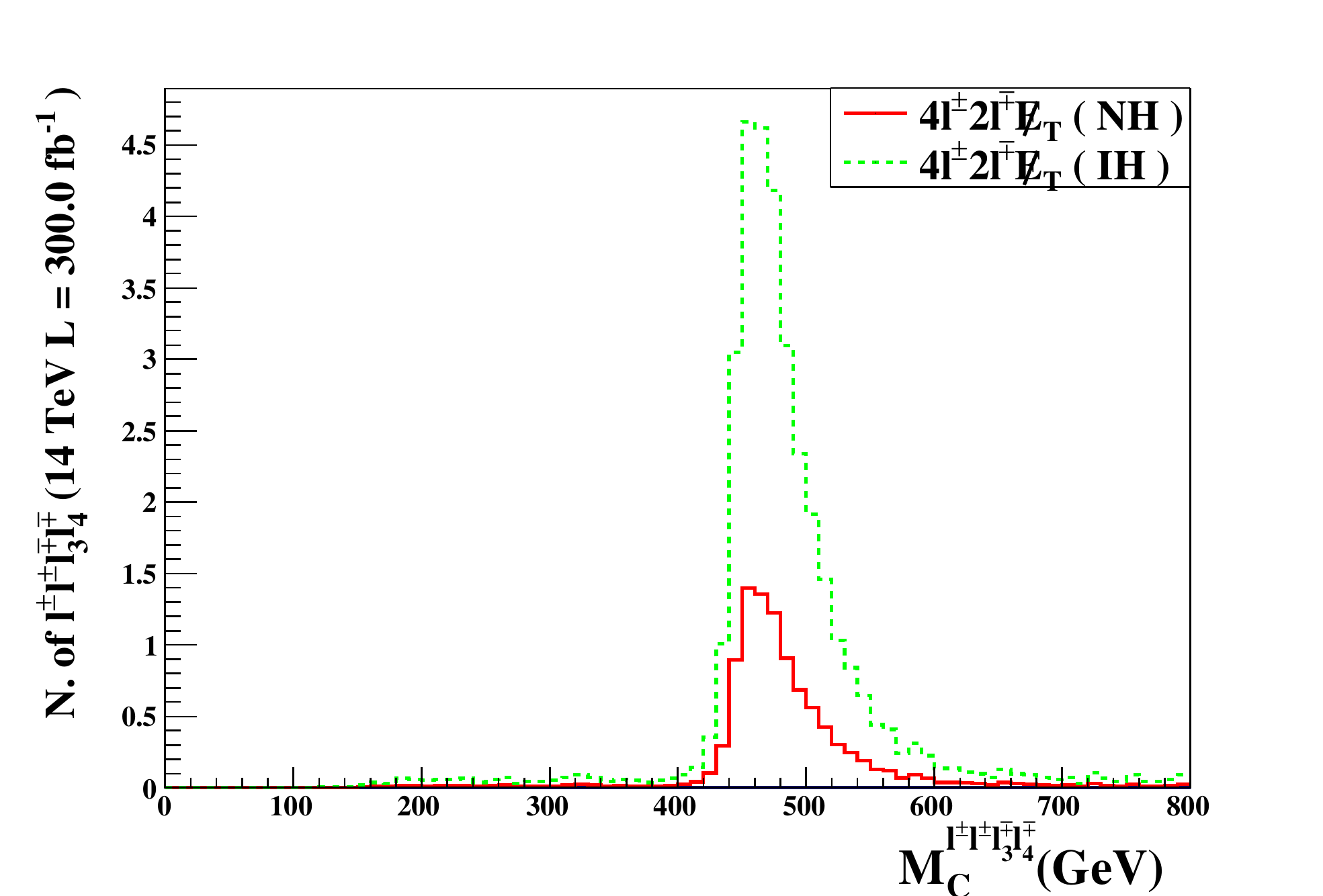}
\end{center}
\caption{Reconstruction of $H^0,~A^0$ via $M_C^{\ell^{\pm}\ell^{\pm}\ell^{\mp}_3\ell^{\mp}_4}$ for $M_{H^0/A^0}=458~\GeV$ at LHC14@300.
\label{fig:6lMCllll}}
\end{figure}

Our signal violates lepton number by two units and thus has no irreducible SM background, but its cross section is also relatively small. For LHC14@300, there are only $11.4$ signal events for NH and $38.0$ for IH. We therefore do not apply any cuts to it. We combine the like-sign dilepton and the two least energetic leptons of opposite charge into a cluster and define the  cluster transverse mass:
\begin{equation}
M_C^{\ell^{\pm}\ell^{\pm}\ell^{\mp}_3\ell^{\mp}_4}=
\sqrt{\Big(\sqrt{p_{T,\ell^{\pm}\ell^{\pm}\ell^{\mp}_3\ell^{\mp}_4}^2
+M^2_{\ell^{\pm}\ell^{\pm}\ell^{\mp}_3\ell^{\mp}_4}}+\cancel{E}_T\Big)^2 -\Big(\vec{p}_{T,\ell^{\pm}\ell^{\pm}\ell^{\mp}_3\ell^{\mp}_4}+\vec{\cancel{E}}_{T}\Big)^2}.
\end{equation}
Its distribution plotted in Fig.~\ref{fig:6lMCllll} can be employed to partially reconstruct the neutral scalars at $M_{H^0/A^0}=458~\GeV$.

\section{LHC signatures in the Negative Scenario}
\label{signal-negative}

In the negative scenario, $M_{H^{\pm\pm}}>M_{H^\pm}>M_{H^0,A^0}$, the more charged scalars can dominantly cascade decay into less charged ones with the radiation of an off-shell $W^\pm$ boson in a significant portion of parameter space. This makes collider search strategies of the scalars very different from the degenerate case, and a systematic study is still lacking in the literature. In this section we shall investigate the issue and when possible compare our results with those in the literature. For illustration, we assume, unless otherwise stated, $M_{H^0,A^0}=130~\GeV$, $\Delta M=30~\GeV$, so that our benchmark point for the scalar masses is
\begin{equation}
\label{massspectrum}
M_{H^{\pm\pm}} = 185~\GeV,\quad M_{H^{\pm}} = 160~\GeV,
\quad M_{H^{0},A^0}=130~\GeV.
\end{equation}

\subsection{Signals for doubly-charged scalars}
\label{negative-Doubly}

The doubly-charged scalars are the heaviest in the negative scenario, and some signal channels for their cascade decays have been briefly discussed in Refs.~\cite{Melfo:2011nx,Aoki:2011pz}. Considering that the cross section for the pair production ($\sigma_{H^{++}H^{--}}\sim 98.5~\fb$ at LHC14) is smaller than that for the associated production ($\sigma_{H^{\pm\pm}H^{\mp}}\sim 223~\fb$), we concentrate on the latter with $\psi^0\to b\bar{b}$~\footnote{This signal can also be produced in the $H^{++}H^{--}$ pair production. A rough estimation based on the ratio of cross sections shows that it enhances the signal events by a factor $1.5$.}:
\begin{equation}
p p \to H^{\pm\pm}H^{\mp} \to W^{\pm*}W^{\pm*}\psi^0+\psi^0W^{\mp*}
    \to \ell^{\pm}\ell^{\pm}\cancel{E}_T b\bar{b}+b\bar{b}jj.
\end{equation}
In this signal, the like-sign $W^{\pm*}$ pair decays leptonically while the other oppositely charged $W^{\mp*}$ decays hadronically. Since the jets from the latter are very soft and difficult to detect at LHC, the visible final states appear as $\ell^{\pm}\ell^{\pm}\cancel{E}_T b\bar{b}+b\bar{b}$. The like-sign dilepton will make the channel suffer from less backgrounds.

We start with some basic cuts:
\begin{eqnarray}
\label{cut:ll4j}
&& p_T^{\ell} > 10 ~\GeV,\;|\eta_{\ell}| < 2.5,\nonumber\\
&& p_T^j > 20 ~\GeV,\;|\eta_j| < 2.5,\;\cancel{E}_T>30\GeV,\nonumber\\
&& \Delta R_{\ell\ell}>0.4,\;\Delta R_{jj}> 0.4,\;\Delta R_{j\ell}> 0.4.
\end{eqnarray}
The leading irreducible background is
\begin{equation}
t\bar{t}W^{\pm} \to W^+bW^-\bar{b}W^{\pm}\to \ell^{\pm}\ell^{\pm}\cancel{E}_T b\bar{b}jj,
\end{equation}
while other backgrounds like $W^{\pm}W^{\pm}jjjj$, $t\bar{b}$ are much smaller~\cite{Perez:2008ha}. To keep the signal rate as much as possible, we do not apply $b$-tagging; instead, we impose the event selection $N_{\ell^{\pm}}=2$ to pick up events with exactly a like-sign lepton pair. Furthermore, since both leptons are from off-shell $W$s, we apply a veto cut on the largest transverse momentum of the lepton pair, $p_T^{\ell_1}<30~\GeV$.

\begin{figure}[!htbp]
\begin{center}
\includegraphics[width=0.45\linewidth]{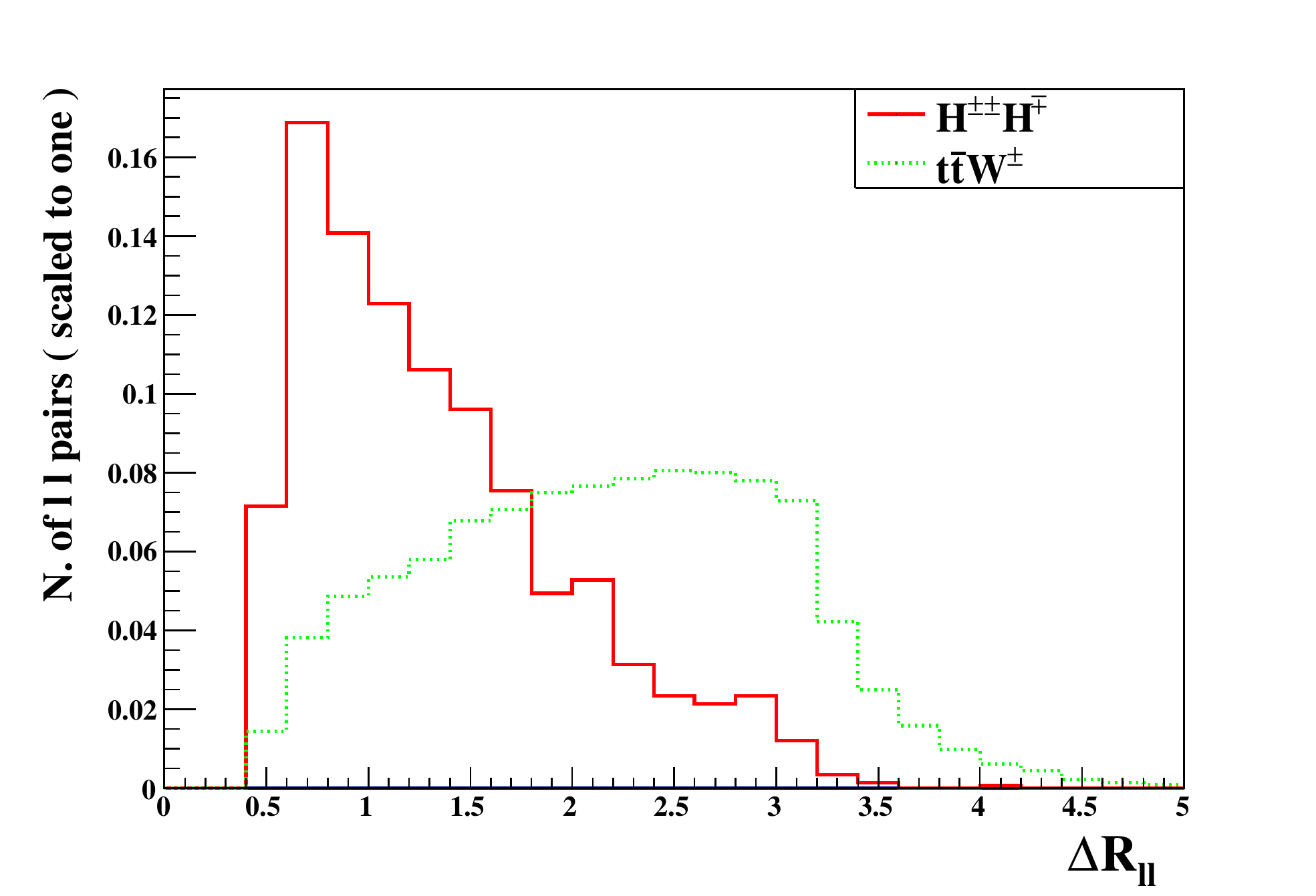}
\includegraphics[width=0.45\linewidth]{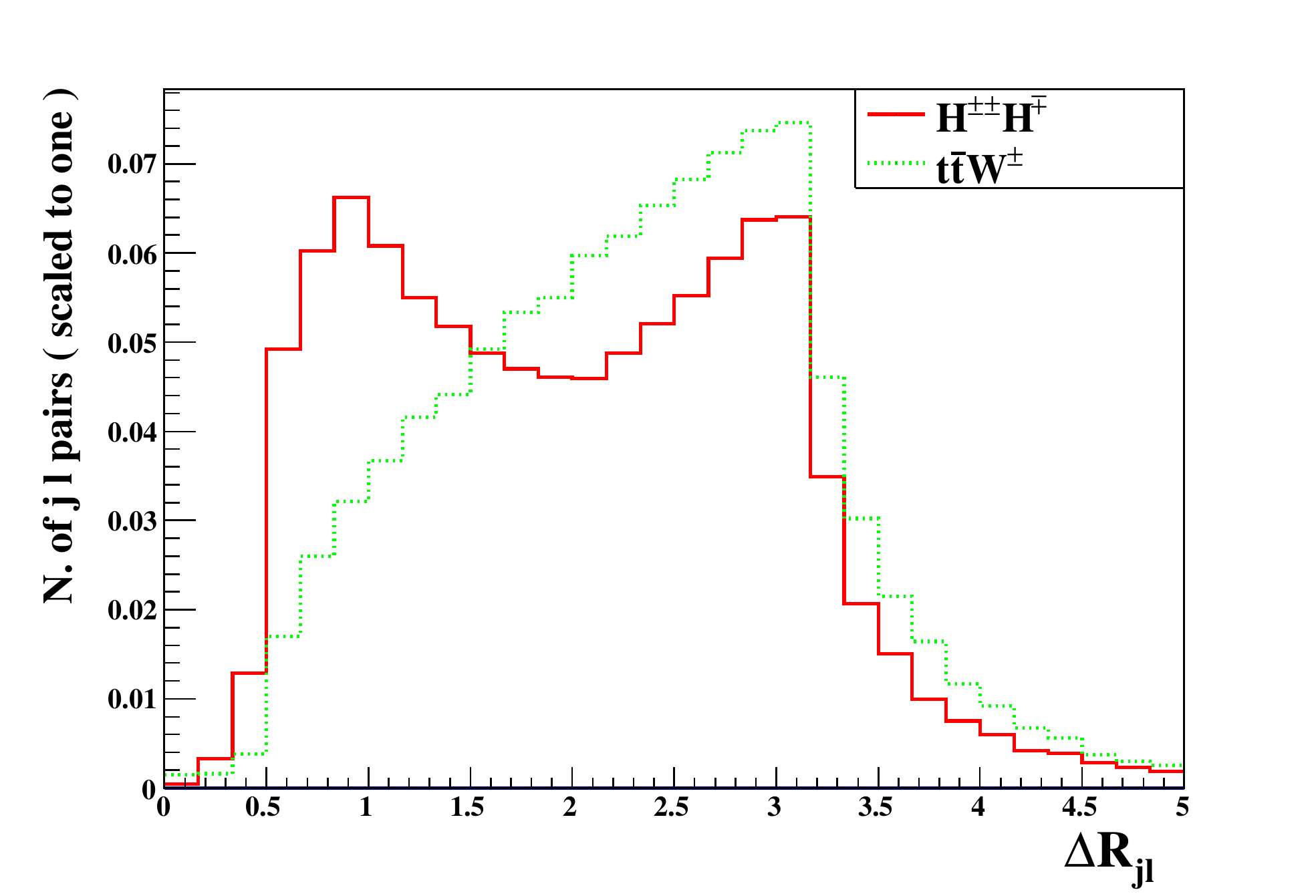}
\end{center}
\caption{Distributions of particle separations $\Delta R_{\ell\ell}$ (left panel) and $\Delta R_{j\ell}$ (right) after imposing the cuts $N_{\ell^{\pm}}=2$, $p_T^{\ell_1}<30~\GeV$, and Eq. (\ref{cut:ll4j}), at LHC14.
\label{fig:h++DRjl}}
\end{figure}

Figure \ref{fig:h++DRjl} displays the distributions of particle separations $\Delta R_{\ell\ell,\;j\ell}$ after imposing above cuts on both signal and $t\bar{t}W^{\pm}$ background. It is evident that the leptons and jets from the $t\bar{t}W^{\pm}$ background are more isolated than those from the signal. We can thus further purify the signal by demanding
\begin{equation}
\Delta R_{\ell^{\pm}\ell^{\pm}} < 2.0,\;\Delta R_{j\ell} < 2.0.
\end{equation}
We list in Table~\ref{tab:ll4jcut} the survival numbers of events, statistical significance, and signal to background ratio upon imposing the cuts. We see that all the cuts chosen here, especially those on $p_T^{\ell_1}$ and $\Delta R$, are efficient enough. The statistical significance can reach $4$, and we have about $16.5$ signal events at LHC14@300.

\begin{table} [!htbp]
\begin{center}
\begin{tabular}{|c|c|c|c|c|}
\hline
Cuts  &  $H^{\pm\pm}H^{\mp}$  &  $t\bar{t}W^{\pm}$  & $S/\sqrt{S+B}$ & $S/B$
\\
\hline
Basic Cuts & 549 & 3682 & 8.45 & 0.1463
\\
$N_{\ell^{\pm}}=2$ & 70.4  & 1588 & 1.73 & 0.0444
\\
~$p_T^{\ell_1} < 30$~GeV~ & 50.0 & 63.8 & 4.69 &0.785
\\
$\Delta R_{\ell^{\pm}\ell^{\pm}}<2.0$ & 40.2 & 23.6 & 5.04 &1.71
\\
$\Delta R_{j\ell}<2.0$ & 16.5 & 1.35 & 3.91 &12.2
\\
\hline
\end{tabular}
\end{center}
\caption{Survival numbers of signal $\ell^{\pm}\ell^{\pm}\cancel{E}_Tjjjj$ and background, statistical significance $S/\sqrt{S+B}$, and signal to background ratio $S/B$ upon imposing each cut sequentially at LHC14@300.}
\label{tab:ll4jcut}
\end{table}

\begin{figure}[!htbp]
\begin{center}
\includegraphics[width=0.45\linewidth]{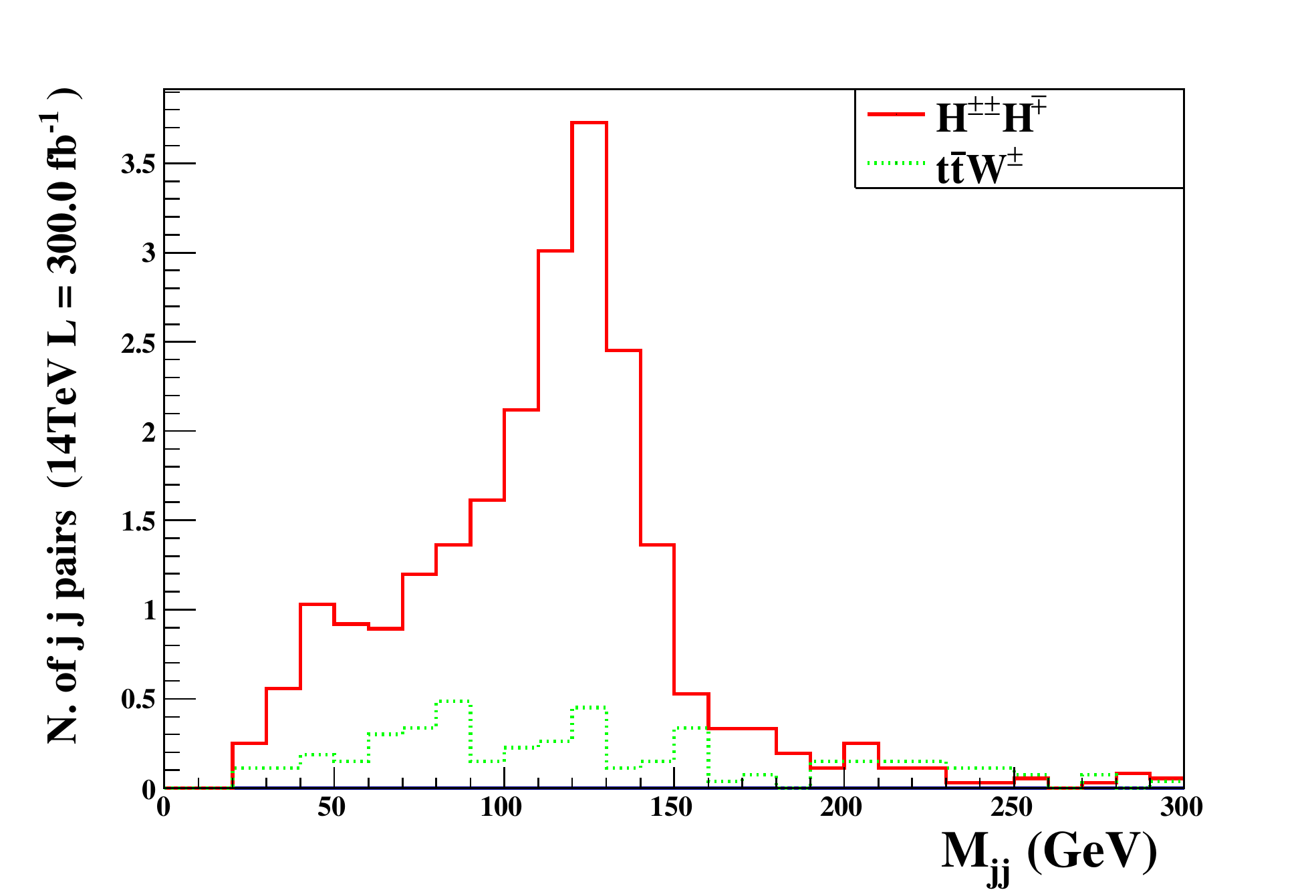}
\includegraphics[width=0.45\linewidth]{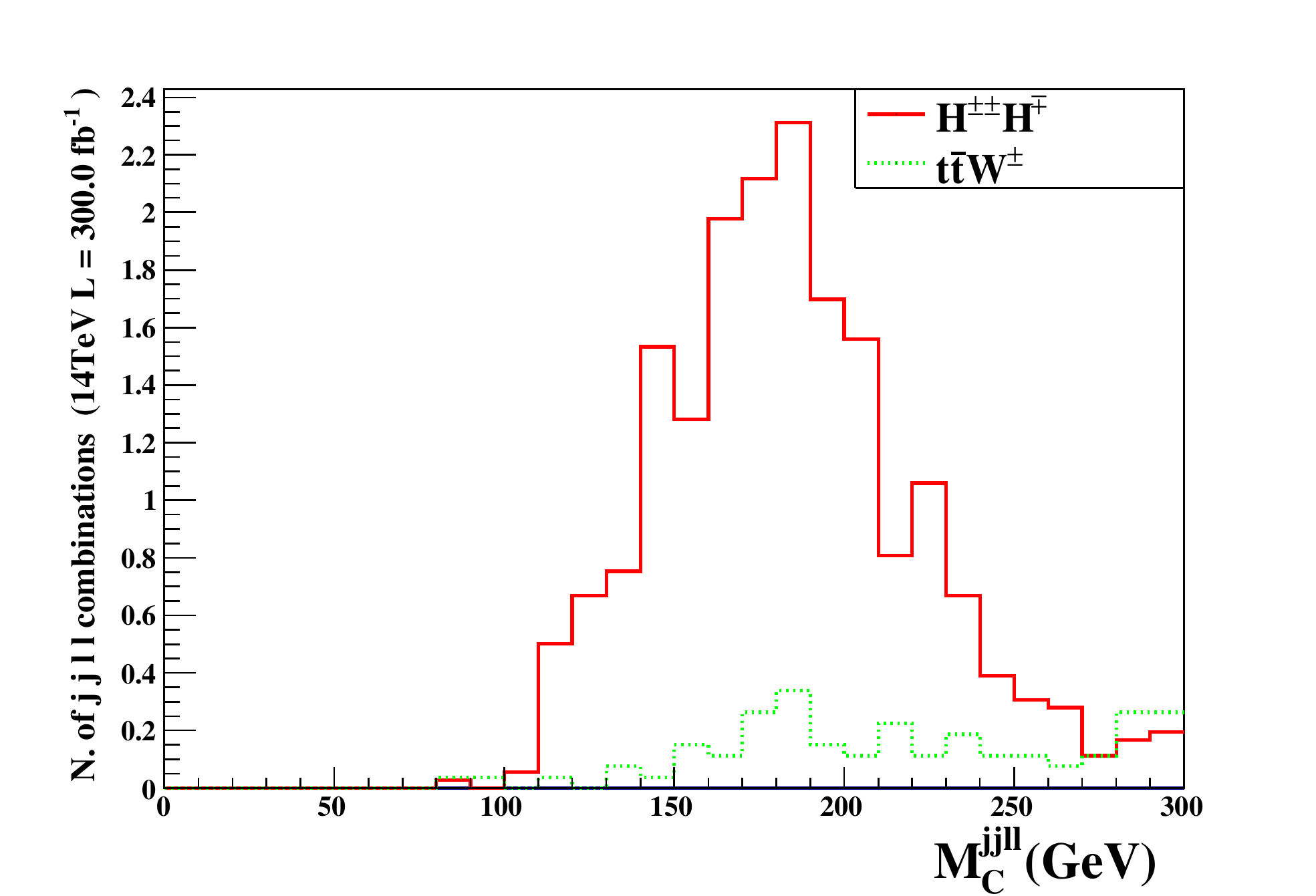}
\end{center}
\caption{Distributions in $M_{jj}$ and $M_C^{jj\ell\ell}$ in signal $\ell^{\pm}\ell^{\pm}\cancel{E}_Tjjjj$ at LHC14@300.
\label{fig:h++Mbb}}
\end{figure}

All new scalars appear in this signal channel. The neutral ones $H^0,~A^0$ decay with no missing particles and can be fully reconstructed by using the two $b$-jet invariant mass $M_{jj}$. The charged scalars $H^{\pm\pm},~H^\pm$ decay with missing energy and can be partially reconstructed with the help of the  cluster transverse mass:
\begin{equation}
M_C^{jj\ell\ell}=\sqrt{\Big(\sqrt{p^2_{T,jj\ell\ell}+M^2_{jj\ell\ell}}+\cancel{E}_T \Big)^2-\Big(\vec{p}_{T,jj\ell\ell}+\vec{\cancel{E}}_T \Big)^2}.
\end{equation}
The distributions of these two variables are plotted in Fig.~\ref{fig:h++Mbb}. One can see the peaks at $M_{H^0,A^0}=130~\GeV$ and $M_{H^{\pm\pm}}=185~\GeV$ respectively, although the numbers of events are limited.

Here we would like to comment on some similarities and differences between cascade decay signal from $H^{\pm\pm}H^{\mp}$ in our study and like-sign diboson signal from $H^{++}H^{--}$ in previous studies \cite{Perez:2008ha,Chiang:2012dk}. First, if we do not use $b$-tagging, the signals are the same $\ell^{\pm}\ell^{\pm}\cancel{E}_Tjjjj$ for both cases. But in practice, the jets are quite different. In the cascade decay, all four jets are $b$-jets, which come from decays of the neutral scalars $H^0,~A^0$, while in the diboson case, all four jets are light ones from decays of like-sign $W^{\pm *}$s. The invariant mass $M_{jj}$ can be used to distinguish between these two signals, since $M_{H^0/A^0}$ is usually heavier than $W^\pm$. Also considered in Ref. \cite{Perez:2008ha} are the contributions from gauge decays of $H^{\pm\pm}H^{\mp}$, which could produce signal $\ell^{\pm}\ell^{\pm}\cancel{E}_Tb\bar{b}jj$. So according to the number of $b$-jets, the original sources of the jets system can be clearly distinguished from each other once $b$-tagging is applied.

\begin{figure}[!htbp]
\begin{center}
\includegraphics[width=0.45\linewidth]{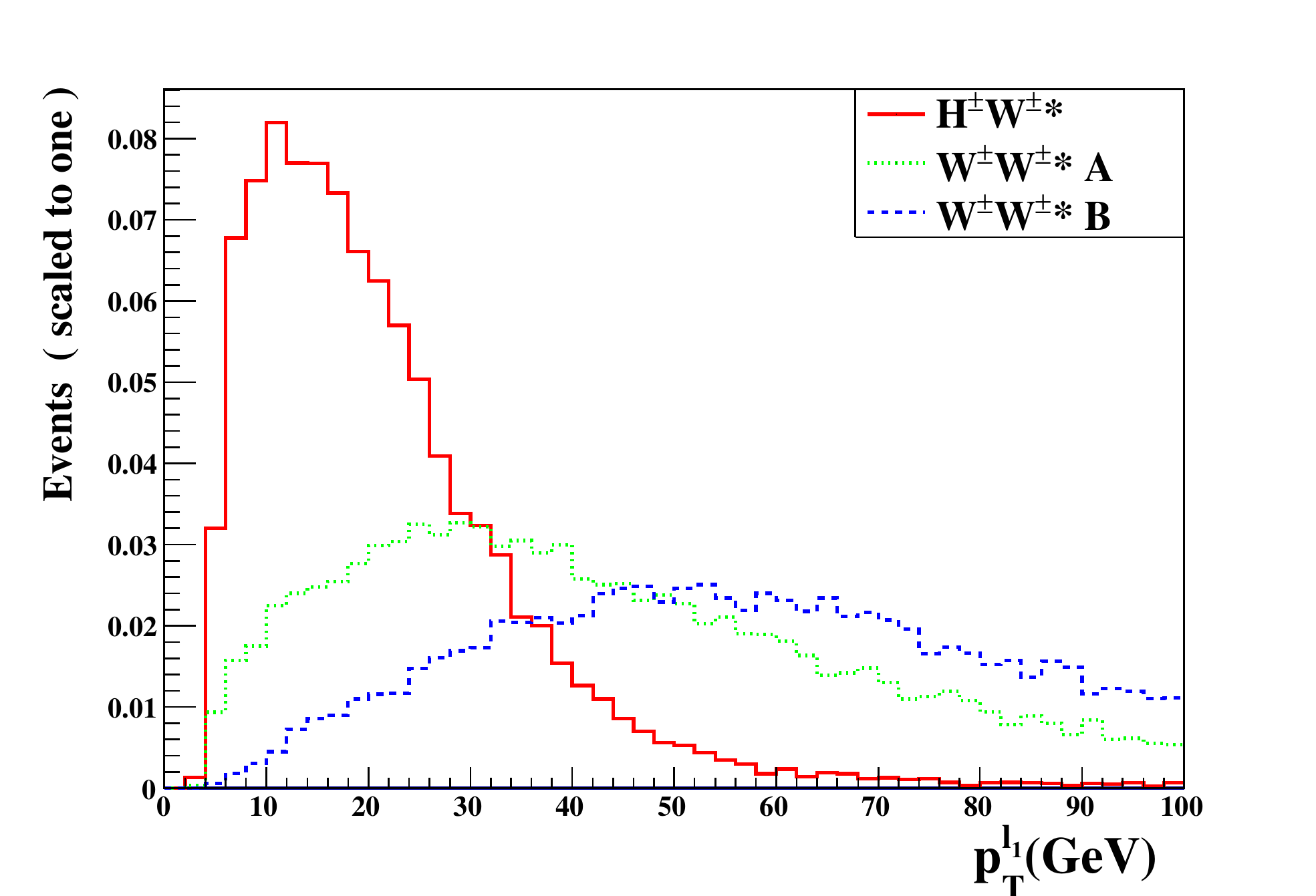}
\includegraphics[width=0.45\linewidth]{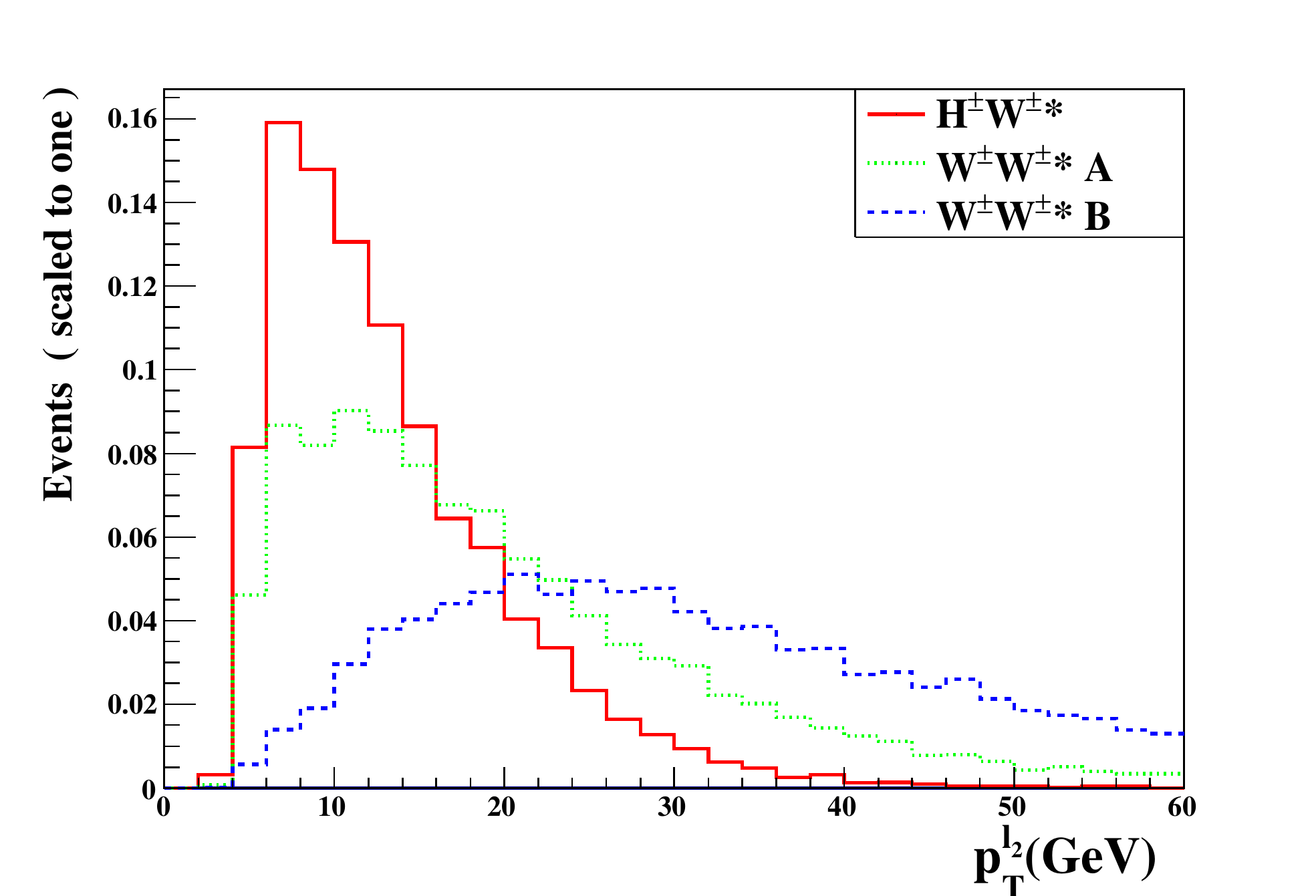}
\includegraphics[width=0.45\linewidth]{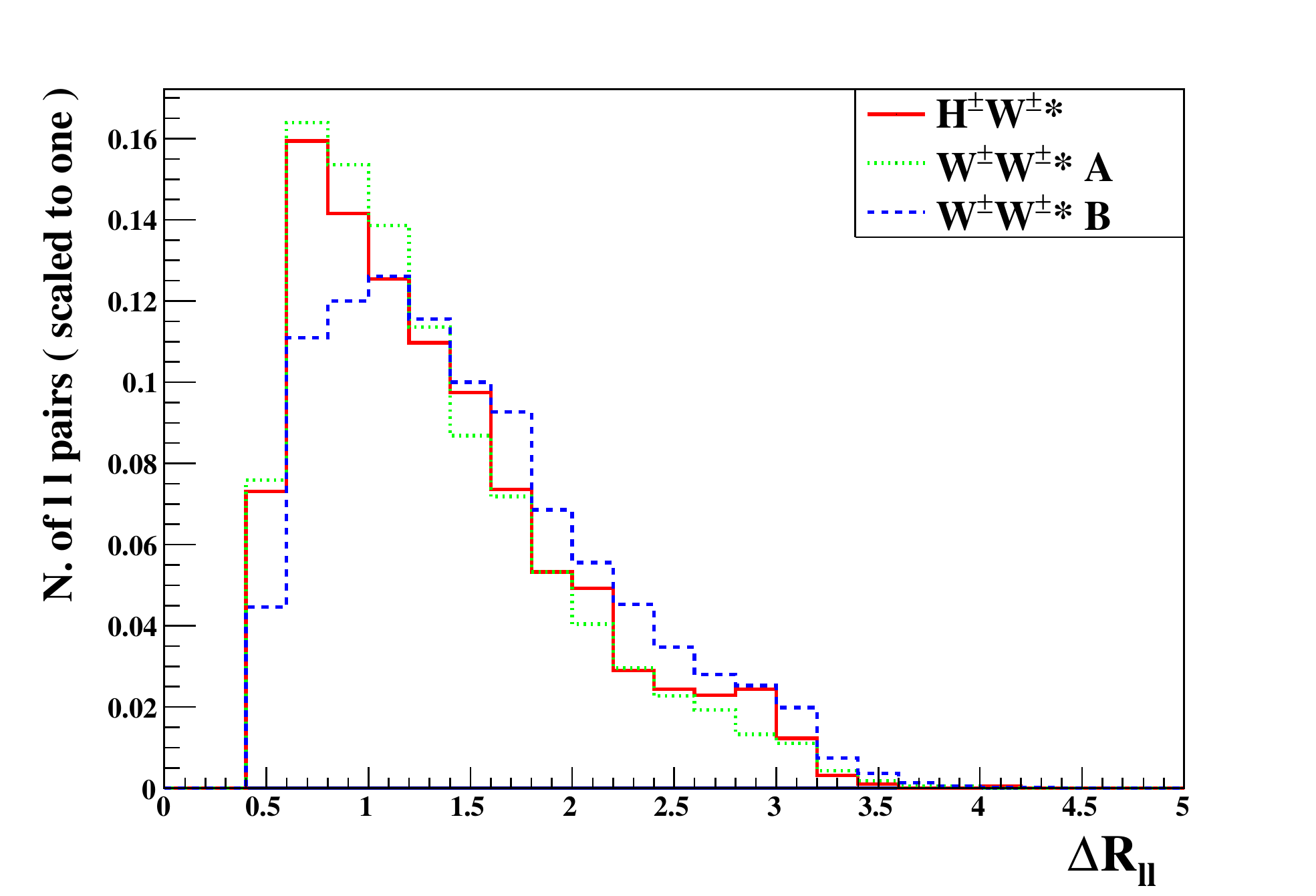}
\includegraphics[width=0.45\linewidth]{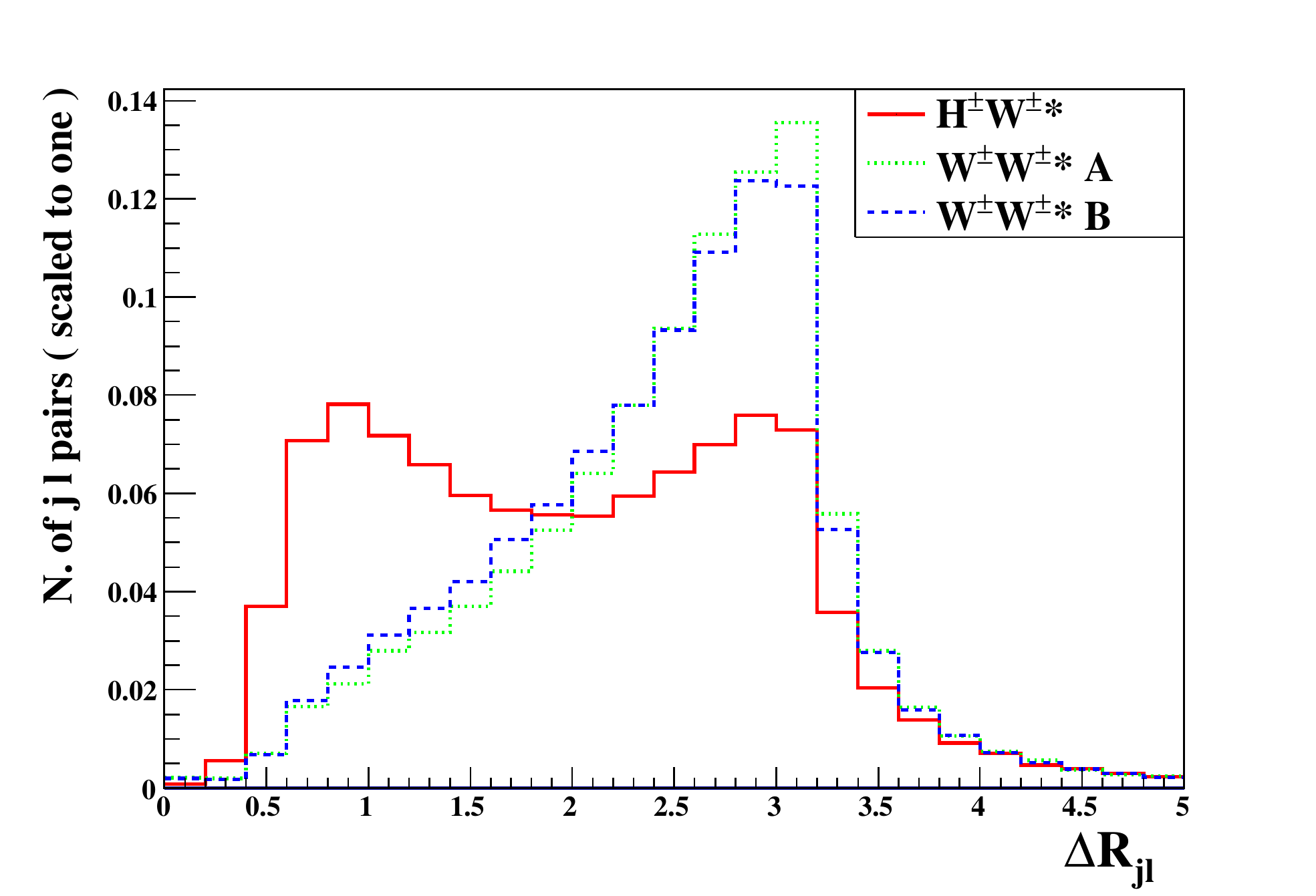}
\includegraphics[width=0.45\linewidth]{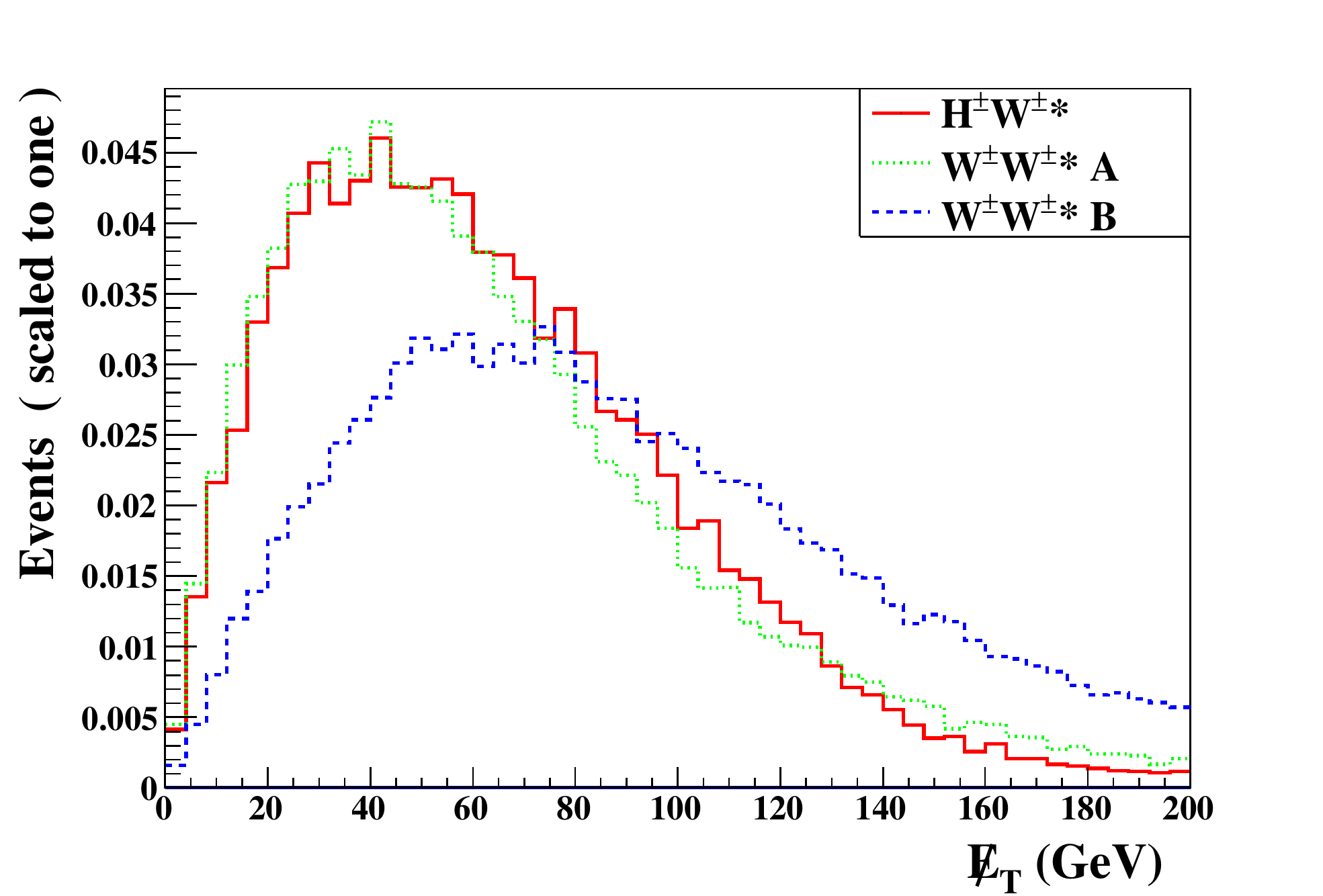}
\includegraphics[width=0.45\linewidth]{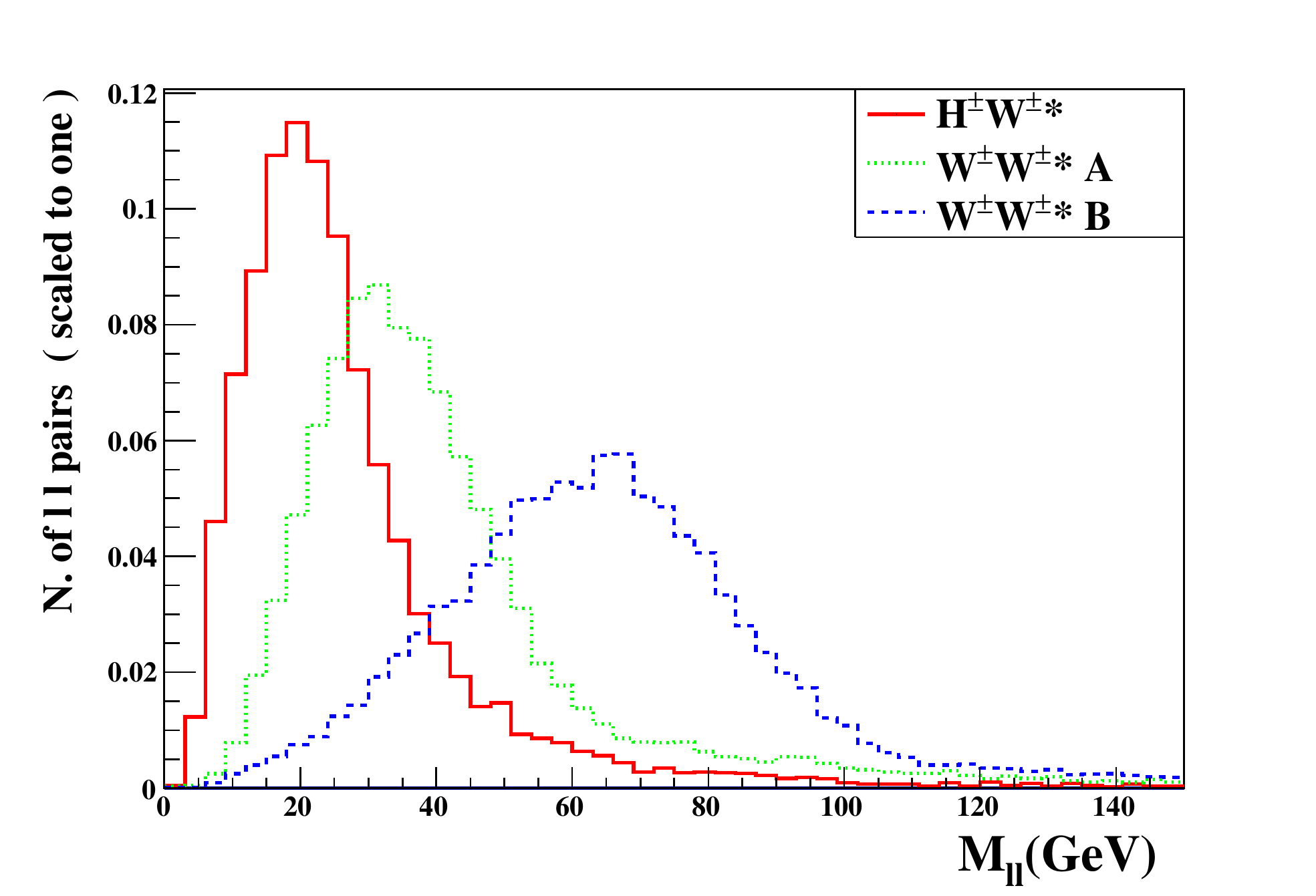}
\end{center}
\caption{Distributions of $p_T^{\ell_1,\ell_2}$, $\Delta R_{\ell\ell,j\ell}$, $\cancel{E}_T$ and $M_{\ell\ell}$ in cascade decay $H^{\pm\pm} \to H^{\pm}W^{\pm*}$ and like-sign diboson decay $H^{\pm\pm} \to W^{\pm}W^{\pm*}$ in the signal event $\ell^{\pm}\ell^{\pm}\cancel{E}_Tjjjj$ at LHC14.
\label{ll4jc}}
\end{figure}

Next, we look more closely into the like-sign dilepton signal. We note that the like-sign dilepton from cascade decays of $H^{\pm\pm}$ in our study is similar to that from leptonic decays of like-sign di-$W$ in the pair production of $H^{\pm\pm}$ but with a lower mass~\cite{Chiang:2012dk}:
\begin{eqnarray}
\textrm{cascade decay: }&& pp\to H^{\pm\pm} H^{\mp} \to H^{\pm}W^{\pm*}+\psi^0W^{\mp*} \to \ell^{\pm}\ell^{\pm}\cancel{E}_Tjj+jj(jj)\label{eq:cascade},
\\
\textrm{diboson decay: }&& pp\to H^{\pm\pm} H^{\mp\mp} \to W^{\pm}W^{\pm*}+W^{\mp}W^{\mp*} \to \ell^{\pm}\ell^{\pm}\cancel{E}_Tjjjj.
\end{eqnarray}
Again, $(jj)$ in Eq.(\ref{eq:cascade}) denotes undetectable soft jets. For a comparative study, we use the masses as shown in Eq. (\ref{massspectrum}) for our cascade decay channel, while in the diboson channel a lower mass $M_{H^{\pm\pm}}=90~\GeV$ (case A) or $M_{H^{\pm\pm}}=150~\GeV$ (case B) is assumed as in Ref. \cite{Chiang:2012dk}. The distributions of some kinematical variables with no cuts are depicted in Fig. \ref{ll4jc}.

We note some features in the distributions. First, the leading lepton transverse momentum $p_T^{\ell_1}$ is quite different in the three cases. In the diboson decay, a heavier $M_{H^{\pm\pm}}$ usually implies a larger $p_T^{\ell_1}$, while $p_T^{\ell_1}$ from cascade decay is generally much softer. For the next-to-leading lepton transverse momentum $p_T^{\ell_2}$, the cascade decay and diboson decay in case A are similar. Second, the distributions of $\Delta R_{\ell\ell}$ are similar in all three cases, but the distributions of $\Delta R_{j\ell}$ can be very different. This difference arises mainly because in cascade decay the dilepton and two of four jets are from one single $H^{\pm\pm}$ and the other two jets from $\psi_0$, resulting in a $\Delta R_{j\ell}$ with two distinct peaks, while the dilepton and jets in diboson decay come from different $H^{\pm\pm}$s and are thus more isolated with a single peak in $\Delta R_{j\ell}$. Third, the distributions in missing transverse momentum $\cancel{E}_T$ are almost the same in cascade decay and diboson decay (case A). Finally, the distributions of dilepton invariant mass $M_{\ell\ell}$ are also distinguishable, with a peak around $20,~30,~70~\GeV$ in cascade decay, case A and B, respectively.
To summarize in a word, the distributions of $p_T^{\ell_1}$, $\Delta R_{j\ell}$, and $M_{\ell\ell}$ can be used to distinguish among these three cases of signals.

\subsection{Signals for singly-charged scalars}
\label{negative-Singly}

The LHC physics of the singly-charged scalars $H^{\pm}$ in type-II seesaw has been
studied in many previous papers but only for the degenerate case, see, e.g., Refs. \cite{Perez:2008ha,Perez:2008zc,delAguila:2008cj}.
In this subsection, we carry out a detailed study for the nondegenerate case in the negative
scenario. As one can see in Fig. \ref{fig:H+decay}, the cascade decay
$H^{\pm} \to \psi^0W^{\pm*}$ ($\psi^0=H^0,~A^0$) dominates in a wide interval of the triplet vev, $10^{-6}~\GeV< v_{\Delta}< 1 ~\GeV$. For $v_{\Delta}\lesssim 10^{-3}~\GeV$, $\psi^0$ decays to $\nu\nu$; more interesting is the case $v_{\Delta}\gtrsim 10^{-3}~\GeV$, when $A^0$ decays mainly to $b\bar{b},~\tau^+\tau^-$, and $H^0$ to $b\bar{b},~\tau^+\tau^-,~W^+W^-$. We therefore consider first the following channel
\footnote{We notice that the same $\ell^{\pm}\cancel{E}_Tb\bar{b}b\bar{b}$ signal could also appear in MSSM via the $H^{\pm}H^0/A^0$ associated production~\cite{Cao:2003tr}:
\begin{equation*}
pp\to H^{\pm}H^0/A^0 \to t\bar{b}/\bar{t}b+b\bar{b} \to W^{\pm}b\bar{b}b\bar{b}\to\ell^{\pm}\cancel{E}_Tb\bar{b}b\bar{b},
\end{equation*}
where in this case $M_{H^{\pm}}$ is heavier than the top quark. Besides different thresholds of $H^{\pm}$, the decay modes of $H^{\pm}$ can also be used to distinguish between the two scenarios. One distinct feature is that the $W$ bosons in our cascade decays are always off-shell while those in MSSM from top decays are on-shell. Another one concerns the separation of the two $b$-jets in $H^{\pm}$ decays. In cascade decays, the $b$-jets originate from $H^0/A^0$ decays, while in MSSM one comes directly from $H^{\pm}$ and the other from tops, resulting in generally more separated two $b$-jets than in cascade decays.}:
\begin{equation}
pp\to H^{\pm} \psi^0 \to W^{\pm*}\psi^0\psi^0 \to \ell^{\pm} \cancel{E}_T+b\bar{b}b\bar{b}.
\label{eq_single1}
\end{equation}
Since the charged lepton and neutrino come from an off-shell $W$, their energies are relatively smaller than those from an on-shell $W$ in the SM background. This feature can be utilized to improve the signal to background ratio efficiently.

The main SM backgrounds we considered are:
\begin{eqnarray}
&t \bar{t}& \to W^+b W^-\bar{b} \to \ell^{\pm}\cancel{E}_T+b\bar{b}jj,\\
&t\bar{t}b\bar{b}& \to W^+b W^-\bar{b} b\bar{b}\to \ell^{\pm}\cancel{E}_T+b\bar{b}b\bar{b}+jj/\ell^{\mp},\\
&W^{\pm}b\bar{b}b\bar{b}& \to  \ell^{\pm} \cancel{E}_T+b\bar{b}b\bar{b},
\end{eqnarray}
where the first two are reducible and the last one is irreducible background. For the $t\bar{t}$ background, we only have to consider the semi-leptonic decay channel with the two light jets mis-tagged as $b$-jets. For $t\bar{t}b\bar{b}$, both semi-leptonic and di-leptonic decays of $t\bar{t}$ are involved. In the di-leptonic decay channel, we require that one of the charged leptons escape detection. The identification efficiencies for $e$ and $\mu$ are taken to be $0.9$ and $0.95$ respectively, and $L=30~\fb^{-1}$ is assumed at 14 TeV LHC (LHC14@30).

\begin{table} [!htbp]
\begin{center}
\begin{tabular}{|c|c|c|c|c|c|c|}
\hline
Cuts  &  $H^{\pm}\psi^0$  &  $t\bar{t}$ &  $t\bar{t}b\bar{b}$ & $W^{\pm}b\bar{b}b\bar{b}$ & $S/\sqrt{S+B}$ & $S/B$
\\
\hline
Basic Cuts & 567& 2039202& 162410 & 1295 & 0.373 & 0.000256
\\
$N_{b}=4,~N_{j}=0,~N_{l}=1$ & 22.2 & 317 & 218 & 3.45 & 0.940 & 0.0412
\\
$p_T^{\ell}<30$~GeV & 18.4 & 69.6 & 53.7 & 0.488 & 1.54 & 0.148
\\
$\cancel{E}_T<30$~GeV & 10.6 & 5.53 & 3.80 & 0.0916 & 2.37 & 1.13
\\
$M_T^{\ell}<30$~GeV & 9.00 & 1.63 & 1.98 & 0.0305 & 2.53 & 2.47
\\
$\Delta R_{b\ell}<2.0$ & 8.94 & 1.30 & 1.32 & 0.0305 & 2.63 & 3.37
\\
\hline
\end{tabular}
\end{center}
\caption{Survival numbers of events, statistical significance $S/\sqrt{S+B}$ and  signal to background ratio $S/B$ after imposing each cut sequentially at LHC14@30.}
\label{tab:lvbbbbcut}
\end{table}

To isolate the signal, we employ the following basic cuts:
\begin{eqnarray}
\label{cut:l4b}
&&p_T^{\ell}>10~\GeV,\;|\eta_{\ell}| < 2.5,
\nonumber\\
&&p_T^j>20~\GeV,\;|\eta_j| < 2.5,
\nonumber\\
&&\Delta R_{\ell\ell}>0.4,\;\Delta R_{j\ell}>0.4,\;\Delta R_{jj}>0.4.
\end{eqnarray}
We found that for $M_{H^{\pm}}-M_{\psi^0}=30~\GeV$, the efficiency of basic cuts for a signal lepton is about $0.77$. To further purify the signal, we first apply the event selection:
\begin{equation}
N_{b}=4,~N_{j}=0,~N_{\ell}=1,
\end{equation}
which means that we only choose the events which exactly contain four $b$-jets and one charged lepton. We found that better $b$-tagging efficiency could potentially suppress the leading background coming from the $t\bar{t}$ channel. The next to leading background is $t\bar{t}b\bar{b}$ channel, while the events from $W^{\pm}b\bar{b}b\bar{b}$ are already smaller than the signal.

\begin{figure}[!htbp]
\begin{center}
\includegraphics[width=0.45\linewidth]{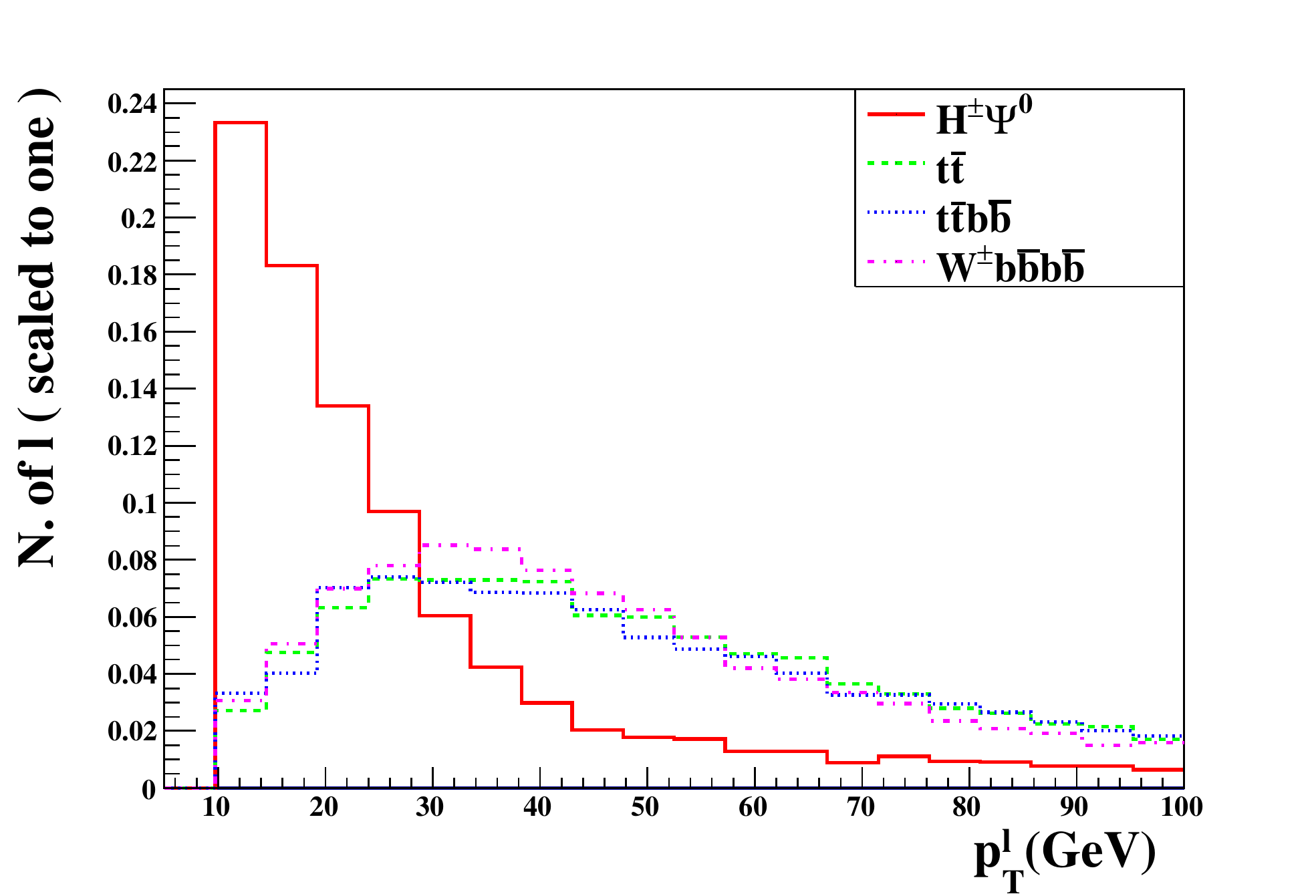}
\includegraphics[width=0.45\linewidth]{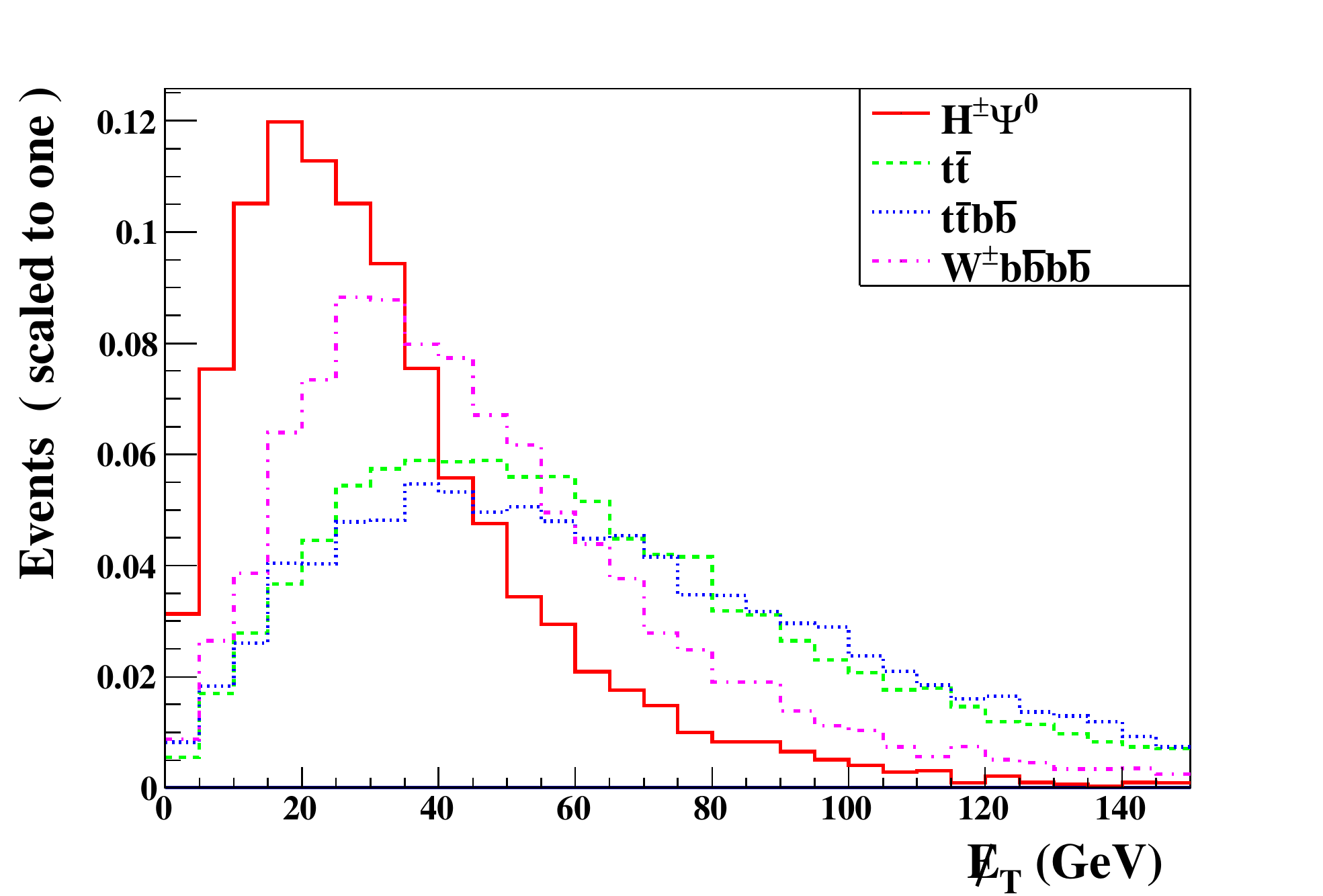}
\includegraphics[width=0.45\linewidth]{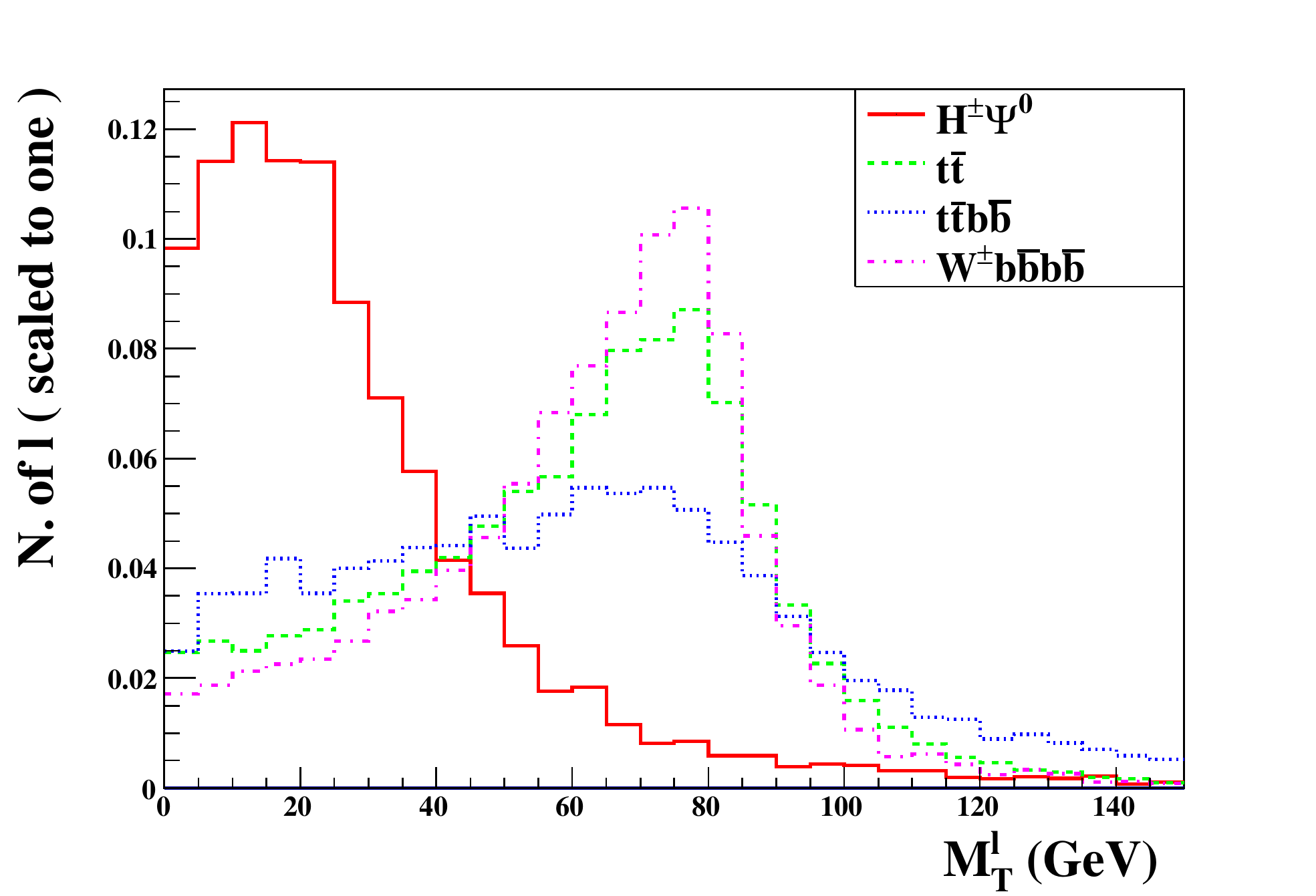}
\includegraphics[width=0.45\linewidth]{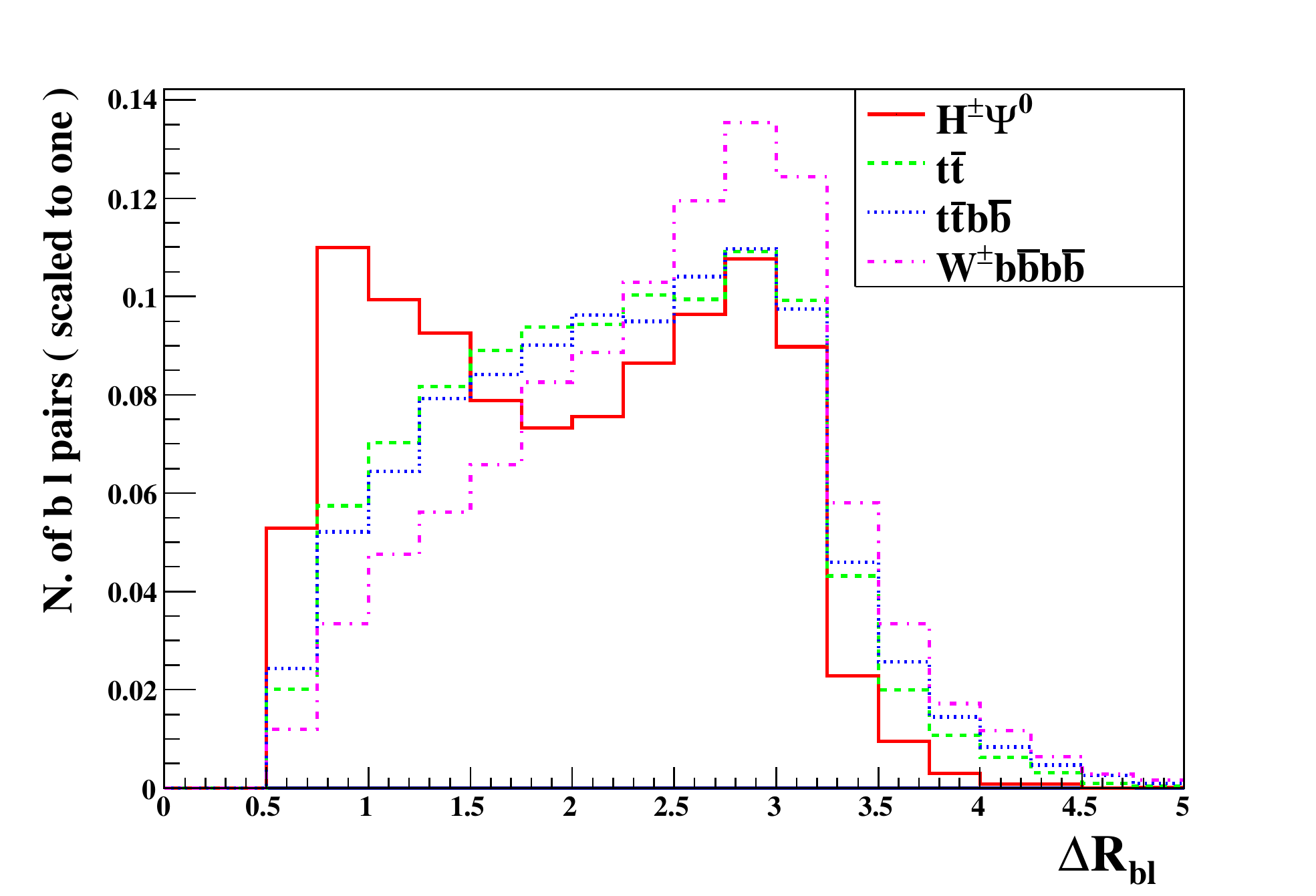}
\end{center}
\caption{Distribution of $p_T^{\ell}$, $\cancel{E}_T$, $M_T^l$, and $\Delta R_{b\ell}$ after imposing basic cuts for signal $\ell^{\pm} \cancel{E}_T b\bar{b}b\bar{b}$ and its backgrounds.
\label{fig:4blvlPT}}
\end{figure}

To suppress further the background, we employ several kinematical cuts that are largely designed on the off-shell nature of the $W$ in the signal versus the on-shell nature in the background. The various distributions are displayed in Fig.\ref{fig:4blvlPT} after imposing the basic cuts. The transverse momentum $p_T^{\ell}$ of leptons in all backgrounds has a peak at about $30\sim40~\GeV$, while $p_T^{\ell}$ in the signal is much softer. The distributions of the missing transverse energy $\cancel{E}_T$ share a similar feature. The off-shell or on-shell origin of leptons and neutrinos can be directly probed in the lepton transverse mass $M_{T}^{\ell}$ distribution. Since the mass splitting between $H^{\pm}$ and $H^0/A^0$ is $30~\GeV$, most signal events have $M_{T}^{\ell}<30~\GeV$. In contrast, the background events have a clear edge around $M_T^{\ell}\sim 80~\GeV$ and only a small number of them falls in the region $M_{T}^{\ell}<30~\GeV$. Finally, the distribution in $\Delta R_{b\ell}$ has two peaks for the signal, one around $0.9$ and the other around $3$. The smaller one most likely corresponds to the case when both particles are from the same cascade decay $H^{\pm}\to W^{\pm *}\psi^0\to  \ell^{\pm}\nu b\bar{b}$, while the larger one has the $b$-jet more likely from the decay of the directly produced $\psi^0$. The separation in the background is usually large and has a peak at $3$. These differences can be used to distinguish between the signal and backgrounds, through the following cuts:
\begin{equation}
p_T^{\ell} < 30 ~\GeV,~\cancel{E}_T<30~\GeV,~M_T^{\ell} < 30~\GeV,
~\Delta R_{b\ell} < 2.0.
\end{equation}

\begin{figure}[!htbp]
\begin{center}
\includegraphics[width=0.45\linewidth]{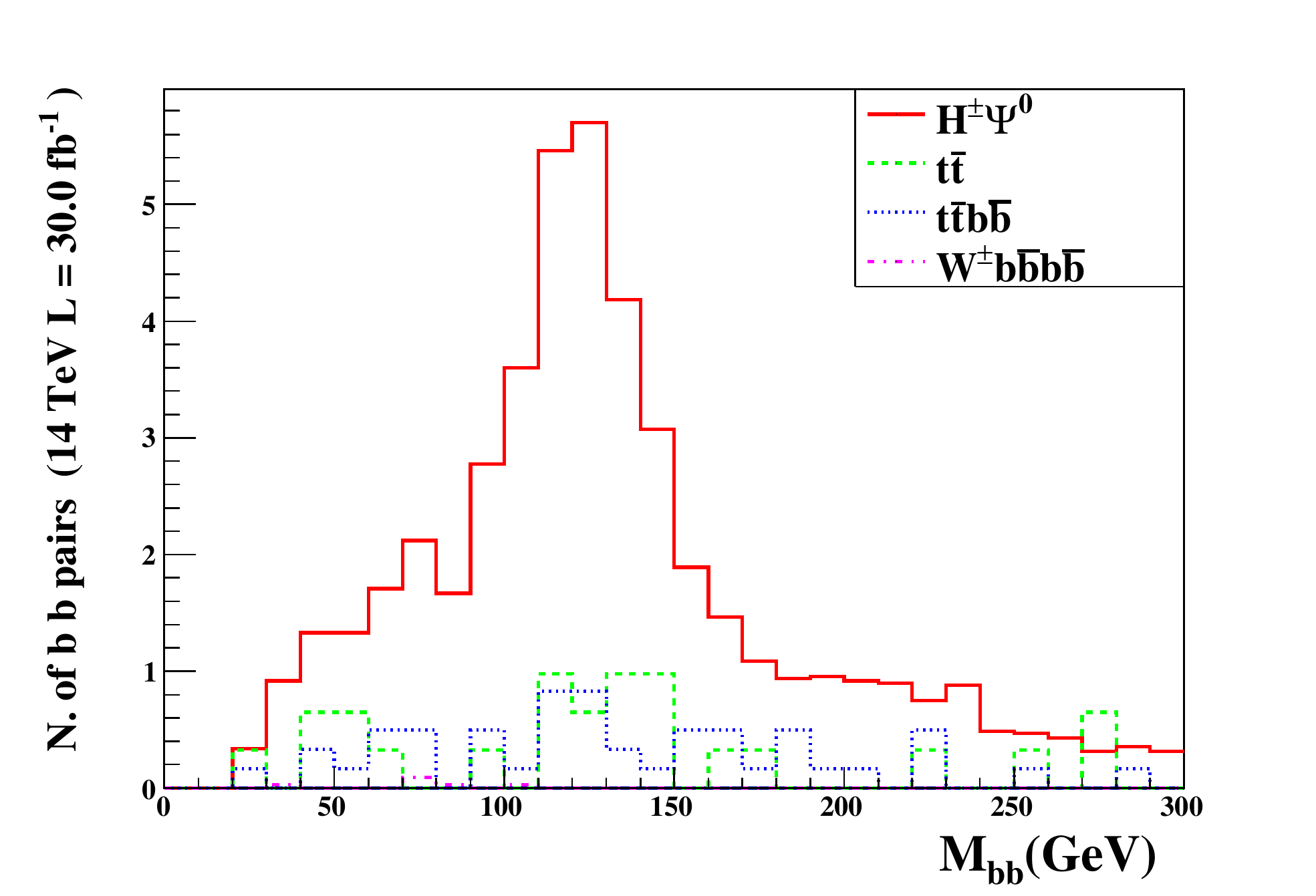}
\includegraphics[width=0.45\linewidth]{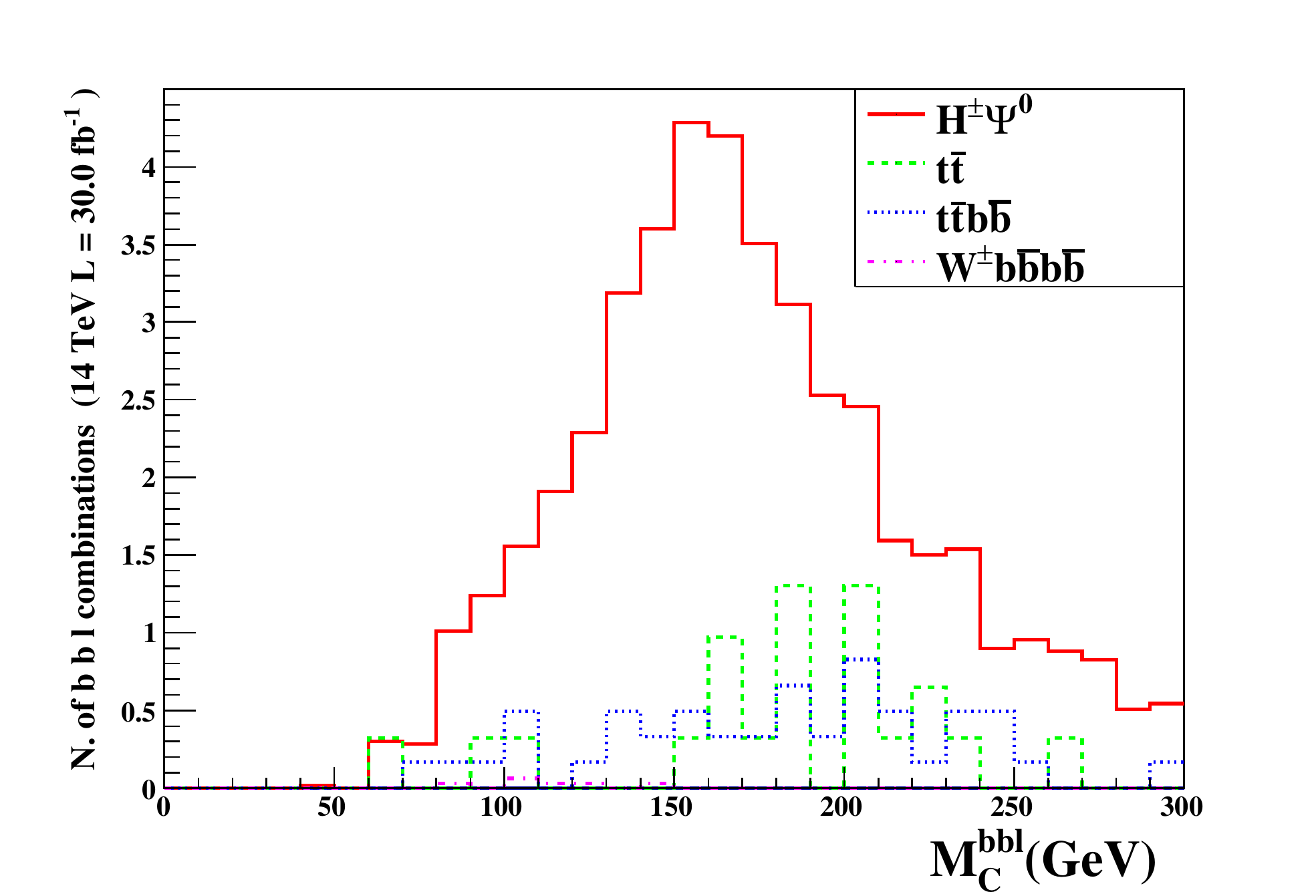}
\end{center}
\caption{Distributions of $M_{bb}$ and $M_C^{bb\ell}$ for $\ell^{\pm} \cancel{E}_T b\bar{b}b\bar{b}$ signal channel at LHC14@30.
\label{fig:h+phiMbb}}
\end{figure}

Table \ref{tab:lvbbbbcut} shows the survival numbers of events for the signal $\ell^{\pm}\cancel{E}_Tb\bar{b}b\bar{b}$ and backgrounds after the basic cuts and further sequential cuts, together with $S/\sqrt{S+B}$ and $S/B$. The cuts are efficient enough in preserving the signal while suppressing the background. At LHC14@30, about $9$ signal events survive with a statistical significance of $2.63$ and a signal to background ratio of $3.37$. Then for a $5\sigma$ reach, the required luminosity is $109~\fb^{-1}$. As shown in Fig. \ref{fig:h+phiMbb}, the mass of $H^0/A^0$ can be fully reconstructed using the invariant mass of the two $b$-jets in the $H^0A^0$ production. Since $H^0$ and $A^0$ are degenerate to good precision, we see effectively one peak at $130~\GeV$ in the distribution of $M_{bb}$. To reconstruct the mass of $H^{\pm}$, we still use the cluster transverse mass:
\begin{equation}
M_C^{bb\ell}=\sqrt{\Big(\sqrt{p^2_{T,bb\ell}+M^2_{bb\ell}}+\cancel{E}_T \Big)^2-\Big(\vec{p}_{T,bb\ell}+\vec{\cancel{E}}_T\Big)^2},
\end{equation}
whose distribution is shown on the right panel of Fig. \ref{fig:h+phiMbb}.

For the associated $H^\pm\psi^0$ production, we consider the second channel:
\begin{equation}
p p \to H^{\pm}\psi^0 \to W^{\pm*}\psi^0\psi^0 \to \ell^{\pm} \cancel{E}_T+b\bar{b}\tau^+\tau^-.
\label{eq_single2}
\end{equation}
This process has a much smaller rate than the previous one in Eq. (\ref{eq_single1}), but its background is also tiny. We thus use LHC14@300 as an illustration. The $b/\tau$-tagging efficiency and mis-tagging rate are assumed to be the same as in our previous analysis. The main backgrounds are:
\begin{eqnarray}
&t\bar{t}\tau^+\tau^-& \to W^+bW^-\bar{b}\tau^+\tau^-
       \to  \ell^{\pm}\cancel{E}_T+b\bar{b}\tau^+\tau^-+jj/\ell^{\mp},\\
&t\bar{t}W^{\pm}& \to  W^{+}b W^-\bar{b} W^{\pm}
        \to   \ell^{\pm}\cancel{E}_T+b\bar{b}\tau^+\tau^-,\\
&W^{\pm}b\bar{b}\tau^+\tau^-& \to \ell^{\pm}\cancel{E}_T+b\bar{b}\tau^+\tau^-.
\end{eqnarray}
For the $t\bar{t}\tau^+\tau^-$ background, we consider both semi-leptonic and di-leptonic decays of $t\bar{t}$. One of the charged leptons in the di-leptonic channel is required to escape detection, while the additional two light jets in the semi-leptonic channel receive no further constraints. The two irreducible backgrounds $t\bar{t}W^{\pm}$ and $W^{\pm}b\bar{b}\tau^+\tau^-$ are also taken into account in our study. As in our previous analysis, we start from basic cuts. We apply the same basic cuts as in Eqn. (\ref{cut:l4b}) for the signal channel $\ell^{\pm}\cancel{E}_Tb\bar{b}b\bar{b}$. The evolution of the event numbers and $S/\sqrt{S+B}$ and $S/B$ upon each cut is summarized in Table \ref{tab:lvbbtatacut}. At the level of the basic cuts, the main background comes from $t\bar{t}\tau^+\tau^-$, while the two irreducible backgrounds are already smaller than the signal.

\begin{table} [!htbp]
\begin{center}
\begin{tabular}{|c|c|c|c|c|c|c|}
\hline
Cuts  &  $H^{\pm}\psi^0$  &  $t\bar{t}\tau^+\tau^-$ &  $t\bar{t}W^{\pm}$ & $W^{\pm}b\bar{b}\tau^+\tau^-$ & $S/\sqrt{S+B}$ & $S/B$
\\
\hline
Basic Cuts & 1905& 9935& 356 & 305 & 16.28 & 0.171
\\
$N_b=2,N_{\tau}=2,N_{\ell}=1$ & 62.8 & 76.1 & 8.22 & 3.36 & 5.12 & 0.717
\\
110~GeV$<M_{bb}<$150~GeV & 39.7 & 14.3 & 1.53 & 0.300 & 5.31 & 2.46
\\
$\Delta R_{bb}<2.0$ & 27.7 & 5.85 & 0.450 & 0.0900 & 4.74  & 4.30
\\
\hline
\end{tabular}
\end{center}
\caption{Survival numbers of signal $\ell^{\pm}\cancel{E}_Tb\bar{b}\tau^+\tau^-$ and its backgrounds, statistical significance $S/\sqrt{S+B}$, and signal to backgrounds ratio $S/B$ after imposing each cut sequentially at LHC14@300.}
\label{tab:lvbbtatacut}
\end{table}

We then apply the event cuts:
\begin{equation}\label{cut:btal}
N_b=2,N_{\tau}=2,N_{\ell}=1,
\end{equation}
to select the signal $\ell^{\pm}\cancel{E}_Tb\bar{b}\tau^+\tau^-$. The numbers of signal and background events are now of the same order. In contrast to our previous signal channel $\ell^{\pm}\cancel{E}_Tb\bar{b}b\bar{b}$, we see that requiring two $\tau$-jets from four jets final states are very useful to remove the backgrounds. When comparing to the signal channel $b\bar{b}\tau^+\tau^-$ in the $H^0A^0$ production, the additional charged lepton is also very helpful to suppress the backgrounds. Except for a smaller production rate, the signal $\ell^{\pm}\cancel{E}_Tb\bar{b}\tau^+\tau^-$ is the cleanest among what we have investigated.

\begin{figure}[!htbp]
\begin{center}
\includegraphics[width=0.45\linewidth]{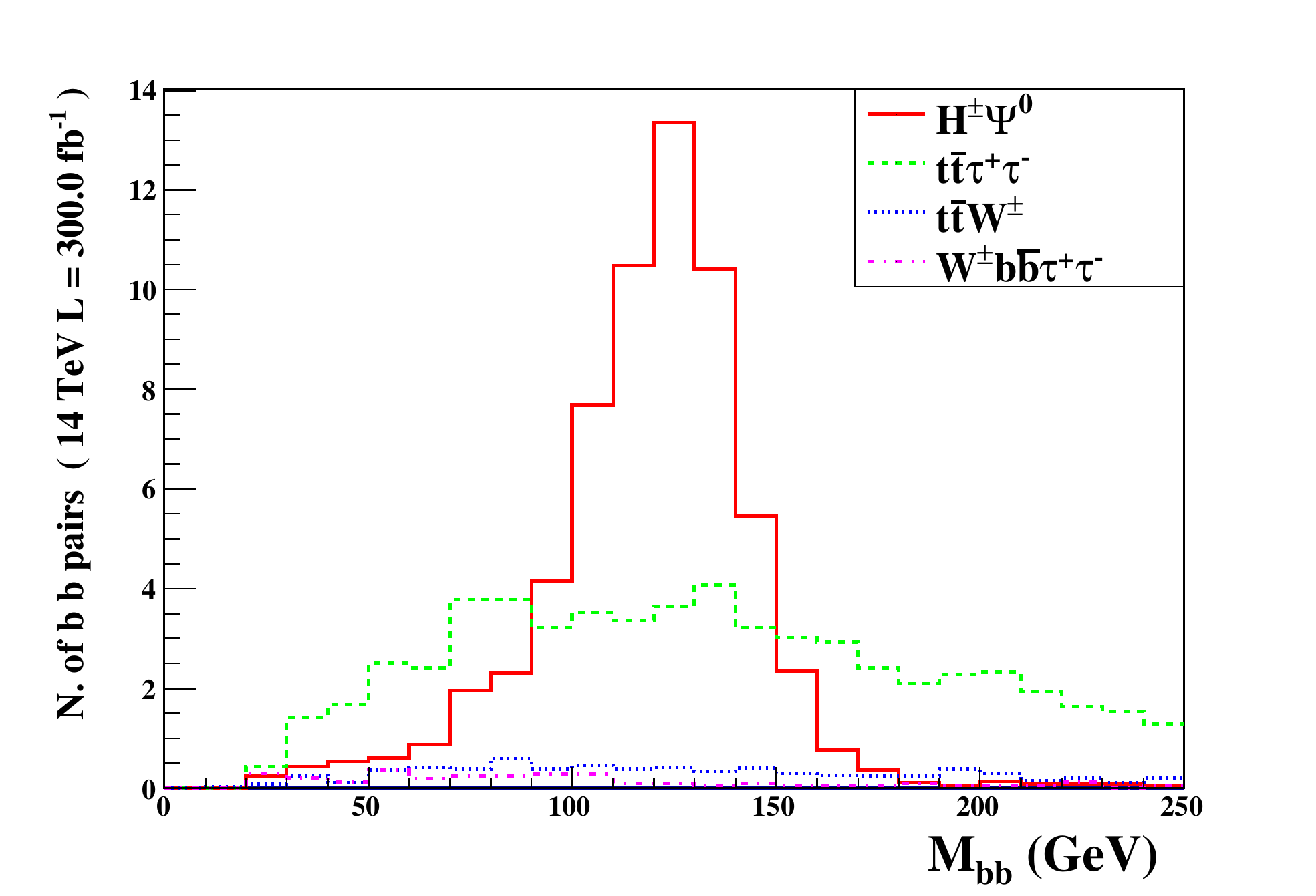}
\includegraphics[width=0.45\linewidth]{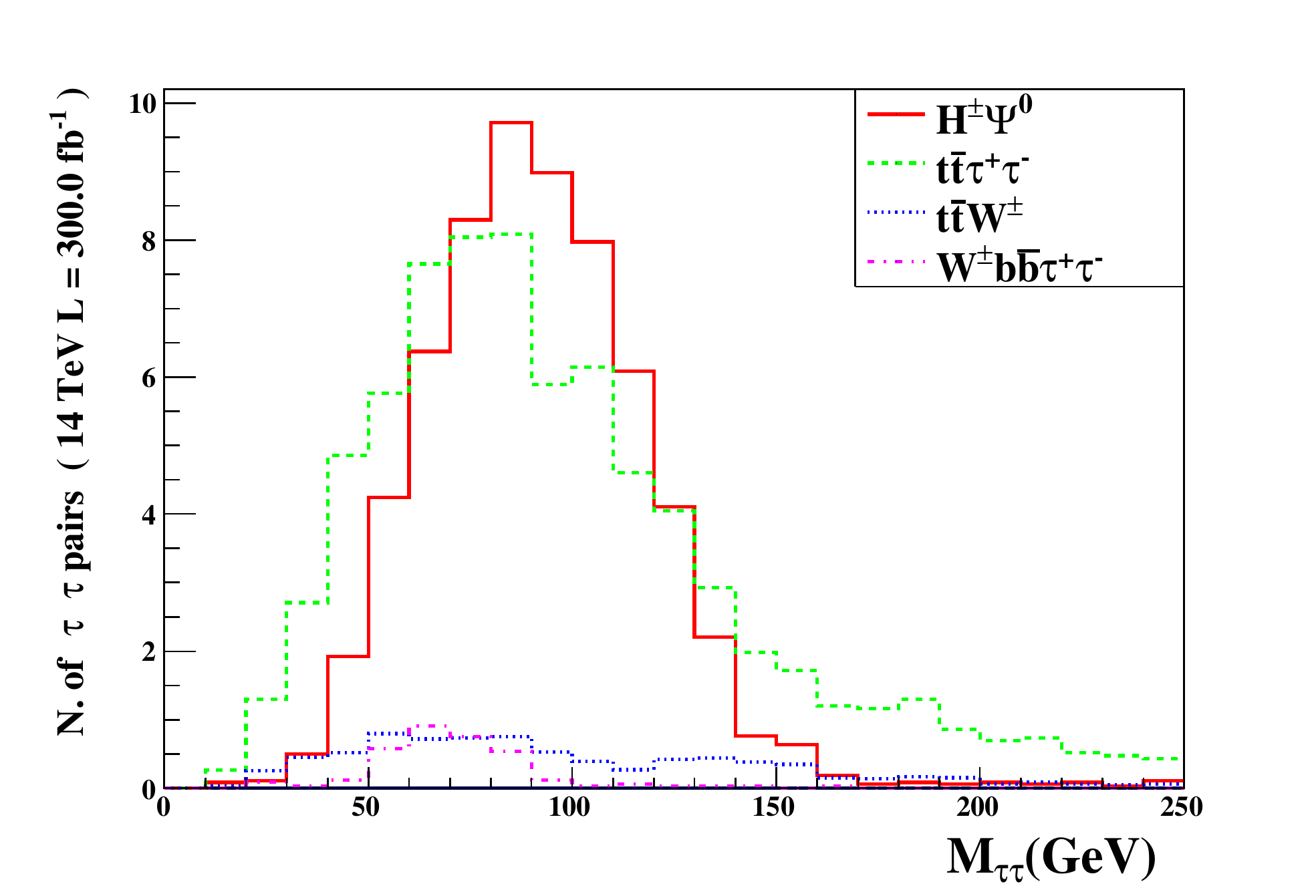}
\end{center}
\caption{Distributions of $M_{bb}$ and $M_{\tau\tau}$ for $\ell^{\pm} \cancel{E}_T b\bar{b}\tau^+\tau^-$ signal after imposing cuts in Eq. (\ref{cut:btal}) at LHC14@300.
\label{fig:bbtatalvlMbb}}
\end{figure}

\begin{figure}[!htbp]
\begin{center}
\includegraphics[width=0.45\linewidth]{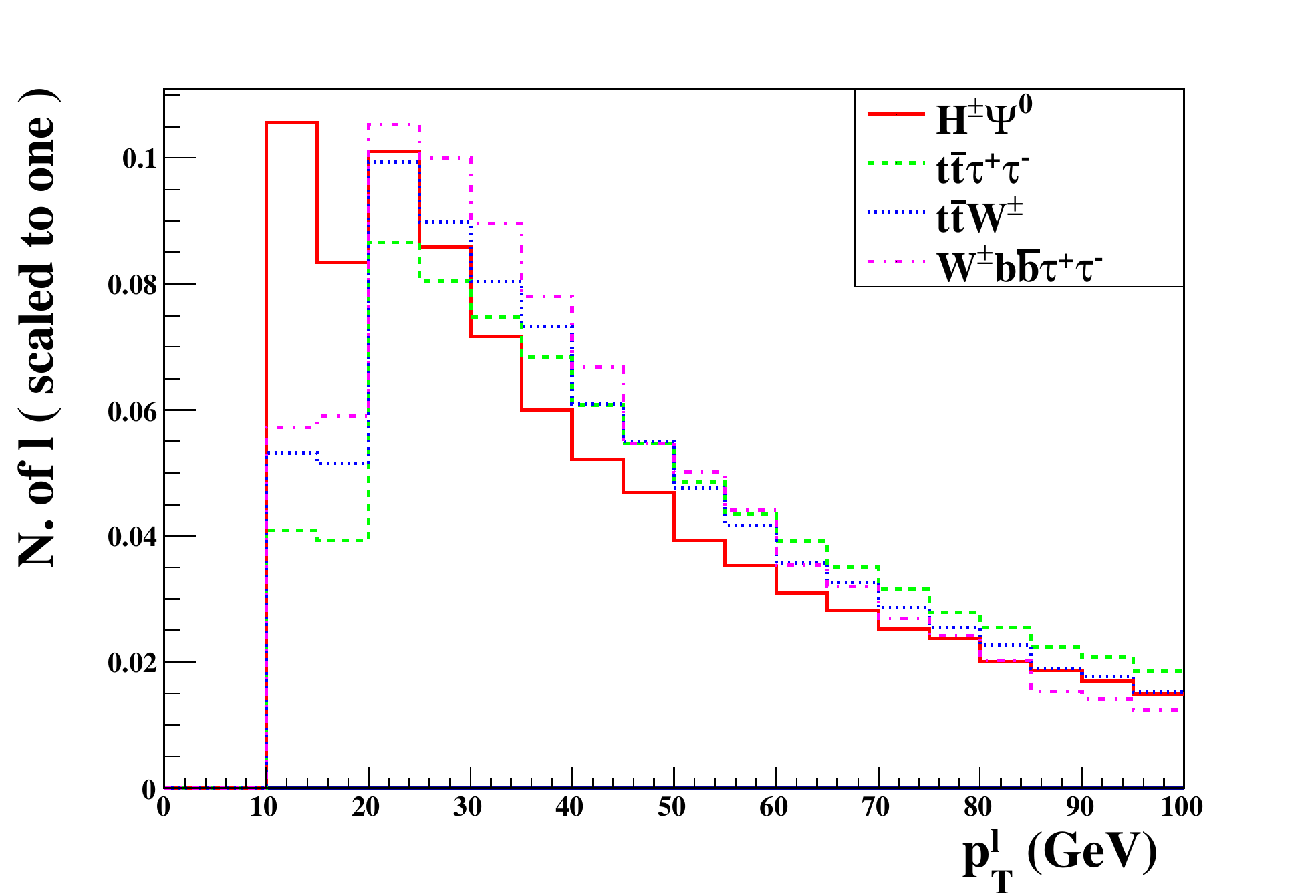}
\includegraphics[width=0.45\linewidth]{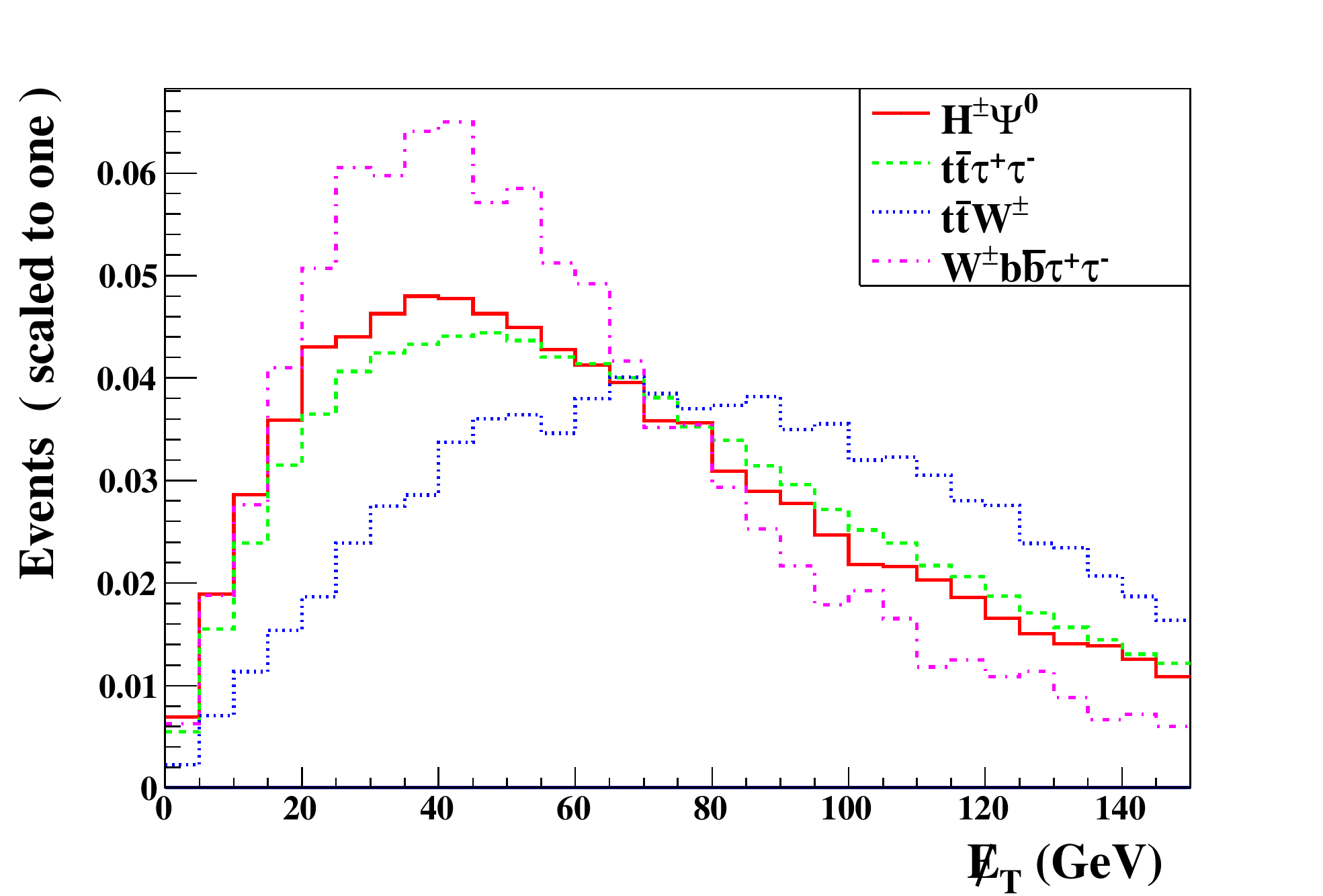}
\includegraphics[width=0.45\linewidth]{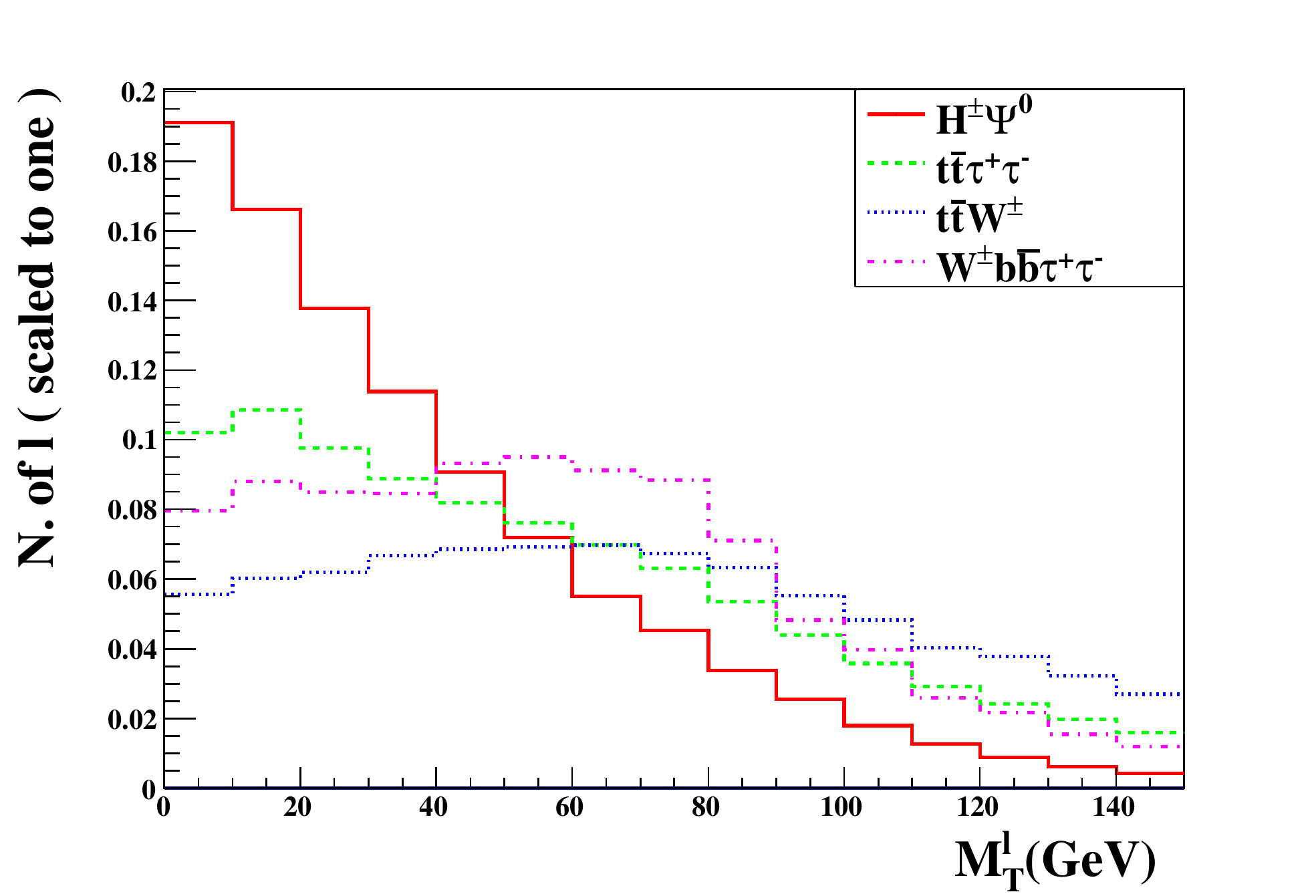}
\includegraphics[width=0.45\linewidth]{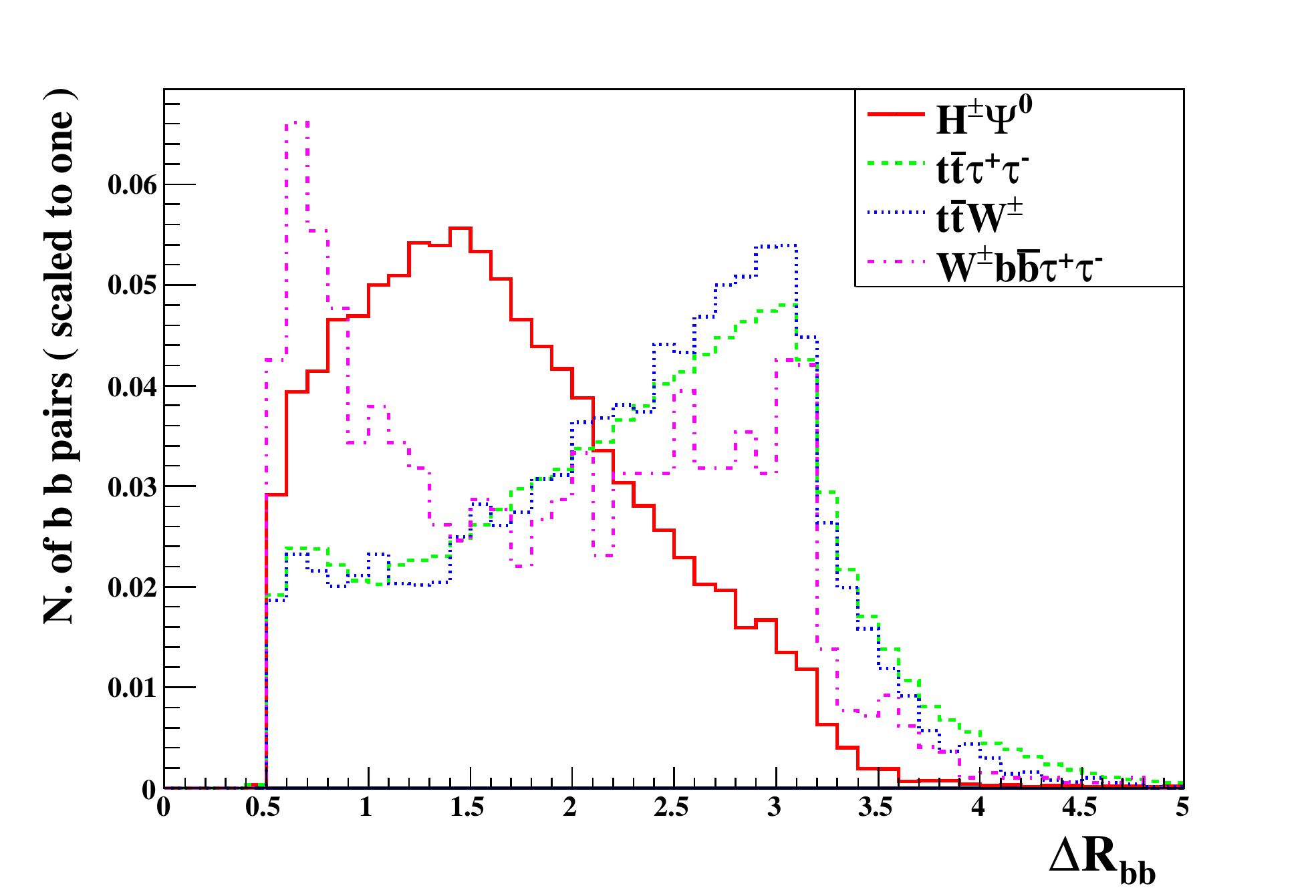}
\end{center}
\caption{Distributions of $p_T^{\ell}$, $\cancel{E}_T$, $M_T^{\ell}$, and $\Delta R_{bb}$ in signal $\ell^{\pm} \cancel{E}_T b\bar{b}\tau^+\tau^-$ at LHC14.
\label{fig:bbtatalvlPTl}}
\end{figure}

In Fig. \ref{fig:bbtatalvlMbb} we show the distributions of invariant masses $M_{bb}$ and $M_{\tau\tau}$ after implementing the cuts (\ref{cut:btal}). There is a clear peak at $130~\GeV$ for the signal in the $M_{bb}$ distribution, which can be employed to reconstruct the neutral scalars $H^0/A^0$. But the $M_{\tau\tau}$ distribution is not distinctive enough. There are two reasons for this difference. First, the $\tau$ pair in the background mainly comes from the on-shell $Z$ decay, whose  mass is close to that of the neutral scalars; second, the peak of $M_{\tau\tau}$ is shifted to the low mass region compared with the original value since we do not renormalize the momentum of the $\tau$ jets. To reconstruct the charged scalar, we need implement more cuts. We first narrow down to a window of $M_{bb}$:
\begin{equation}
110~\GeV< M_{bb} < 150 ~\GeV.
\end{equation}
Next we concentrate on the off-shell $W$ boson again. In Fig. \ref{fig:bbtatalvlPTl}, the distributions in $p_T^\ell$, $\cancel{E}_T$ and $M_T^\ell$ are plotted. They are changed significantly from those of the previous signal $\ell^{\pm}\cancel{E}_Tb\bar{b}b\bar{b}$. The leptonic decays of the $\tau$ lepton contribute to the final state leptons $\ell$. These fake leptons have a much harder transverse momentum $p_T^{\ell}$, and behave just like the backgrounds when $p_T^{\ell}>20~\GeV$.
The missing transverse momentum $\cancel{E}_T$ is also much larger, since additional (anti-)neutrinos carry away part of the energy in the hadronic decays of the $\tau$ leptons. This also alters the distribution of the lepton's transverse mass $M_T^{\ell}$. Even worse is that the distribution of $\cancel{E}_T$ happens to be similar to that for the largest $t\bar{t}\tau^+\tau^-$ background. Therefore, the cuts applied on the off-shell $W^{\pm*}$ would not be very effective.

Also shown in Fig. \ref{fig:bbtatalvlPTl} is the distribution in separation $\Delta R_{bb}$. The two $b$-jets in the signal come from the decay of one heavy neutral scalar while those in the background originate from the decays of $t\bar{t}$ respectively. This results in a smaller $\Delta R_{bb}$ for the signal than for the background. So we apply one more cut:
\begin{equation}
\Delta R_{bb} < 2.0.
\end{equation}
After all the cuts, we still have about $28$ signal events and $6$ background events, with a statistical significance around $4.7$ and a signal to background ratio $4.3$. This signature is therefore also promising at LHC14@300.

\begin{figure}[!htbp]
\begin{center}
\includegraphics[width=0.5\linewidth]{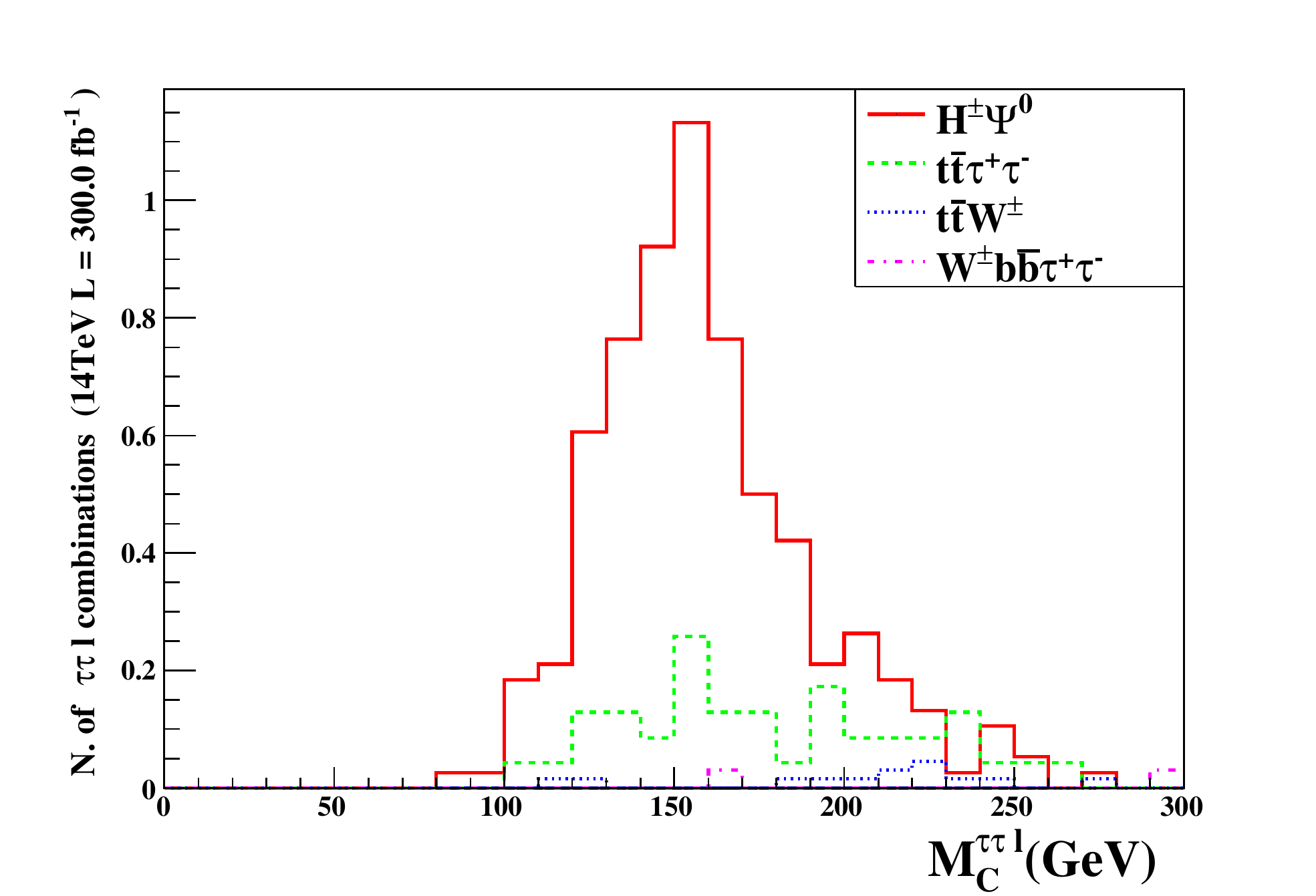}
\end{center}
\caption{Distributions of $M_C^{\tau\tau \ell}$ for $\ell^{\pm} \cancel{E}_T b\bar{b}\tau^+\tau^-$ signal channel at LHC14@300.
\label{fig:bbtatalvlMtatal}}
\end{figure}

Theoretically, the mass of $H^{\pm}$ could be reconstructed using some transverse variables of the $bb\ell$ or $\tau\tau\ell$ system. But since $\cancel{E}_T$ has additional contributions from hadronic $\tau$ decays,  the $bb\ell$ system is not useful here. We therefore employ the cluster transverse mass for the $\tau\tau\ell$ system:
\begin{equation}
M_C^{\tau\tau\ell}=\sqrt{\Big(\sqrt{p^2_{T,\tau\tau\ell}+M^2_{\tau\tau\ell}}+\cancel{E}_T \Big)^2-\Big(\vec{p}_{T,\tau\tau\ell}+\vec{\cancel{E}}_T \Big)^2},
\end{equation}
whose distribution shown in Fig.~\ref{fig:bbtatalvlMtatal} displays a clear peak albeit with a small number of events.

Finally, we would like to discuss briefly some other interesting signals in associated $H^{\pm}\psi^0$ production. As is well known, the CP-even $H^0$ can decay into $W^+W^-$, while the CP-odd $A^0$ cannot. The off-shell $W^{\pm*}$ from cascade $H^{\pm}$ decays must have the same charge as one of the two $W$s from $H^0$. The leptonic decays of these two like-sign $W$s result in the like-sign dilepton processes:
\begin{eqnarray}
&pp&\to H^{\pm}H^0 \to W^{\pm*}\psi^0 W^+W^- \to \ell^{\pm}\ell^{\pm}\cancel{E}_T b\bar{b}jj,\\
&pp& \to H^{\pm}A^0 \to W^{\pm*}H^0 A^0\to W^{\pm*}W^+W^- b\bar{b} \to \ell^{\pm}\ell^{\pm}\cancel{E}_T b\bar{b}jj.
\end{eqnarray}
From Fig. \ref{fig:Phidecay} we read off $\Br(H^0\to W^+W^-)\approx 0.3$, thus the theoretical cross section for this signal is about $1.2 ~\fb$ from $H^{\pm}H^0$ and $0.6~\fb$ from $H^{\pm}A^0$. The signals from $H^{\pm}H^0$ resemble very much the background $t\bar{t}W^{\pm}$ since the like-sign $W$s are usually well separated. The signals from $H^{\pm}A^0$ can be different from the background $t\bar{t}W^{\pm}$, but it is negligible compared to the same signals from $H^{\pm\pm}H^{\mp}(5~\fb)$ that we discussed in Sec.~\ref{negative-Doubly}. If there are two $H^0$s, they can both decay into $W^+W^-$, inducing a like-sign trilepton signal,
\begin{eqnarray}
pp\to H^{\pm}H^0 \to W^{\pm*}H^0 H^0 \to W^{\pm*}W^+W^- W^+W^- \to \ell^{\pm}\ell^{\pm}\ell^{\pm}\cancel{E}_Tjjjj.
\end{eqnarray}
Although this signal is very distinct and background free, its production cross section is too tiny ($0.037~\fb$) to be feasible.

\subsection{Signals for neutral scalars}
\label{negative-neutral}

The Higgs pair production is an important process in SM as it can be used to probe the electroweak symmetry breaking sector. The dominant production mechanism at a hadron collider is through gluon fusion~\cite{Glover:1987nx,Plehn:1996wb,Dawson:1998py,Djouadi:1999rca}. At LHC14, its cross section is about $18~\fb$ at leading order, and is enhanced to about $33~\fb$ by the next-to-leading-order QCD corrections~\cite{Dawson:1998py}. Recently, the next-to-next-to-leading-order QCD corrections are available, resulting in a further increase of the cross section to about $40~\fb$~\cite{deFlorian:2013jea}. We refer to Refs.~\cite{Baglio:2012np,Baglio:HH} for a more detailed discussion on the theoretical status of Higgs couplings at LHC. Physics beyond SM could alter Higgs pair production dramatically. The neutral Higgs pair production has been investigated in a variety of models including, for instance, MSSM~\cite{Plehn:1996wb}, NMSSM~\cite{Cao:2013si,Ellwanger:2013ova}, 2HDM~\cite{Asakawa:2010xj,Arhrib:2008pw,Arhrib:2009hc,Hespel:2014sla}, extended colored scalars~\cite{Asakawa:2010xj,Kribs:2012kz,Heng:2013cya}, extended colored fermions~\cite{Asakawa:2010xj,Dawson:2012mk,Chen:2014xwa}, Little Higgs~\cite{Dib:2005re}, Higgs portal~\cite{Dolan:2012ac,No:2013wsa} and composite Higgs models~\cite{Grober:2010yv,Gillioz:2012se}. There are also model independent approaches based on effective operators~\cite{Pierce:2006dh,Contino:2012xk,Liu:2013woa}. All these studies concentrate on the gluon fusion mechanism, since it is the dominant contribution.

The $H^0A^0$ associated production via the Drell-Yan process through an $s$-channel exchange of the $Z$ boson has received less attention in previous studies due to several reasons. The single production of neutral scalars proceeds through gluon fusion and thus dominates in a large portion of parameter space. The associated production of neutral scalars through a $Z$ boson exchange is much smaller than the gluon fusion channel in most models that we mentioned.
And the charged scalars usually have more distinct features that may help to detect them at a collider. The circumstance could be modified significantly in the negative scenario of type II seesaw. Since the triplet vev $v_\Delta$ is constrained by the $\rho$ parameter to be small, the mixing between the doublet and triplet scalars and thus the couplings of $H^0/A^0$ to quarks are suppressed. As a consequence, the single and pair or associated production of the neutral heavy scalars through gluon fusion is also suppressed~\cite{Godunov:2014waa}. Since in the negative scenario more charged scalars are heavier than less charged ones, this makes associated production of neutral scalars comparable to those involving charged scalars. 
The matter becomes even more interesting when all neutral scalars are nearly degenerate, i.e., $M_{H^0}\approx M_{A^0}\approx M_{h}$. We therefore have chosen $M_{H^0,A^0}=130~\GeV$ in Eq. (\ref{massspectrum}) as our benchmark point, and in addition we shall assume $v_{\Delta}=10^{-2}~\GeV$. The cross section for the associated $H^0A^0$ production through the $Z$ boson at LHC14 is $350~\fb$ at leading order, which is about $10$ times as large as the SM Higgs pair production.

Some recent studies~\cite{Baglio:2012np,Dolan:2012rv,Gouzevitch:2013qca,Papaefstathiou:2012qe,Goertz:2013kp,
deLima:2014dta} have demonstrated the potential of Higgs pair production in the signal channels $b\bar{b}\gamma\gamma$, $b\bar{b}W^+W^-$, $b\bar{b}\tau^+\tau^-$, and $b\bar{b}b\bar{b}$. With $L=3000~\fb^{-1}$ at LHC14, the trilinear coupling of the Higgs bosons could be measured at an accuracy of $\sim 40\%$ \cite{Barger:2013jfa}. With a much higher production cross section, we have carried out the simulation of $H^0A^0$ production at LHC14@300. To illustrate the potential, we shall consider the channel~\footnote{A systematic analysis on probing associated production of neutral scalars at LHC will be presented elsewhere~\cite{han:2015xxx}.}:
\begin{eqnarray}
p p \to Z^* \to H^0 A^0 \to b \bar{b} \tau^+\tau^-.
\end{eqnarray}
Signals like $b\bar{b}\gamma\gamma,~b\bar{b}W^+W^-$ are also interesting, and will be studied in the future. An important part of this analysis depends on the ability to reconstruct the $b$ pair and $\tau$ pair. In our simulation, the $b$-tagging efficiency is assumed to be $0.7$, and the misidentification rate of a $c$- and light-jet as a $b$-jet is taken to be $0.1$ and $0.01$ respectively. We consider hadronic decays of $\tau$ for its tagging, and assume an efficiency of $0.8$ with a negligible fake rate.

The main SM backgrounds are:
\begin{eqnarray}
&p p&\to b\bar{b}Z/\gamma^*/h \to b\bar{b}\tau^+\tau^-,
\\
&p p&\to b\bar{b}W^+W^- \to b\bar{b}\tau^+ \nu_{\tau} \tau^- \bar{\nu}_{\tau},
\\
&p p&\to Zh \to b\bar{b}\tau^+\tau^-.
\end{eqnarray}
The irreducible background comes from $b\bar{b}\tau^+\tau^-$, where the $\tau$-pair originates from the decay of $Z/\gamma^*/h$. Since the hadronic decays of $\tau$ always have neutrinos, we also include the background $b\bar{b}W^+W^-$, which contributes to the $b\bar{b}\tau^+ \nu_{\tau} \tau^- \bar{\nu}_{\tau}$ final state. The $b\bar{b}W^+W^-$ background dominantly stems from the $t\bar{t}$ production with subsequent decays $t\to b W$ and $W\to \tau \nu_{\tau}$. In our simulation, we also include the associated $Zh$ production with subsequent decays $h\to b\bar{b}$ and $Z\to\tau^+\tau^-$ or vice versa.

\begin{table} [!htbp]
\begin{center}
\begin{tabular}{|c|c|c|c|c|c|c|}
\hline
Cuts  &  $H^0A^0$  &  $b\bar{b}\tau^+\tau^-$ & $b\bar{b}W^+W^-$  & $Zh$ &  $S/\sqrt{S+B}$ & $S/B$
\\
\hline
Basic Cuts & 6797 & 1529101 &  952937 & 2617 & 4.31 & 0.00274
\\
$N_b=2,~N_{\tau}=2$ & 297 & 20058 & 34959  & 102 & 1.26 & 0.00539
\\
$\Delta R_{bb}<2.0,\Delta R_{\tau\tau}<2.0,\Delta R_{b\tau}>1.5$ &
                    147 & 2285 & 3088 & 35.4 & 1.97 & 0.0272
\\
$p_T^{b_1}>80$~GeV,~ $p_T^{b_2}>40$~GeV &
                    108 & 537 & 763&  21.3 & 2.86 & 0.0817
\\
$p_T^{\tau_1}>70$~GeV,~ $p_T^{\tau_2}>30$~GeV &
                    68.5 & 221 & 93.7  & 11.4 & 3.45 & 0.210
\\
 $H_T>250$~GeV  & 44.0 &  103 & 30.3  & 5.82 & 3.25 & 0.317
\\
95~GeV$<M_{\tau\tau}<135$~GeV & 16.0 & 7.94 & 7.42  & 0.426 & 2.84 & 1.02
\\
110~GeV$<M_{bb}<150$~GeV & 12.41 & 2.65 & 2.86 & 0.284 & 2.91 & 2.14
\\
\hline
\end{tabular}
\end{center}
\caption{Survival numbers of signal $b\bar{b}\tau^+\tau^-$ and its backgrounds, statistical significance $S/\sqrt{S+B}$, and signal to background ratio $S/B$ after imposing each cut sequentially at LHC14@300.}
\label{tab:bbtatacut}
\end{table}

\begin{figure}[!htbp]
\begin{center}
\includegraphics[width=0.45\linewidth]{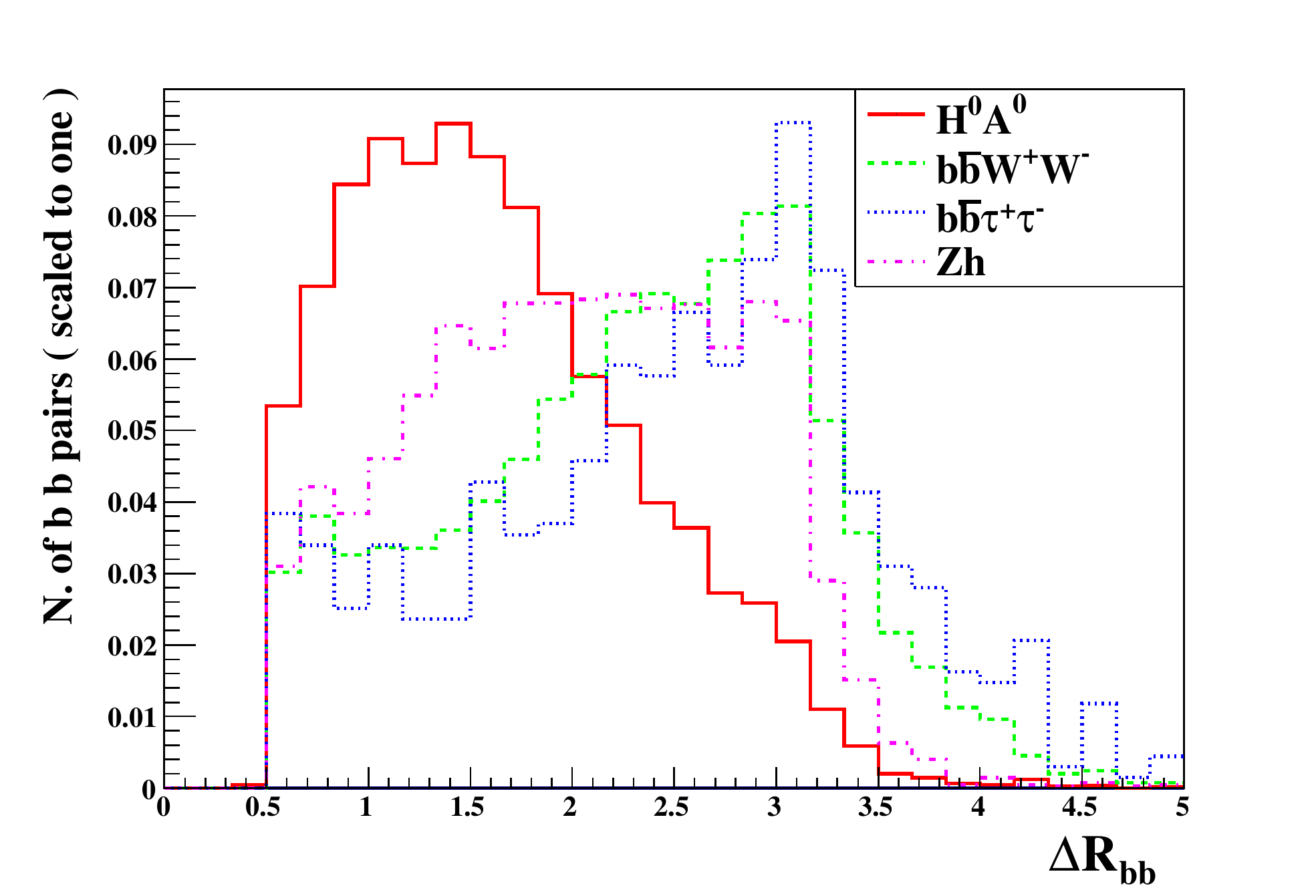}
\includegraphics[width=0.45\linewidth]{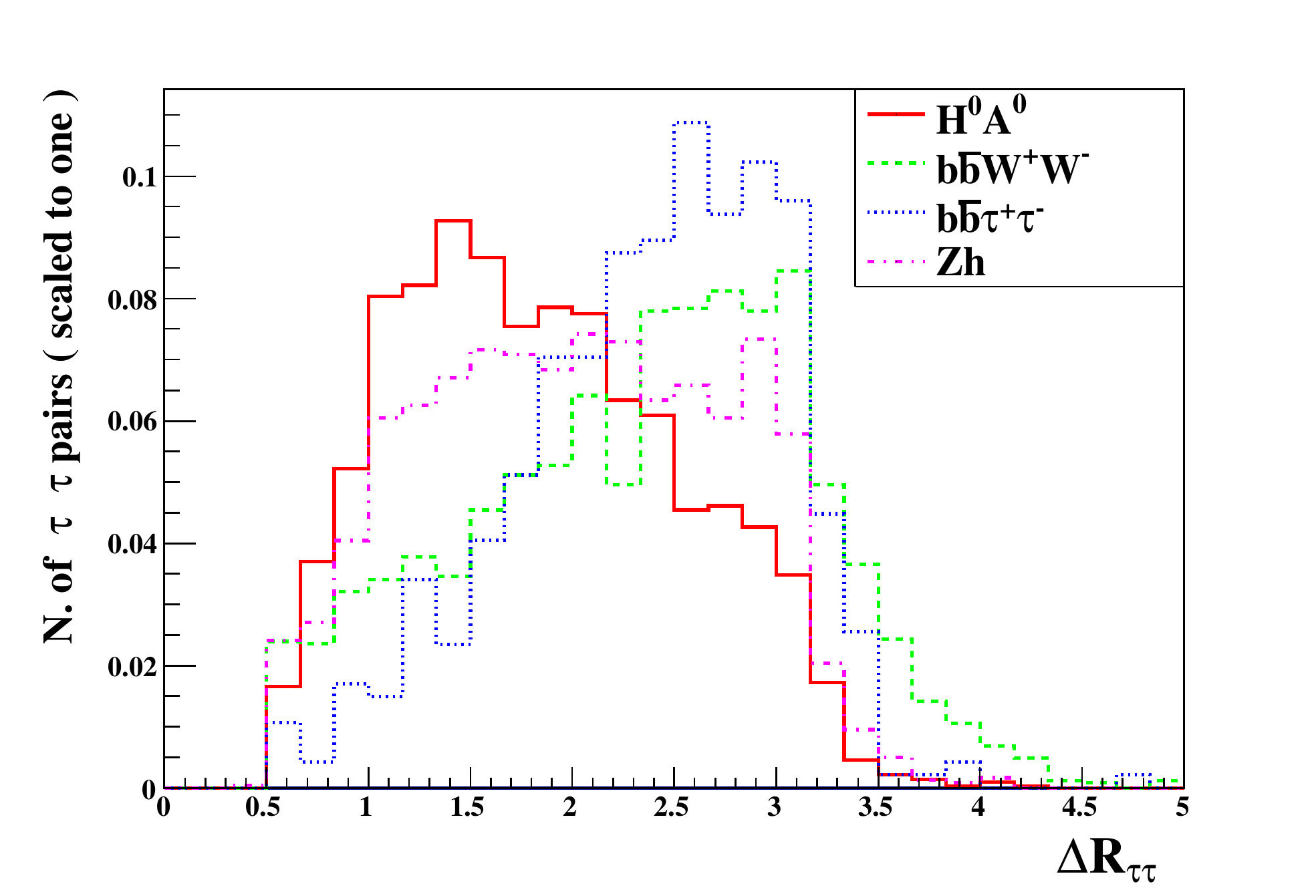}
\includegraphics[width=0.45\linewidth]{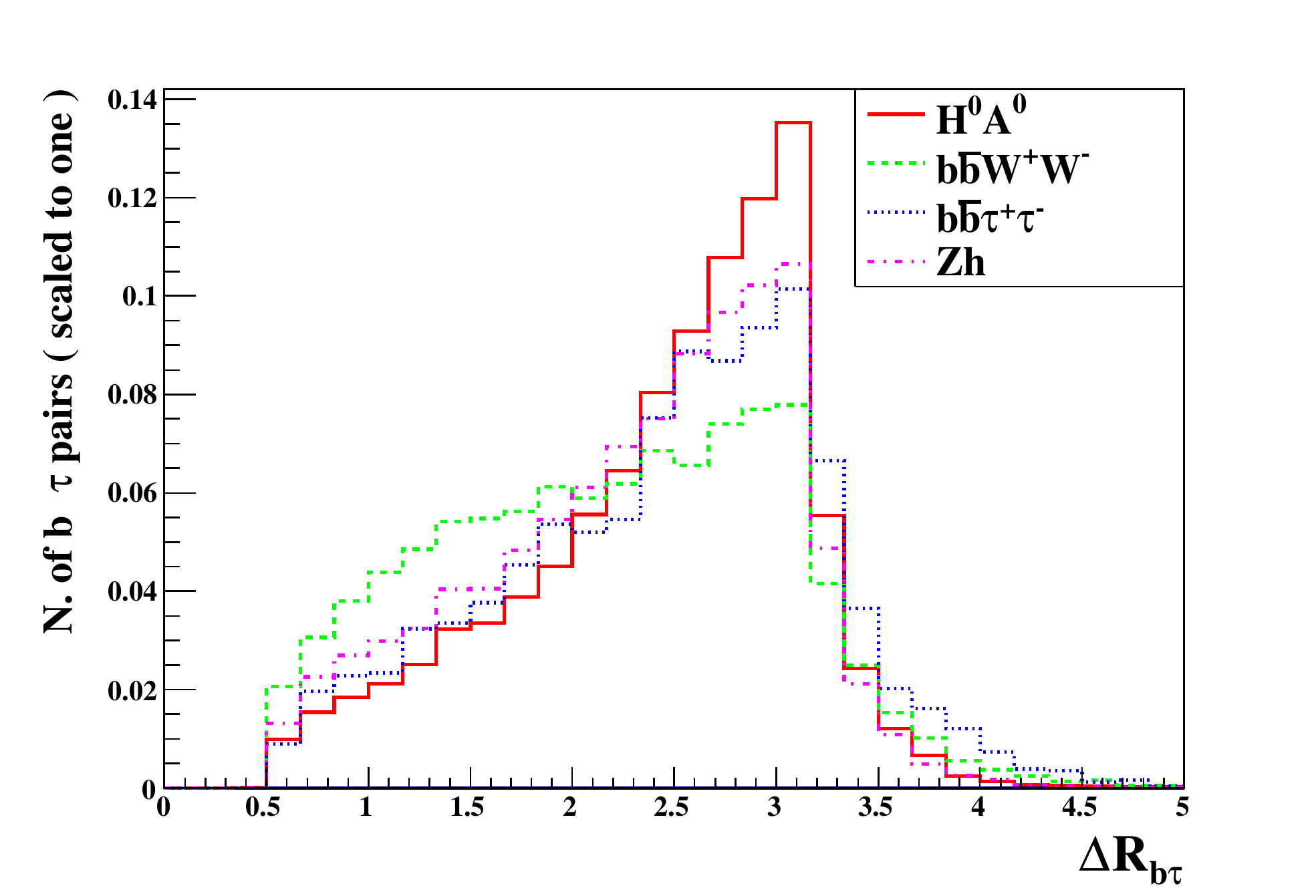}
\includegraphics[width=0.45\linewidth]{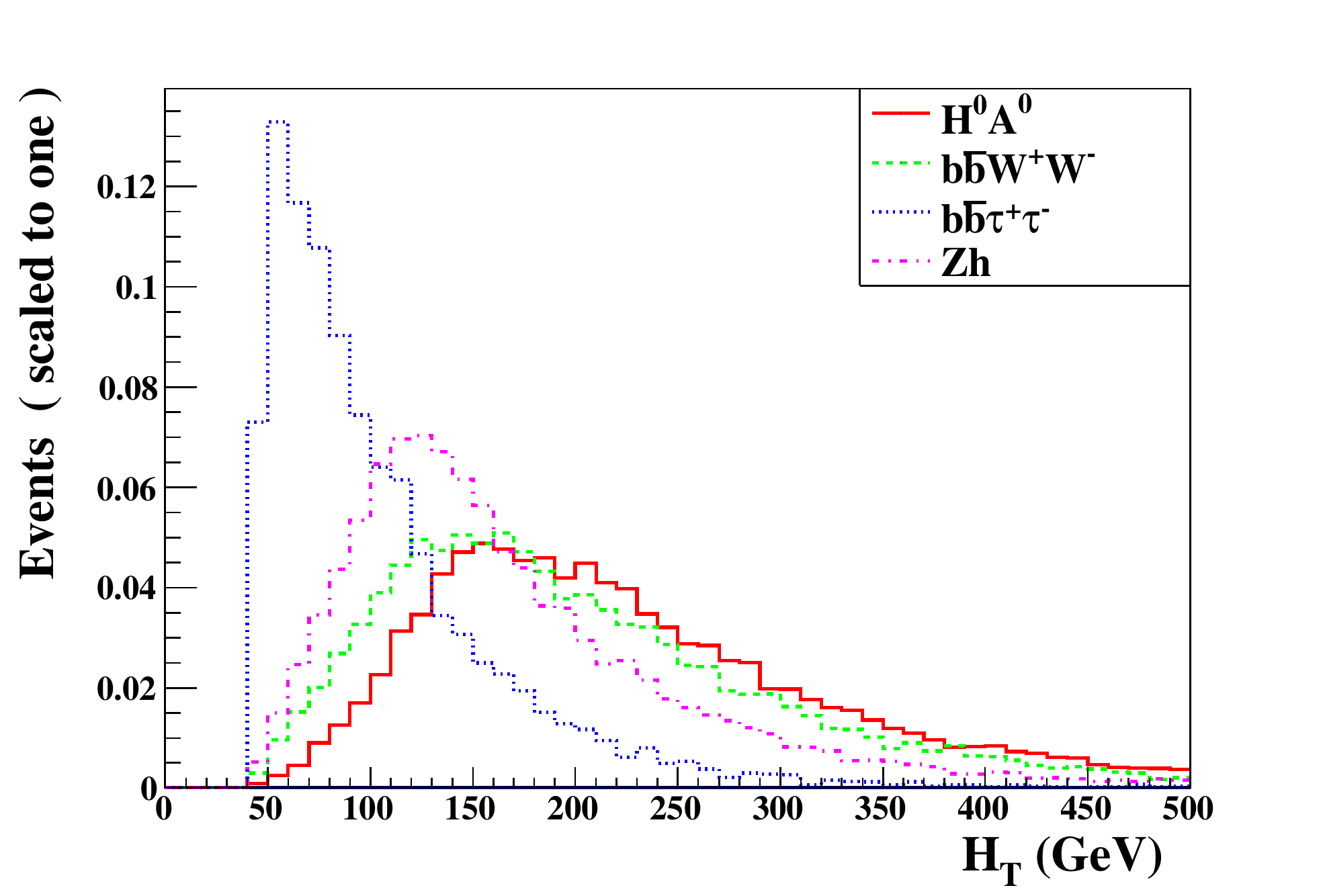}
\includegraphics[width=0.45\linewidth]{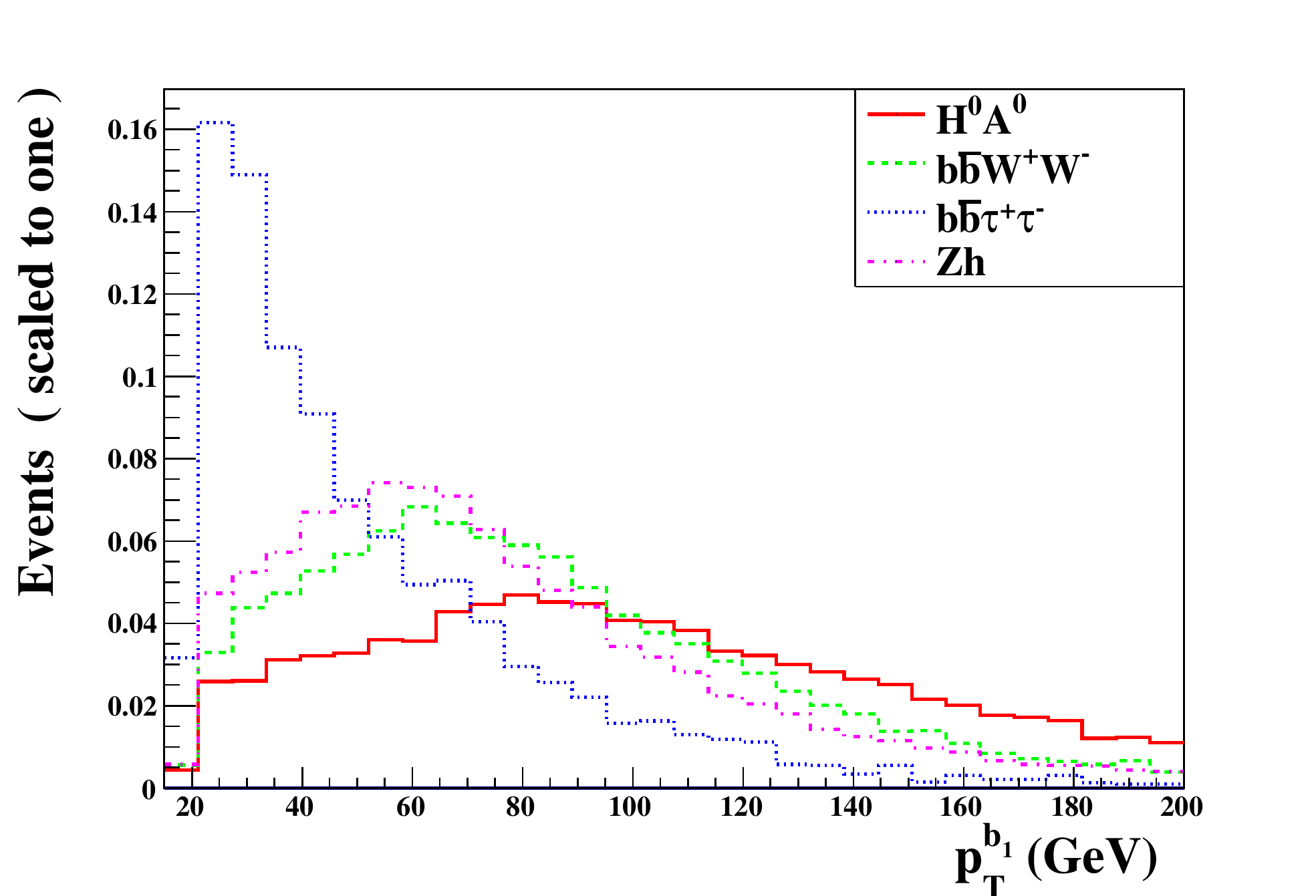}
\includegraphics[width=0.45\linewidth]{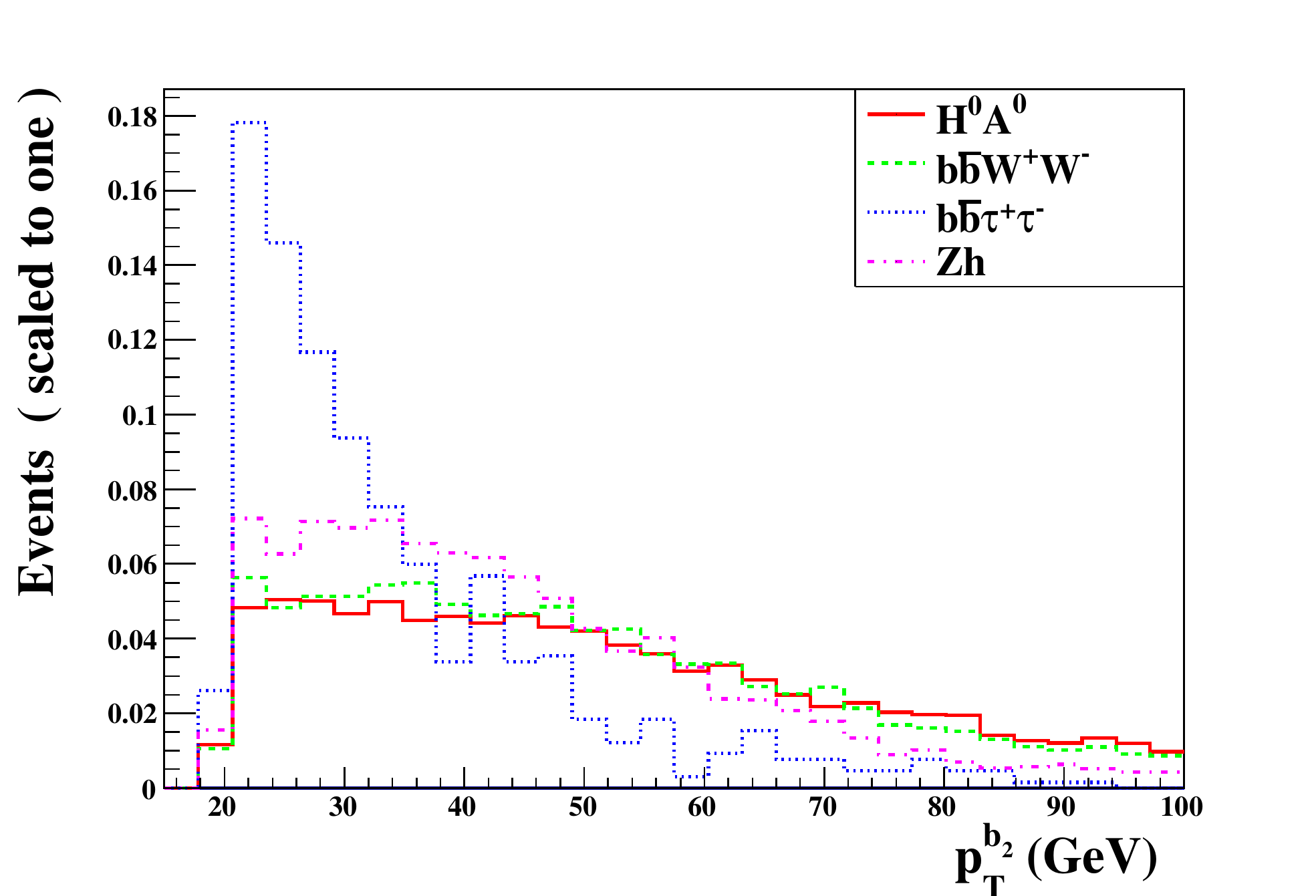}
\includegraphics[width=0.45\linewidth]{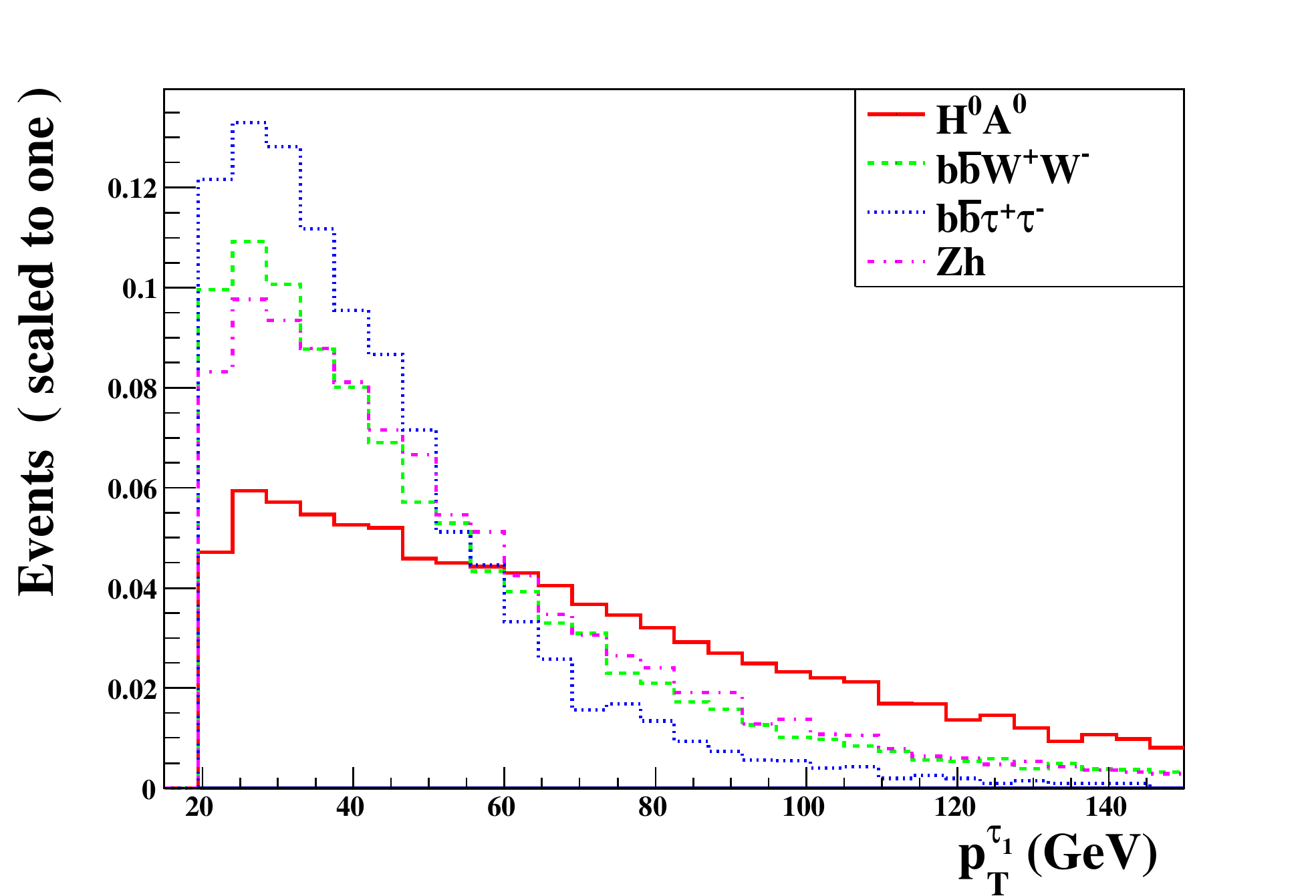}
\includegraphics[width=0.45\linewidth]{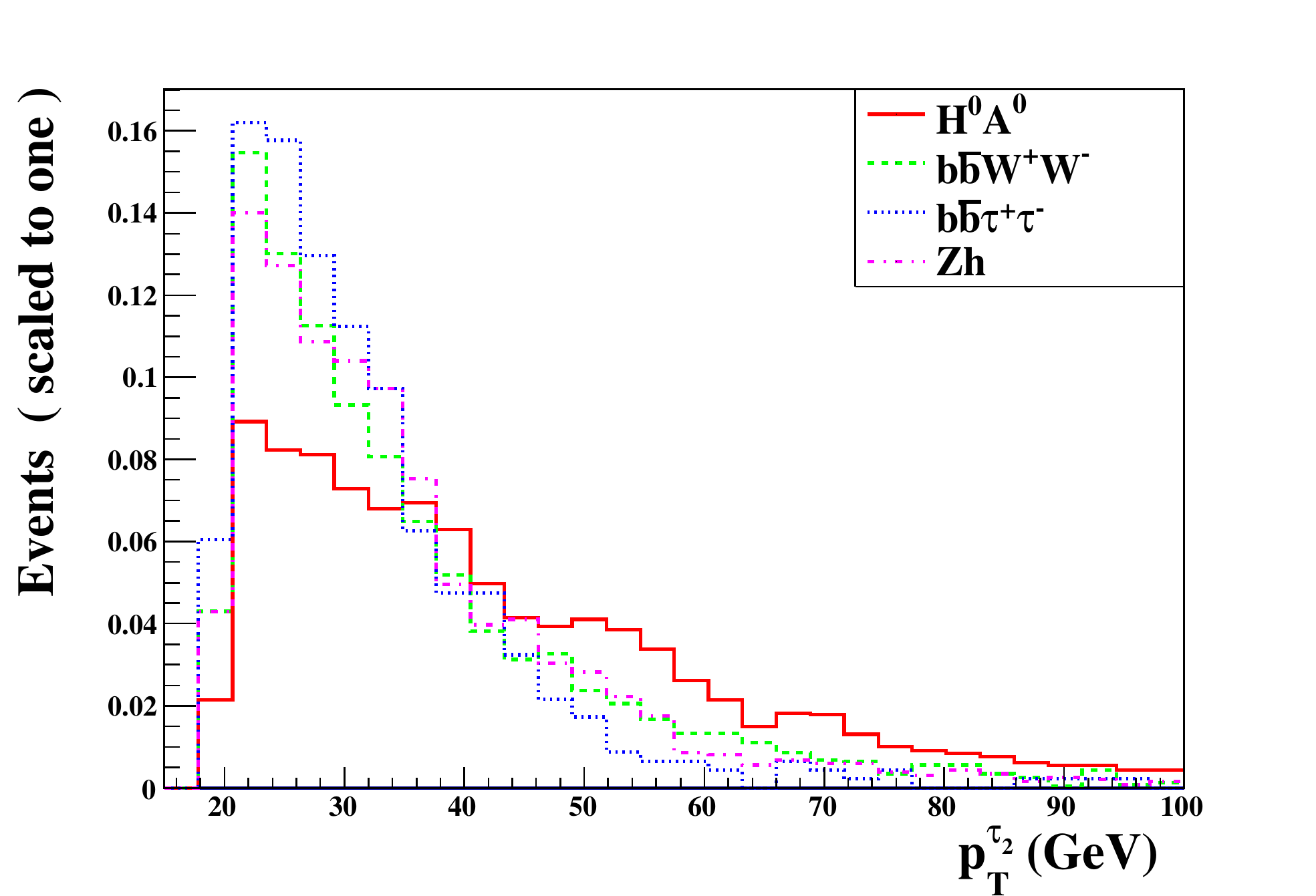}
\end{center}
\caption{Distributions of various separations $\Delta R_{bb,\tau\tau,b\tau}$ and transverse
momenta $H_T,~p_T^{b_1},~p_T^{b_2},~p_T^{\tau_1},~p_T^{\tau_2}$ after imposing basic cuts. See text for notations.
\label{fig:bbtataDRbb}}
\end{figure}

We first employ some basic cuts for the selection of events:
\begin{eqnarray}
p_T^j > 20 ~\mbox{GeV},~\Delta R_{jj} > 0.4,~|\eta_j| < 2.5.
\end{eqnarray}
Here the jets include both $b$- and $\tau$-jets.
The numbers of the signal and background events after imposing the cuts are summarized
in Table \ref{tab:bbtatacut}. The last two columns in the table show the statistical significance $S/\sqrt{S+B}$ and the signal to background ratio $S/B$. Then we apply the cut:
\begin{equation}
N_{b}=2,~N_{\tau}=2,
\end{equation}
so that all four jets are tagged in this channel. At this stage, the background is about 200 times larger than the signal, since both $b\bar{b}\tau^+\tau^-$ and $b\bar{b}W^+W^-$ receive large QCD contributions although the fully electroweak $Zh$ background is about $1/3$ of the signal. To further suppress the background, we shall apply several kinematical cuts. In Fig. \ref{fig:bbtataDRbb} we show the distributions in the eight kinematical variables: the separations $\Delta R_{bb,\tau\tau,b\tau}$, the total hadronic transverse momentum $H_T$, the leading (next-to-leading) $b$-jet transverse momentum $p_T^{b_1(b_2)}$, and the leading (next-to-leading) $\tau$-jet transverse momentum $p_T^{\tau_1(\tau_2)}$.

Because the two signal $b$-jets come from decays of heavy neutral scalars, their separation
$\Delta R_{bb}$ is usually small with a clear peak around $1.4$. In contrast, the two $b$-jets
in the $b\bar{b}\tau^+\tau^-$ and $b\bar{b}W^+W^-$ background respectively are mainly from
the gluon fusion and the $t\bar t$ decays, and thus tend to be separated. The same features are shared by the separation $\Delta R_{\tau\tau}$ for the signal and the $b\bar{b}\tau^+\tau^-$ and $b\bar{b}W^+W^-$ background.
For the $Zh$ background, the separations $\Delta R_{bb}$ and $\Delta R_{\tau\tau}$ have a relatively large peak coverage from 1.5 to 3.
Since the $b$- and $\tau$-jets in the $b\bar{b}W^+W^-$ background both come from $t/\bar{t}$,
their separation $\Delta R_{b\tau}$ has more chances to be smaller than from other sources,
as is seen in Fig. \ref{fig:bbtataDRbb}. These differences between the signal and backgrounds
suggest the following comprehensive cuts:
\begin{equation}
\Delta R_{bb}<2.0,~\Delta R_{\tau\tau}<2.0,~\Delta R_{b\tau}>1.5.
\end{equation}

The distributions in the transverse momentum $p_T^j$ of various jets also provide useful information. First, both $p_T^{b_1}$ and $p_T^{b_2}$ in $b\bar{b}\tau^+\tau^-$ background
are usually much softer, and the leading $b$-jet transverse momentum $p_T^{b_1}$ in the other
two backgrounds is also small compared to the signal. Second, the $\tau$-jets in the $H^0A^0$
signal are harder than in the backgrounds, since they come from the decays of heavier
neutral scalars. We thus apply the following cuts on the transverse momentum of various jets:
\begin{equation}
p_T^{b_1}>80~\GeV,~p_T^{b_2}>40~\GeV,
~p_T^{\tau_1}>70~\GeV,~p_T^{\tau_2}>30~\GeV.
\end{equation}
Since (anti-)neutrinos carry away part of energy in the hadronic decays of $\tau$, the $\tau$-jets in the final states will be softer than the $b$-jets. Therefore, the $p_T$ cuts applied on the $\tau$-jets are $10~\GeV$ smaller than on the $b$-jets. Till this point, the QCD backgrounds are still about five times larger than the signal. Since the total hadronic transverse momentum $H_T$ in the signal is usually larger than in the background, we further apply the cut:
\begin{equation}
H_T > 250 ~\GeV.
\end{equation}
After this cut, the survival $b\bar{b}W^+W^-$ events become smaller than the signal, while the $b\bar{b}\tau^+\tau^-$ background is still more than twice as large as the signal.

\begin{figure}[!htbp]
\begin{center}
\includegraphics[width=0.45\linewidth]{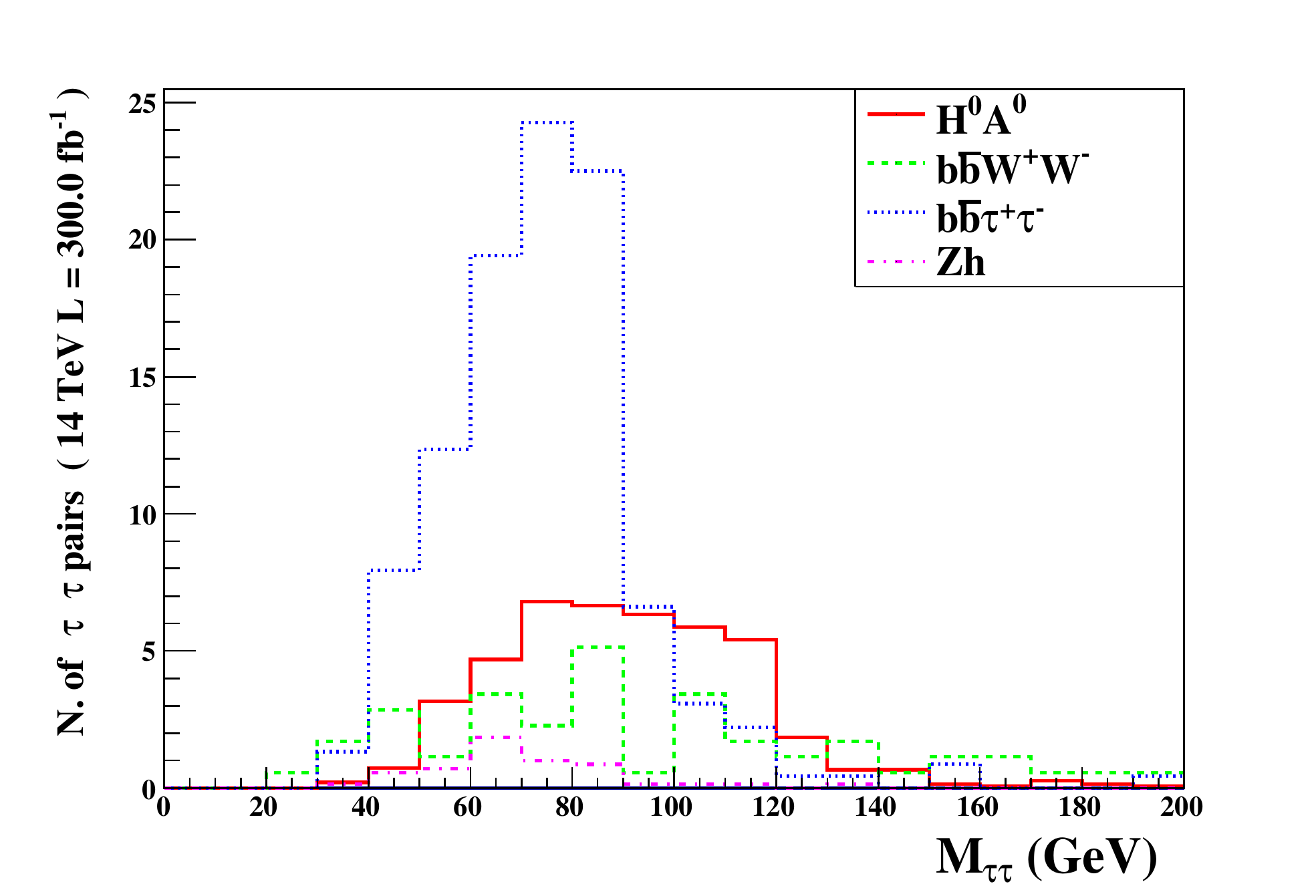}
\includegraphics[width=0.45\linewidth]{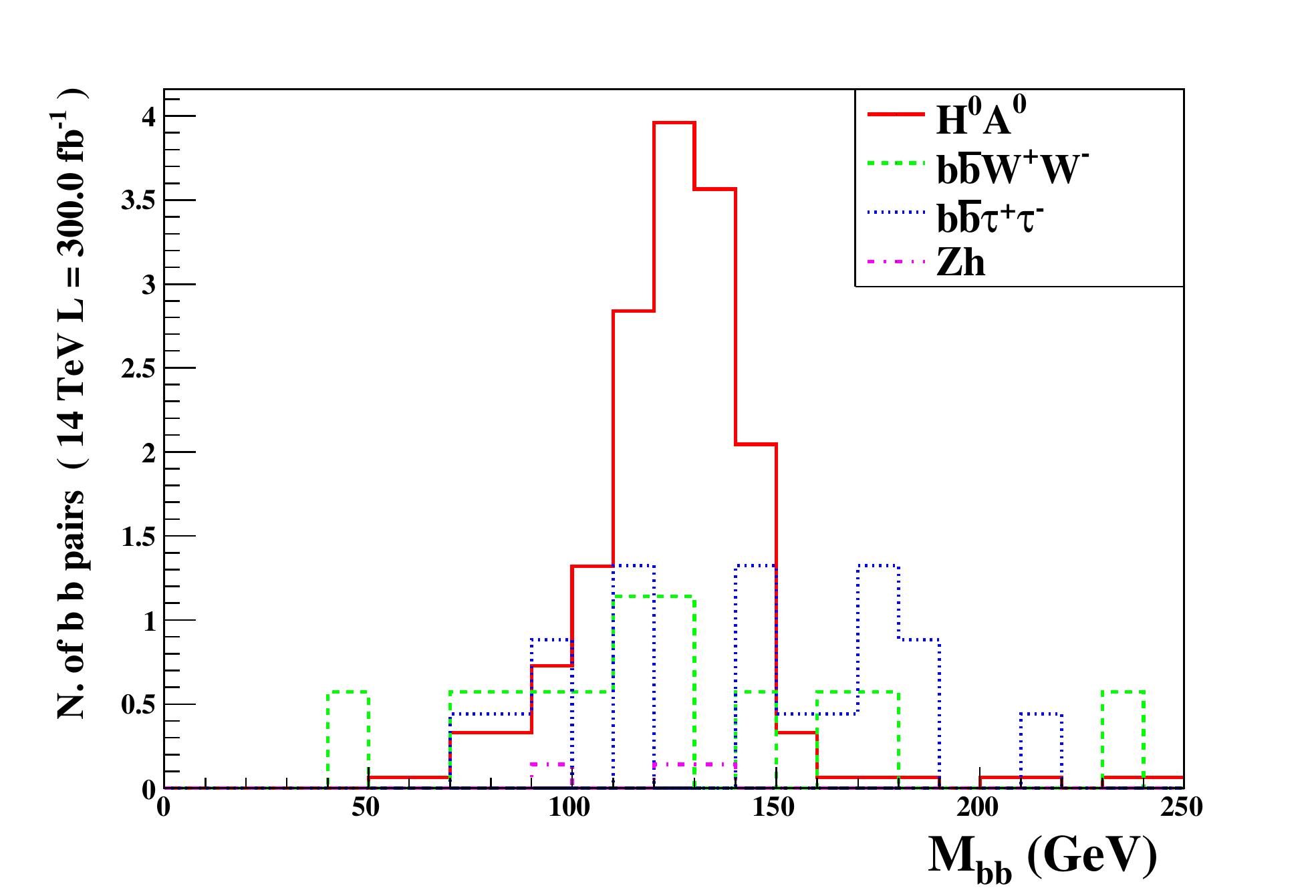}
\end{center}
\caption{Distributions of invariant masses in signal $b\bar{b}\tau^+\tau^-$ at LHC14@300. Left panel: $M_{\tau\tau}$ before cut (\ref{cut:Mtata}); right panel: $M_{bb}$ after cut (\ref{cut:Mtata}).
\label{fig:bbtataMbb}}
\end{figure}

Now we consider to utilize the invariant mass distributions to further suppress the background. The left panel of Fig. \ref{fig:bbtataMbb} shows the invariant mass $M_{\tau\tau}$
of the $\tau$ pair. The $b\bar{b}\tau^+\tau^-$ background has a peak around $M_{\tau\tau}=70~\GeV$, which is the shifted peak of the $Z$ boson pole at $M_Z=90~\GeV$, and also has a sharp edge at $M_Z$. In contrast, $M_{\tau\tau}$ in the signal has a much fatter peak. We therefore impose a cut on $M_{\tau\tau}$:
\begin{equation}\label{cut:Mtata}
95~\GeV<M_{\tau\tau}<135~\GeV.
\end{equation}
This cut turns out to be very efficient. The reason that we apply a cut on $M_{\tau\tau}$ first is due to the fact that $M_{\tau\tau}$ is not only fatter than $M_{bb}$ but also shifted from the original value of $M_{H^0/A^0}$ according to our analysis. In the right panel of Fig. \ref{fig:bbtataMbb}, we show the invariant mass of the $bb$ pair. A sharp peak around $130~\GeV$ is evident, so that the mass of $H^0/A^0$ can be reconstructed in this channel, although the CP nature of these scalars remains to be discerned. We then select the mass window of $M_{bb}$:
\begin{equation}
110~\GeV<M_{bb}<150~\GeV.
\end{equation}
After all these cuts, about $12$ signal events survive at LHC14@300, with a statistical significance of $2.91$ and a signal to background ratio of $2.14$. So this channel is expected to be discovered with higher luminosity.

\section{Discussions and Conclusions}
\label{Conclusions}

The new scalars in type-II seesaw are generically nondegenerate, when cascade decays between scalars proceed with the radiation of a $W$ boson. This can modify significantly their signatures at LHC even if the mass splitting is not large so that the radiated $W$ boson is far off-shell. In this work, we have presented a systematic and comprehensive analysis on this issue, by simulating all potentially interesting signals to the detector level and elaborately designing specific cuts to purify the signals. Our results may serve as a reference that encourages more detailed simulations by experimental collaborations.

The positive scenario with doubly-charged scalars being the lightest is stringently constrained by LHC limits, and the scalars must be heavier than about $400~\GeV$. In contrast, the negative scenario is less restricted by current experiments. A much lighter spectrum is allowed in this case, and in particular, the new neutral scalars can be nearly degenerate with the standard model Higgs boson. This makes it very appealing from the point of view of collider phenomenology.

Our main results are summarized as follows. In the positive scenario, the four-lepton signal is still the most promising channel to discover $H^{\pm\pm}$. At LHC14@300, we can potentially probe up to a mass $M_{H^{\pm\pm}}\sim600~\GeV$ for NH and $M_{H^{\pm\pm}}\sim700~\GeV$ for IH.
The five-lepton signal is crucial to determine the spectrum of triplet scalars; for instance, a $5\sigma$ significance can be reached with $L=76~\fb^{-1}$ at $M_{H^{\pm\pm}}=400~\GeV,~M_{H^{\pm}}=430~\GeV$ for IH. The LNV signal from cascade decays of neutral scalars $H^0/A^0$ is heavily affected by their mass difference $\Delta M_{H^0/A^0}$ and total decay widths $\Gamma_{H^0/A^0}$ due to the significant interference between them. For degenerate $H^0$ and $A^0$, our signal receives an enhancement factor of two while the like-sign four-lepton signal considered in the previous literature tends to diminish. But its discovery at LHC14 would require a large integrated luminosity due to small cross section.

For the negative scenario, the associated production of neutral scalars $H^0A^0$ can give the same signals ($b\bar{b}\tau^+\tau^-$, $b\bar{b}W^+W^-$, and $b\bar{b}\gamma\gamma$) as the SM Higgs pair production, but with a ten times as large cross section. We could reach $3\sigma$ significance in its $b\bar{b}\tau^+\tau^-$ signal channel at LHC14@300. The most promising signal is $\ell^{\pm}\cancel{E}_Tb\bar{b}b\bar{b}$ from associated $H^{\pm}\psi^0$ production: a $5\sigma$ significance can be reached at $L=109~\fb^{-1}$ for  LHC14. Another signal from $H^{\pm}\psi^0$ is $\ell^{\pm}\cancel{E}_Tb\bar{b}\tau^+\tau^-$, which has a cleaner background and could be detected for $L\sim300~\fb^{-1}$. The doubly-charged scalars $H^{\pm\pm}$ being the heaviest are relatively hard to probe. We investigated in detail the $\ell^\pm\ell^\pm\cancel{E}_T jjjj$ signal from associated $H^{\pm\pm}H^{\mp}$ production, and found that a $5\sigma$ significance would require an integrated luminosity of $L\sim500~\fb^{-1}$.

\section*{Acknowledgement}
RD thanks Liang-Liang Zhang for help on data analysis and Wan-Peng Xie for useful discussion on interference effect and narrow width approximation. This work was supported in part by the grant NSFC-11025525 and by the CAS Center for Excellence in Particle Physics (CCEPP).



\begin{thebibliography}{000}

\bibitem{Weinberg:1979sa}
  S.~Weinberg,
  Phys.\ Rev.\ Lett.\  {\bf 43} (1979) 1566.

\bibitem{Liao:2010ku}
  Y.~Liao,
  Phys.\ Lett.\ B {\bf 694}, 346 (2011)  [arXiv:1009.1692 [hep-ph]].

\bibitem{Ma:1998dn}
  E.~Ma,
  Phys.\ Rev.\ Lett.\  {\bf 81} (1998) 1171
  [arXiv:hep-ph/9805219].

\bibitem{type1}
P.~Minkowski,
Phys.\ Lett.\ B {\bf 67} (1977) 421;
T.~Yanagida, proceedings of the {\em Workshop on Unified Theories
and Baryon Number in the Universe}, Tsukuba, 1979, eds.
A. Sawada, A. Sugamoto;
S.~Glashow, in {\em Cargese 1979, Proceedings, Quarks and
Leptons}
(1979) ;
M.~Gell-Mann, P.~Ramond, R.~Slansky, proceedings of the
{\em Supergravity Stony Brook Workshop}, New York, 1979,
eds. P. Van Niewenhuizen, D. Freeman;
R.~N.~Mohapatra, G.~Senjanovi\' c,
Phys.~Rev.~Lett.~{\bf 44} (1980) 912.

\bibitem{type2}
  M.~Magg and C.~Wetterich,
  Phys.\ Lett.\  B {\bf 94}, 61 (1980).
 T.~P.~Cheng and L.~F.~Li,
 Phys.\ Rev.\  D {\bf 22}, 2860 (1980);
 J.~Schechter and J.~W.~F.~Valle,
 Phys.\ Rev.\  D {\bf 22}, 2227 (1980);
  G.~Lazarides, Q.~Shafi and C.~Wetterich,
  Nucl.\ Phys.\  B {\bf 181}, 287 (1981);
  R.~N.~Mohapatra and G.~Senjanovic,
  Phys.\ Rev.\  D {\bf 23}, 165 (1981);

\bibitem{Foot:1988aq}
  R.~Foot, H.~Lew, X.~G.~He and G.~C.~Joshi,
  Z.\ Phys.\ C {\bf 44} (1989) 441.

\bibitem{Type1-3}
  E.~Ma,
  Phys.\ Rev.\ Lett.\  {\bf 81} (1998) 1171
  [arXiv:hep-ph/9805219];
  B.~Bajc and G.~Senjanovi\'c,
  JHEP {\bf 0708} (2007) 014
  [arXiv:hep-ph/0612029];
  P.~Fileviez~P\'erez,
  Phys.\ Lett.\  B {\bf 654} (2007) 189
  [arXiv:hep-ph/0702287];
  P.~Fileviez~P\'erez,
  Phys.\ Rev.\  D {\bf 76} (2007) 071701
  [arXiv:0705.3589 [hep-ph]].

\bibitem{LR}
  R.~N.~Mohapatra and J.~C.~Pati,
  Phys.\ Rev.\  D {\bf 11} (1975) 2558;
  G.~Senjanovi\'c and R.~N.~Mohapatra,
  Phys.\ Rev.\  D {\bf 12} (1975) 1502;
  G.~Senjanovi\'c,
  Nucl.\ Phys.\  B {\bf 153} (1979) 334.

\bibitem{Zee-Babu}
  A.~Zee,
  Phys.\ Lett.\  B {\bf 93}, 389 (1980)
  [Erratum-ibid.\  B {\bf 95}, 461 (1980)];
  A.~Zee,
  Nucl.\ Phys.\  B {\bf 264}, 99 (1986);
  K.~S.~Babu,
  Phys.\ Lett.\  B {\bf 203}, 132 (1988);
  L.~M.~Krauss, S.~Nasri and M.~Trodden,
  Phys.\ Rev.\  D {\bf 67}, 085002 (2003)
  [arXiv:hep-ph/0210389];
  K.~Cheung and O.~Seto,
  Phys.\ Rev.\  D {\bf 69}, 113009 (2004)
  [arXiv:hep-ph/0403003];
  M.~Aoki, S.~Kanemura and O.~Seto,
  Phys.\ Rev.\ Lett.\  {\bf 102}, 051805 (2009)
  [arXiv:0807.0361 [hep-ph]];
  Phys.\ Rev.\  D {\bf 80}, 033007 (2009)
  [arXiv:0904.3829 [hep-ph]];
  E.~Ma,
  Phys.\ Rev.\ D {\bf 73}, 077301 (2006)
  [hep-ph/0601225];
  P.~Fileviez Perez and M.~B.~Wise,
  Phys.\ Rev.\ D {\bf 80}, 053006 (2009)
  [arXiv:0906.2950 [hep-ph]];
  A.~Ahriche, S.~Nasri and R.~Soualah,
  Phys.\ Rev.\ D {\bf 89} (2014) 095010
  [arXiv:1403.5694 [hep-ph]];
  A.~Ahriche, C.~S.~Chen, K.~L.~McDonald and S.~Nasri,
  Phys.\ Rev.\ D {\bf 90} (2014) 015024
  [arXiv:1404.2696 [hep-ph]].

\bibitem{Babu:2001ex}
  K.~S.~Babu and C.~N.~Leung,
  Nucl.\ Phys.\  B {\bf 619}, 667 (2001)
  [arXiv:hep-ph/0106054];
  E.~Ma,
  Phys.\ Rev.\  D {\bf 66}, 037301 (2002)
  [arXiv:hep-ph/0204013];
  I.~Gogoladze, N.~Okada and Q.~Shafi,
  Phys.\ Lett.\  B {\bf 672}, 235 (2009)
  [arXiv:0809.0703 [hep-ph]];
  A.~de Gouvea and J.~Jenkins,
  Phys.\ Rev.\  D {\bf 77}, 013008 (2008)
  [arXiv:0708.1344 [hep-ph]];
  W.~Grimus, L.~Lavoura and B.~Radovcic,
  Phys.\ Lett.\  B {\bf 674}, 117 (2009)
  [arXiv:0902.2325 [hep-ph]];
  P.~H.~Gu, H.~J.~He, U.~Sarkar and X.~Zhang,
  Phys.\ Rev.\  D {\bf 80}, 053004 (2009)
  [arXiv:0906.0442 [hep-ph]];
  Z.~z.~Xing and S.~Zhou,
  Phys.\ Lett.\  B {\bf 679}, 249 (2009)
  [arXiv:0906.1757 [hep-ph]];
  F.~Bonnet, D.~Hernandez, T.~Ota and W.~Winter,
  JHEP {\bf 0910}, 076 (2009)
  [arXiv:0907.3143 [hep-ph]];
  I.~Picek and B.~Radovcic,
  Phys.\ Lett.\  B {\bf 687}, 338 (2010)
  [arXiv:0911.1374 [hep-ph]];
  Y.~Liao,
  JHEP {\bf 1106}, 098 (2011)
  [arXiv:1011.3633 [hep-ph]];
  K.~L.~McDonald,
  JHEP {\bf 1307}, 020 (2013)
  [arXiv:1303.4573 [hep-ph]];
  S.~S.~C.~Law and K.~L.~McDonald,
  Phys.\ Rev.\ D {\bf 87}, 113003 (2013)
  [arXiv:1303.4887 [hep-ph]].

\bibitem{Han:2006ip}
  T.~Han and B.~Zhang,
  Phys.\ Rev.\ Lett.\  {\bf 97}, 171804 (2006)
  [hep-ph/0604064];
  P.~Fileviez Perez, T.~Han and T.~Li,
  Phys.\ Rev.\ D {\bf 80}, 073015 (2009)
  [arXiv:0907.4186 [hep-ph]];
  A.~Atre, T.~Han, S.~Pascoli and B.~Zhang,
  JHEP {\bf 0905}, 030 (2009)
  [arXiv:0901.3589 [hep-ph]];
  F.~del Aguila, J.~A.~Aguilar-Saavedra and R.~Pittau,
  JHEP {\bf 0710}, 047 (2007)  [hep-ph/0703261];
  J.~Kersten and A.~Y.~.Smirnov,
  Phys.\ Rev.\ D {\bf 76} (2007) 073005
  [arXiv:0705.3221 [hep-ph]];
  P.~S.~Bhupal Dev, A.~Pilaftsis and U.~K.~Yang,
  Phys.\ Rev.\ Lett.\  {\bf 112}, 081801 (2014)
  [arXiv:1308.2209 [hep-ph]];
  A.~Das, P.~S.~Bhupal Dev and N.~Okada,
  Phys.\ Lett.\ B {\bf 735}, 364 (2014)  [arXiv:1405.0177 [hep-ph]];
  A.~Das and N.~Okada,
  Phys.\ Rev.\ D {\bf 88} (2013) 113001
  [arXiv:1207.3734 [hep-ph]].

\bibitem{Han:2007bk}
  T.~Han, B.~Mukhopadhyaya, Z.~Si and K.~Wang,
  Phys.\ Rev.\ D {\bf 76}, 075013 (2007)  [arXiv:0706.0441 [hep-ph]].

\bibitem{Chiang:2012dk}
  C.~W.~Chiang, T.~Nomura and K.~Tsumura,
  Phys.\ Rev.\ D {\bf 85}, 095023 (2012)
  [arXiv:1202.2014 [hep-ph]].

\bibitem{Perez:2008ha}
  P.~Fileviez Perez, T.~Han, G.~-y.~Huang, T.~Li and K.~Wang,
  Phys.\ Rev.\ D {\bf 78}, 015018 (2008)  [arXiv:0805.3536 [hep-ph]].

\bibitem{Garayoa:2007fw}
  J.~Garayoa and T.~Schwetz,
  JHEP {\bf 0803}, 009 (2008)  [arXiv:0712.1453 [hep-ph]].

\bibitem{Akeroyd:2007zv}
  A.~G.~Akeroyd, M.~Aoki and H.~Sugiyama,
  Phys.\ Rev.\ D {\bf 77}, 075010 (2008)  [arXiv:0712.4019 [hep-ph]].

\bibitem{delAguila:2008cj}
  F.~del Aguila and J.~A.~Aguilar-Saavedra,
  Nucl.\ Phys.\ B {\bf 813}, 22 (2009)
  [arXiv:0808.2468 [hep-ph]].

\bibitem{Perez:2008zc}
  P.~Fileviez Perez, T.~Han, G.~Y.~Huang, T.~Li and K.~Wang,
  Phys.\ Rev.\ D {\bf 78}, 071301 (2008)  [arXiv:0803.3450 [hep-ph]].

\bibitem{Franceschini:2008pz}
  R.~Franceschini, T.~Hambye and A.~Strumia,
  Phys.\ Rev.\ D {\bf 78} (2008) 033002
  [arXiv:0805.1613 [hep-ph]];
  T.~Li and X.~-G.~He,
  Phys.\ Rev.\ D {\bf 80}, 093003 (2009)  [arXiv:0907.4193 [hep-ph]].

\bibitem{Arhrib:2009mz}
  A.~Arhrib, B.~Bajc, D.~K.~Ghosh, T.~Han, G.~-Y.~Huang, I.~Puljak and G.~Senjanovic,
  Phys.\ Rev.\ D {\bf 82}, 053004 (2010)
  [arXiv:0904.2390 [hep-ph]].

\bibitem{Han:2012vk}
  T.~Han, I.~Lewis, R.~Ruiz and Z.~-g.~Si,
  Phys.\ Rev.\ D {\bf 87}, 035011 (2013)
  [Erratum-ibid.\ D {\bf 87}, no. 3, 039906 (2013)]
  [arXiv:1211.6447 [hep-ph]].

\bibitem{Dev:2013oxa}
  C.~H.~Lee, P.~S. Bhupal Dev and R.~N.~Mohapatra,
  Phys.\ Rev.\ D {\bf 88}, 093010 (2013)
  [arXiv:1309.0774 [hep-ph]].

\bibitem{FileviezPerez:2010ch}
  P.~Fileviez Perez, T.~Han, S.~Spinner and M.~K.~Trenkel,
  JHEP {\bf 1101}, 046 (2011)  [arXiv:1010.5802 [hep-ph]].

\bibitem{Ding:2014nga}
  R.~Ding, Z.~L.~Han, Y.~Liao, H.~J.~Liu and J.~Y.~Liu,
  Phys.\ Rev.\ D {\bf 89} (2014) 115024
  [arXiv:1403.2040 [hep-ph]];
  C.~-S.~Chen and Y.~-J.~Zheng,
  arXiv:1312.7207 [hep-ph].

\bibitem{Gunion:1996pq}
  J.~F.~Gunion, C.~Loomis and K.~T.~Pitts,
  eConf C {\bf 960625}, LTH096 (1996)
  [hep-ph/9610237].

\bibitem{Akeroyd:2010ip}
  A.~G.~Akeroyd, C.~W.~Chiang and N.~Gaur,
  JHEP {\bf 1011}, 005 (2010)
  [arXiv:1009.2780 [hep-ph]].

\bibitem{Chun:2003ej}
  E.~J.~Chun, K.~Y.~Lee and S.~C.~Park,
  Phys.\ Lett.\ B {\bf 566}, 142 (2003)
  [hep-ph/0304069].

\bibitem{Akeroyd:2011zza}
  A.~G.~Akeroyd and H.~Sugiyama,
  Phys.\ Rev.\ D {\bf 84}, 035010 (2011)
  [arXiv:1105.2209 [hep-ph]].

\bibitem{Chun:2012zu}
  E.~J.~Chun and P.~Sharma,
  JHEP {\bf 1208}, 162 (2012)
  [arXiv:1206.6278 [hep-ph]].

\bibitem{Chun:2013vma}
  E.~J.~Chun and P.~Sharma,
  Phys.\ Lett.\ B {\bf 728}, 256 (2014)
  [arXiv:1309.6888 [hep-ph]].

\bibitem{Akeroyd:2012nd}
  A.~G.~Akeroyd, S.~Moretti and H.~Sugiyama,
  Phys.\ Rev.\ D {\bf 85}, 055026 (2012)
  [arXiv:1201.5047 [hep-ph]].

\bibitem{Akeroyd:2005gt}
  A.~G.~Akeroyd and M.~Aoki,
  Phys.\ Rev.\ D {\bf 72}, 035011 (2005)
  [hep-ph/0506176].

\bibitem{Aoki:2011pz}
  M.~Aoki, S.~Kanemura and K.~Yagyu,
  Phys.\ Rev.\ D {\bf 85}, 055007 (2012)
  [arXiv:1110.4625 [hep-ph]].

\bibitem{Yagyu:2012qp}
  K.~Yagyu,
  arXiv:1204.0424 [hep-ph].

\bibitem{Arhrib:2011uy}
  A.~Arhrib, R.~Benbrik, M.~Chabab {\it et al.},
  Phys.\ Rev.\ D {\bf 84}, 095005 (2011)
  [arXiv:1105.1925 [hep-ph]].

\bibitem{Aoki:2012jj}
  M.~Aoki, S.~Kanemura, M.~Kikuchi and K.~Yagyu,
  Phys.\ Rev.\ D {\bf 87}, 015012 (2013)
  [arXiv:1211.6029 [hep-ph]].

\bibitem{Chun:2012jw}
  E.~J.~Chun, H.~M.~Lee and P.~Sharma,
  JHEP {\bf 1211}, 106 (2012)
  [arXiv:1209.1303 [hep-ph]].

\bibitem{Tortola:2012te}
  D.~V.~Forero, M.~Tortola and J.~W.~F.~Valle,
  Phys.\ Rev.\ D {\bf 86}, 073012 (2012)
  [arXiv:1205.4018 [hep-ph]].

\bibitem{Lesgourgues:2012uu}
  J.~Lesgourgues and S.~Pastor,
  Adv.\ High Energy Phys.\  {\bf 2012}, 608515 (2012)
  [arXiv:1212.6154 [hep-ph]].

\bibitem{Ade:2013zuv}
  P.~A.~R.~Ade {\it et al.}  [Planck Collaboration],
  Astron.\ Astrophys.\  (2014)
  [arXiv:1303.5076 [astro-ph.CO]].

\bibitem{Bilenky:2012qi}
  S.~M.~Bilenky and C.~Giunti,
  Mod.\ Phys.\ Lett.\ A {\bf 27}, 1230015 (2012)
  [arXiv:1203.5250 [hep-ph]].

\bibitem{Auger:2012ar}
  M.~Auger {\it et al.}  [EXO Collaboration],
  Phys.\ Rev.\ Lett.\  {\bf 109}, 032505 (2012)
  [arXiv:1205.5608 [hep-ex]].

\bibitem{Aseev:2011dq}
  V.~N.~Aseev {\it et al.}  [Troitsk Collaboration],
  Phys.\ Rev.\ D {\bf 84}, 112003 (2011)
  [arXiv:1108.5034 [hep-ex]].

\bibitem{Drexlin:2013lha}
  G.~Drexlin, V.~Hannen, S.~Mertens and C.~Weinheimer,
  Adv.\ High Energy Phys.\  {\bf 2013}, 293986 (2013)
  [arXiv:1307.0101 [physics.ins-det]].

\bibitem{Abada:2007ux}
  A.~Abada, C.~Biggio, F.~Bonnet, M.~B.~Gavela and T.~Hambye,
  JHEP {\bf 0712}, 061 (2007)
  [arXiv:0707.4058 [hep-ph]].

\bibitem{Akeroyd:2009nu}
  A.~G.~Akeroyd, M.~Aoki and H.~Sugiyama,
  Phys.\ Rev.\ D {\bf 79}, 113010 (2009)
  [arXiv:0904.3640 [hep-ph]].

\bibitem{Fukuyama:2009xk}
  T.~Fukuyama, H.~Sugiyama and K.~Tsumura,
  JHEP {\bf 1003}, 044 (2010)
  [arXiv:0909.4943 [hep-ph]].

\bibitem{Adam:2013mnn}
  J.~Adam {\it et al.}  [MEG Collaboration],
  Phys.\ Rev.\ Lett.\  {\bf 110}, no. 20, 201801 (2013)
  [arXiv:1303.0754 [hep-ex]].

\bibitem{Kanemura:2012rs}
  S.~Kanemura and K.~Yagyu,
  Phys.\ Rev.\ D {\bf 85}, 115009 (2012)
  [arXiv:1201.6287 [hep-ph]].

\bibitem{Akeroyd:2012ms}
  A.~G.~Akeroyd and S.~Moretti,
  Phys.\ Rev.\ D {\bf 86}, 035015 (2012)
  [arXiv:1206.0535 [hep-ph]].

\bibitem{Arbabifar:2012bd}
  F.~Arbabifar, S.~Bahrami and M.~Frank,
  Phys.\ Rev.\ D {\bf 87}, 015020 (2013)
  [arXiv:1211.6797 [hep-ph]].

\bibitem{Dev:2013ff}
  P.~S.~Bhupal Dev, D.~K.~Ghosh, N.~Okada and I.~Saha,
  JHEP {\bf 1303}, 150 (2013)
  [Erratum-ibid.\  {\bf 1305}, 049 (2013)]
  [arXiv:1301.3453].

\bibitem{Chabab:2014ara}
  M.~Chabab, M.~C.~Peyranere and L.~Rahili,
  arXiv:1407.1797 [hep-ph].

\bibitem{Aad:2014eha}
  G.~Aad {\it et al.}  [ ATLAS Collaboration],
  arXiv:1408.7084 [hep-ex].

\bibitem{Khachatryan:2014ira}
  V.~Khachatryan {\it et al.}  [CMS Collaboration],
  arXiv:1407.0558 [hep-ex].

\bibitem{ATLAS:2012hi}
  G.~Aad {\it et al.}  [ATLAS Collaboration],
  Eur.\ Phys.\ J.\ C {\bf 72}, 2244 (2012)
  [arXiv:1210.5070 [hep-ex]].

\bibitem{Chatrchyan:2012ya}
  S.~Chatrchyan {\it et al.}  [CMS Collaboration],
  Eur.\ Phys.\ J.\ C {\bf 72}, 2189 (2012)
  [arXiv:1207.2666 [hep-ex]].

\bibitem{ATLAS:2014kca}
  G.~Aad {\it et al.}  [ ATLAS Collaboration],
  arXiv:1412.0237 [hep-ex].

\bibitem{Kanemura:2013vxa}
  S.~Kanemura, K.~Yagyu and H.~Yokoya,
  Phys.\ Lett.\ B {\bf 726}, 316 (2013)
  [arXiv:1305.2383 [hep-ph]].

\bibitem{Kanemura:2014goa}
  S.~Kanemura, M.~Kikuchi, K.~Yagyu and H.~Yokoya,
  arXiv:1407.6547 [hep-ph].

\bibitem{Kanemura:2014ipa}
  S.~Kanemura, M.~Kikuchi, K.~Yagyu and H.~Yokoya,
  arXiv:1412.7603 [hep-ph].

\bibitem{Schael:2006cr}
  S.~Schael {\it et al.}  [ALEPH and DELPHI and L3 and OPAL and LEP Working Group for Higgs Boson Searches Collaborations],
  Eur.\ Phys.\ J.\ C {\bf 47} (2006) 547
  [hep-ex/0602042].

\bibitem{Abbiendi:2013hk}
  G.~Abbiendi {\it et al.}  [ALEPH and DELPHI and L3 and OPAL and LEP Collaborations],
  Eur.\ Phys.\ J.\ C {\bf 73} (2013) 2463
  [arXiv:1301.6065 [hep-ex]].

\bibitem{Melfo:2011nx}
  A.~Melfo, M.~Nemevsek, F.~Nesti, G.~Senjanovic and Y.~Zhang,
  Phys.\ Rev.\ D {\bf 85}, 055018 (2012)
  [arXiv:1108.4416 [hep-ph]].

\bibitem{feynrules}
  N.~D.~Christensen and C.~Duhr,
  Comput.\ Phys.\ Commun.\  {\bf 180}, 1614 (2009)
  [arXiv:0806.4194 [hep-ph]];
  A.~Alloul, N.~D.~Christensen, C.~Degrande, C.~Duhr and B.~Fuks,
  Comput.\ Phys.\ Commun.\  {\bf 185}, 2250 (2014)
  [arXiv:1310.1921 [hep-ph]].

\bibitem{Degrande:2011ua}
  C.~Degrande, C.~Duhr, B.~Fuks, D.~Grellscheid, O.~Mattelaer and T.~Reiter,
  Comput.\ Phys.\ Commun.\  {\bf 183}, 1201 (2012)
  [arXiv:1108.2040 [hep-ph]].

\bibitem{MG5}
  J.~Alwall, M.~Herquet, F.~Maltoni, O.~Mattelaer and T.~Stelzer,
  JHEP {\bf 1106}, 128 (2011)
  [arXiv:1106.0522 [hep-ph]];
  J.~Alwall, R.~Frederix, S.~Frixione, V.~Hirschi, F.~Maltoni, O.~Mattelaer, H.-S.~Shao and T.~Stelzer {\it et al.},
  JHEP {\bf 1407}, 079 (2014)
  [arXiv:1405.0301 [hep-ph]].

\bibitem{Sjostrand:2006za}
  T.~Sjostrand, S.~Mrenna and P.~Z.~Skands,
  JHEP {\bf 0605}, 026 (2006)
  [hep-ph/0603175].

\bibitem{Delphes}
  S.~Ovyn, X.~Rouby and V.~Lemaitre,
  arXiv:0903.2225 [hep-ph];
  J.~de Favereau {\it et al.}  [DELPHES 3 Collaboration],
  JHEP {\bf 1402}, 057 (2014)
  [arXiv:1307.6346 [hep-ex]].

\bibitem{Conte:2012fm}
  E.~Conte, B.~Fuks and G.~Serret,
  Comput.\ Phys.\ Commun.\  {\bf 184}, 222 (2013)
  [arXiv:1206.1599 [hep-ph]].

\bibitem{Nadolsky:2008zw}
  P.~M.~Nadolsky, H.~-L.~Lai, Q.~-H.~Cao, J.~Huston, J.~Pumplin, D.~Stump, W.~-K.~Tung and C.~-P.~Yuan,
  Phys.\ Rev.\ D {\bf 78} (2008) 013004
  [arXiv:0802.0007 [hep-ph]].

\bibitem{Muhlleitner:2003me}
  M.~Muhlleitner and M.~Spira,
  Phys.\ Rev.\ D {\bf 68}, 117701 (2003)
  [hep-ph/0305288].

\bibitem{Kadastik:2007yd}
  M.~Kadastik, M.~Raidal and L.~Rebane,
  Phys.\ Rev.\ D {\bf 77}, 115023 (2008)
  [arXiv:0712.3912 [hep-ph]].

\bibitem{Fuchs:2014ola}
  E.~Fuchs, S.~Thewes and G.~Weiglein,
  arXiv:1411.4652 [hep-ph].

\bibitem{Cao:2003tr}
  Q.~H.~Cao, S.~Kanemura and C.~P.~Yuan,
  Phys.\ Rev.\ D {\bf 69}, 075008 (2004)
  [hep-ph/0311083].

\bibitem{Glover:1987nx}
  E.~W.~N.~Glover and J.~J.~van der Bij,
  Nucl.\ Phys.\ B {\bf 309}, 282 (1988).

\bibitem{Plehn:1996wb}
  T.~Plehn, M.~Spira and P.~M.~Zerwas,
  Nucl.\ Phys.\ B {\bf 479}, 46 (1996)
  [Erratum-ibid.\ B {\bf 531}, 655 (1998)]
  [hep-ph/9603205].

\bibitem{Dawson:1998py}
  S.~Dawson, S.~Dittmaier and M.~Spira,
  Phys.\ Rev.\ D {\bf 58}, 115012 (1998)
  [hep-ph/9805244].

\bibitem{Djouadi:1999rca}
  A.~Djouadi, W.~Kilian, M.~Muhlleitner and P.~M.~Zerwas,
  Eur.\ Phys.\ J.\ C {\bf 10}, 45 (1999)
  [hep-ph/9904287].

\bibitem{deFlorian:2013jea}
  D.~de Florian and J.~Mazzitelli,
  Phys.\ Rev.\ Lett.\  {\bf 111}, 201801 (2013)
  [arXiv:1309.6594 [hep-ph]].

\bibitem{Baglio:2012np}
  J.~Baglio, A.~Djouadi, R.~Gr\"{o}ber, M.~M.~M\"{u}hlleitner, J.~Quevillon and M.~Spira,
  JHEP {\bf 1304}, 151 (2013)
  [arXiv:1212.5581 [hep-ph]].

\bibitem{Baglio:HH}
  J.~Baglio,
  Pos DIS {\bf 2014}, 120 (2014)
  [arXiv:1407.1045 [hep-ph]];
  J.~Baglio,
  arXiv:1408.6066 [hep-ph].

\bibitem{Cao:2013si}
  J.~Cao, Z.~Heng, L.~Shang, P.~Wan and J.~M.~Yang,
  JHEP {\bf 1304}, 134 (2013)
  [arXiv:1301.6437 [hep-ph]].

\bibitem{Ellwanger:2013ova}
  U.~Ellwanger,
  JHEP {\bf 1308}, 077 (2013)
  [arXiv:1306.5541, arXiv:1306.5541 [hep-ph]].

\bibitem{Asakawa:2010xj}
  E.~Asakawa, D.~Harada, S.~Kanemura, Y.~Okada and K.~Tsumura,
  Phys.\ Rev.\ D {\bf 82}, 115002 (2010)
  [arXiv:1009.4670 [hep-ph]].

\bibitem{Arhrib:2008pw}
  A.~Arhrib, R.~Benbrik, R.~B.~Guedes and R.~Santos,
  Phys.\ Rev.\ D {\bf 78}, 075002 (2008)
  [arXiv:0805.1603 [hep-ph]].

\bibitem{Arhrib:2009hc}
  A.~Arhrib, R.~Benbrik, C.~-H.~Chen, R.~Guedes and R.~Santos,
  JHEP {\bf 0908}, 035 (2009)
  [arXiv:0906.0387 [hep-ph]].

\bibitem{Hespel:2014sla}
  B.~Hespel, D.~Lopez-Val and E.~Vryonidou,
  arXiv:1407.0281 [hep-ph].

\bibitem{Kribs:2012kz}
  G.~D.~Kribs and A.~Martin,
  Phys.\ Rev.\ D {\bf 86}, 095023 (2012)
  [arXiv:1207.4496 [hep-ph]].

\bibitem{Heng:2013cya}
  Z.~Heng, L.~Shang, Y.~Zhang and J.~Zhu,
  JHEP {\bf 1402}, 083 (2014)
  [arXiv:1312.4260 [hep-ph]].

\bibitem{Dawson:2012mk}
  S.~Dawson, E.~Furlan and I.~Lewis,
  Phys.\ Rev.\ D {\bf 87}, 014007 (2013)
  [arXiv:1210.6663 [hep-ph]].

\bibitem{Chen:2014xwa}
  C.~-Y.~Chen, S.~Dawson and I.~M.~Lewis,
  [arXiv:1406.3349 [hep-ph]].

\bibitem{Dib:2005re}
  C.~O.~Dib, R.~Rosenfeld and A.~Zerwekh,
  JHEP {\bf 0605}, 074 (2006)
  [hep-ph/0509179].

\bibitem{Dolan:2012ac}
  M.~J.~Dolan, C.~Englert and M.~Spannowsky,
  Phys.\ Rev.\ D {\bf 87}, no. 5, 055002 (2013)
  [arXiv:1210.8166 [hep-ph]].

\bibitem{No:2013wsa}
  J.~M.~No and M.~Ramsey-Musolf,
  Phys.\ Rev.\ D {\bf 89}, 095031 (2014)
  [arXiv:1310.6035 [hep-ph]].

\bibitem{Grober:2010yv}
  R.~Grober and M.~Muhlleitner,
  JHEP {\bf 1106}, 020 (2011)
  [arXiv:1012.1562 [hep-ph]].

\bibitem{Gillioz:2012se}
  M.~Gillioz, R.~Grober, C.~Grojean, M.~Muhlleitner and E.~Salvioni,
  JHEP {\bf 1210}, 004 (2012)
  [arXiv:1206.7120 [hep-ph]].

\bibitem{Pierce:2006dh}
  A.~Pierce, J.~Thaler and L.~-T.~Wang,
  JHEP {\bf 0705}, 070 (2007)
  [hep-ph/0609049].

\bibitem{Contino:2012xk}
  R.~Contino, M.~Ghezzi, M.~Moretti, G.~Panico, F.~Piccinini and A.~Wulzer,
  JHEP {\bf 1208}, 154 (2012)
  [arXiv:1205.5444 [hep-ph]].

\bibitem{Liu:2013woa}
  J.~Liu, X.~-P.~Wang and S.~-H.~Zhu,
  arXiv:1310.3634 [hep-ph].

\bibitem{Godunov:2014waa}
  S.~I.~Godunov, M.~I.~Vysotsky and E.~V.~Zhemchugov,
  arXiv:1408.0184 [hep-ph].

\bibitem{Dolan:2012rv}
  M.~J.~Dolan, C.~Englert and M.~Spannowsky,
  JHEP {\bf 1210}, 112 (2012)
  [arXiv:1206.5001 [hep-ph]].

\bibitem{Gouzevitch:2013qca}
  M.~Gouzevitch, A.~Oliveira, J.~Rojo, R.~Rosenfeld, G.~P.~Salam and V.~Sanz,
  JHEP {\bf 1307}, 148 (2013)
  [arXiv:1303.6636 [hep-ph]].

\bibitem{Papaefstathiou:2012qe}
  A.~Papaefstathiou, L.~L.~Yang and J.~Zurita,
  Phys.\ Rev.\ D {\bf 87}, 011301 (2013)
  [arXiv:1209.1489 [hep-ph]].

\bibitem{Goertz:2013kp}
  F.~Goertz, A.~Papaefstathiou, L.~L.~Yang and J.~Zurita,
  JHEP {\bf 1306}, 016 (2013)
  [arXiv:1301.3492 [hep-ph]].

\bibitem{deLima:2014dta}
  D.~E.~Ferreira de Lima, A.~Papaefstathiou and M.~Spannowsky,
  [arXiv:1404.7139 [hep-ph]].

\bibitem{Barger:2013jfa}
  V.~Barger, L.~L.~Everett, C.~B.~Jackson and G.~Shaughnessy,
  Phys.\ Lett.\ B {\bf 728}, 433 (2014)
  [arXiv:1311.2931 [hep-ph]].

\bibitem{han:2015xxx}
  Z.~L.~Han, R.~Ding, and Y.~Liao, in preparation.

\end{thebibliography}
\end{document}